\documentclass[fleqn,useAMS,usenatbib]{mnras}

\usepackage{graphicx,amssymb,multirow,units,float,lpic}
\usepackage{lscape}
\usepackage{comment}
\usepackage{tabularx, adjustbox, booktabs, array}
\usepackage{graphics}
\usepackage{amsmath}
\usepackage{afterpage,rotating,url}
\usepackage{color}
\usepackage{caption}
\usepackage[T1]{fontenc}
\usepackage{ae,aecompl}

\definecolor{purple}{rgb}{0.75,0.0,0.75}

\newcommand{\var}{ \mathrm{var} }
\newcommand{\msun}{\mbox{$\rm M_{\odot}$}}        
\newcommand{\lsub}{\mbox{$L_{850}$}} 
\newcommand{\lcoa}{\mbox{$L^{\prime}_{\rm CO}$}}
\newcommand{\lcob}{\mbox{$L^{\prime}_{21}$}}
\newcommand{\lci}{\mbox{$L^{\prime}_{\rm C\,I}$}}
\newcommand{\Md}{\mbox{$M_{\rm dust}$}}
\newcommand{\Ms}{\mbox{$M_{\ast}$}}
\newcommand{\Mh}{\mbox{$M_{\rm H2}$}}

\newcommand{\Mmol}{\mbox{$M_{\rm mol}$}}
\newcommand{\Lir}{\mbox{$L_{\rm IR}$}}
\newcommand{\COa}{\mbox{$^{12}$CO(1--0)}}

\newcommand{\CI}{\mbox{C\,{\sc i}}}
\newcommand{\HI}{\mbox{H\,{\sc i}}}
\newcommand{\mol}{\mbox{H$_2$}}
\newcommand{\CIfull}{\mbox{[C\,{\sc i}]($^3P_1$--$^3P_0$)}}
\newcommand{\Xci}{\mbox{$X_{\rm C\,I}$}}

\newcommand{\XciR}{\mbox{$X^{\rm R}_{\rm C\,I}$}}
\newcommand{\kd}{\mbox{$\kappa_{850}$}}
\newcommand{\kh}{\mbox{$\kappa_{\rm H}$}}
\newcommand{\cs}{\mbox{$\sigma^{\prime}_{850}$}}

\newcommand{\kms}{\mbox{km\,s$^{-1}$}}
\newcommand{\td}{\mbox{$T_{\rm d}$}}
\newcommand{\tx}{\mbox{$T_{\rm x}$}}
\newcommand{\mwtd}{\mbox{$T_{\rm mw}$}}

\newcommand{\aco}{\mbox{$\alpha_{\rm CO}$}}
\newcommand{\acoR}{\mbox{$\alpha^{\rm R}_{\rm CO}$}}

\newcommand{\aci}{\mbox{$\alpha_{\rm C\,I}$}}
\newcommand{\asub}{\mbox{$\alpha_{850}$}}

\newcommand{\gdr}{\mbox{$\delta_{\rm GDR}$}}
\newcommand{\gdrR}{\mbox{$\delta^{\rm R}_{\rm GDR}$}}
\newcommand{\Q}{\mbox{$Q_{10}$}}
\newcommand{\tk}{\mbox{$T_{\rm k}$}}
\newcommand{\zcr}{\mbox{$\zeta_{\rm CR}$}}

\newcommand{\aunit}{\mbox{M$_{\odot}$\,({\sc k}\,km\,s$^{-1}$\,pc$^2$)$^{-1}$}}
\newcommand{\kunit}{\mbox{m$^2$\,kg$^{-1}$}}
\newcommand{\khunit}{\mbox{kg\,m$^{-2}$}}
\newcommand{\asunit}{\mbox{W\,Hz$^{-1}$\,M$_{\odot}^{-1}$}}
\newcommand{\Lhi}{\mbox{log $L_{\rm IR}>11$}}
\newcommand{\Llo}{\mbox{log $L_{\rm IR}<11$}}

\newcommand{\fhi}{\mbox{$f_{\rm H\,I}$}}

\newcommand{\CIcor}{\mbox{C\,{\sc i}$^{\rm cor}$}}

\newcommand{\ms}{\rm MS}

\defcitealias{Draine2007_kappa}{DL07}	
\newcommand{\mic}{$\mu$m}
\newcommand{\cc}{\mbox{cm$^{-3}$}}

\newcommand{\Ntot}{407}
\newcommand{\Nad}{326}
\newcommand{\Nadhi}{240}
\newcommand{\NXa}{109}
\newcommand{\NXahi}{97}
\newcommand{\NXd}{140}
\newcommand{\NXdhi}{128}
\newcommand{\NdaX}{101}
\newcommand{\NdaXhi}{90}

\newcommand{\xcir}{1.6}

\newcommand{\khr}{1990}

\newcommand{\acir}{17.0}
\newcommand{\acorm}{4.0}
\newcommand{\asubr}{6.9}
\newcommand{\eacir}{0.3}
\newcommand{\eacorm}{0.1}
\newcommand{\easubr}{0.1}

\newcommand{\tmwMS}{23.0}
\newcommand{\etmwMS}{0.4}
\newcommand{\tmwSMG}{30.1}
\newcommand{\etmwSMG}{0.7}

\newcolumntype{@}{>{\global\let\currentrowstyle\relax}}
\newcolumntype{^}{>{\currentrowstyle}}

\title[Cross-calibration of global $\rm H_2$ gas mass tracers]{Dust, CO and [C\,{\sc i}]: Cross-calibration
  of molecular gas mass tracers in metal-rich galaxies across cosmic time}

\author[L.~Dunne et al.]{L.~Dunne,$^{\! 1}$
S.\,J.~Maddox,$^{\! 1}$
P.\,P.~Papadopoulos,$^{\! 3, 2, 1}$ R.\,J.~Ivison$^{4}$
and H.\,L.~Gomez$^{1}$\thanks{E-mail:GomezH@cardiff.ac.uk} 
\vspace*{1mm}\\
$^1$School of Physics \&\ Astronomy, Cardiff University, Queens Buildings, The Parade, Cardiff CF24~3AA\\
$^2$Dept of Physics, Section of Astrophysics, Astronomy and Mechanics, Aristotle University of Thessaloniki,
GR-54124, Greece\\
$^3$Research Center for Astronomy, Academy of Athens, Soranou Efesiou 4, GR-11527 Athens, Greece\\
$^4$European Southern Observatory, Karl-Schwarzschild-Strasse~2, D-85748 Garching, Germany
}

\date{
Submitted to MNRAS Main Journal, 2022 Xxxxxx; Manuscript ID: MN-22-XXXX-MJ
}

\pubyear{2022}

\begin{document}
\label{firstpage}
\pagerange{\pageref{firstpage}--\pageref{lastpage}} 
\maketitle

\begin{abstract}
  We present a self-consistent cross-calibration of the three main
  molecular gas mass tracers in galaxies, namely the \COa,
  \CIfull\ lines, and the submm dust continuum emission, using a
  sample of 407 galaxies, ranging from local disks to
  submillimetre-selected galaxies (SMGs) up to $z\approx 6$.  A
  Bayesian statistical method is used to produce galaxy-scale universal
  calibrations of these molecular gas indicators, 
  that hold over 3--4 orders of magnitude in infrared
  luminosity, \Lir.  Regarding the dust continuum, we use a
  mass-weighted dust temperature,
  \mwtd, determined using new empirical relations between
  temperature and luminosity.  We find the average L/\Mmol\ gas mass
  conversion factors (including He) to be
  $\asub=\asubr\times10^{12}\,\rm W\,Hz^{-1}\,M_{\odot}^{-1}$, \aco
  =\acorm\,\aunit\ and \aci=\acir\,\aunit, based on the assumption that the mean dust properties of the sample (\kh = gas-to-dust ratio/dust emissivity) will be similar to those of local metal rich galaxies and the Milky Way. The tracer with the least intrinsic
  scatter is [\CI](1--0), while CO(1--0) has the
  highest. The conversion factors show a weak but significant
  correlation with \Lir\ which is not apparent when \mwtd\ is held
  constant. Assuming dust properties typical of metal-rich galaxies,
  we infer a neutral carbon abundance $\Xci=\rm [C^0/\mol]=1.6\times 10^{-5}$,
  similar to that in the Milky Way.  We find no evidence for
  bi-modality of \aco\ between main-sequence (MS)
  galaxies and those with extreme star-formation intensity,
  i.e.\ ultraluminous infrared galaxies (ULIRGs) and SMGs. The means of the
  three conversion factors are found to be similar between MS galaxies and
  ULIRGs/SMGs, to within 10--20 per cent. The overarching conclusion of our
  work is that, for metal-rich galaxies, near-universal average values
  for \aco, \Xci\ and \kh\ are adequate for global molecular gas
  estimates within the expected uncertainties. The 1$\sigma$ scatter in our optimised values for \aco, \Xci\ and \kh\ are 0.14, 0.11 and 0.15 dex respectively.

  \end{abstract}


\section{Introduction}

The cosmic star-formation rate (SFR) density has declined by more than
an  order of  magnitude during  the  past $\approx  8$\,Gyr of  cosmic
history  \citep{Lilly1996,Madau1996,Madau2014}.  The  driver  of  star
formation is the molecular gas  supply in galaxies,
and indeed  the SFR--stellar mass (SFR--$M_\star$)  relationship known
as  the galaxy  main  sequence  (MS) is  purely  a  by-product of  the
relationship between SFR  and molecular gas \citep[e.g.][]{baker2022},
for  unperturbed  galaxies  with  significant gas  reserves.  A  major
observational goal  is to produce  a combined census of  the molecular
gas --  the `potential for future  star formation' -- and  the stellar
content --  the `record of  past star  formation' -- over  this period
\citep[e.g.][]{Keres2003,Dunne2003,Dunne2011,Zwaan2004, 
  Zafar2013,Walter2014,Decarli2016,Saintonge2017,Driver2018,Rhee2018,Decarli2019,
  Riechers2019}.

The  molecular gas  fraction of  a galaxy  is a  crucial component  in
models of  galaxy formation \citep[e.g.][]{Obreschkow2009,Popping2014,
  Lagos2015,Chen2018} and  thus measurements of $\rm  H_2$ and stellar
mass over large representative galaxy samples are key requirements for
understanding  how  galaxies  have  transformed  from  clouds  of  gas
residing  in dark  matter haloes  into the  regular agglomerations  of
stars  we  see  in  the  local  Universe.   While  it  is  clear  that
CO(1--0)-luminous gas is the phase linked  with star formation
\citep[e.g.][]{Wong2002},  observations  of  molecules  with  higher  critical  densities
(e.g.\ HCN) revealed that it is the dense H$_2$ gas phase
($n> 10^4$\,cm$^{-3}$) that correlates most tightly and  linearly with tracers of star-formation \citep{Gao2004}.

Atomic hydrogen (\HI), on the other hand, constitutes a longer-term gas
reservoir  for star formation, where under certain conditions of pressure, far-UV radiation field, density and metallicity, a phase transition \HI\ $\rightarrow$  H$_2$ takes place, catalysed by dust grains \citep[e.g.][]{ElmegreenH21993,PPP2002,Blitz2006}: a picture supported
by      numerous      observations      \cite[e.g.][]{Honma1995,Leroy2008,Bigiel2008,Schruba2011}. This  transition   occurs in the  inner  \HI\ distribution of galaxies, in the cold neutral medium (CNM: $n\sim  50$--100\,cm$^{-3}$,   $T_{\rm kin}\sim
100$--200\,{\sc k}), meanwhile pure \HI\ gas often  extends many
optical     radii     beyond     the     luminous     stellar     disk
\citep[e.g.][]{peroux2020}, where it can be found 
  concomitant with cold dust (e.g. \citealp{thomas2002}). 

Unlike \HI\  and its  hyperfine line  emission at 21\,cm, the  H$_2$ molecule in its S(0):$J=2-0$ transition at 28$\mu$m (the least excitation-demanding $\rm H_2$ line) is essentially invisible at temperatures typical of giant molecular clouds (10--20\,{\sc k}). This is because its $\Delta E/k_{\rm B}$$\sim $$510\,{\sc k}$, limits its  excitation and detetection only to shocked regions of molecular clouds, where gas temperatures  can rise past  $\sim 1000$\,{\sc k}, for small  ($\sim 1$--2 per cent)  gas mass fractions. Even then, to observe this \mol\ line at 28\,$\mu$m requires space-borne telescopes.
   
For these reasons  the rotational  transitions  of CO  (the  next most  abundant
molecule with $\rm [CO/H_2]\sim 10^{-4}$)  are commonly used to trace $\rm H_2$
gas, with the  lowest transition ($^{12}$CO $J=1$--0) being  the most established
tracer.  Its $E_{10}/k_{\rm B}\sim 5.5$\,{\sc k} ensures a well-populated
upper level even in the coldest gas, while its low critical density,
$n_{\rm cr}\sim  400$\,cm$^{-3}$,  ensures  its   excitation  even  at  low
densities\footnote{Because  the  CO(1--0)  line  is  typically
  optically  thick,  with  $\tau_{10}\sim  5$--10 (\citealp[e.g.][]{BS1996,Papadopoulos2012}: their Eqn. 11),  the  {\it effective}
  critical density is lower still: $n_{\rm cr}(\beta_{10})= \beta_{10}
  n_{\rm crit}\sim     40$--80\,cm$^{-3}$,  where $\beta_{10}=(1-e^{-\tau
    _{10}})/\tau_{10}$ is the line escape probability.}.

The CO(1--0) line has significant optical  depths in the typically macro-turbulent $\rm  H_2$ gas, though these arise locally within the velocity-coherent gas cells allowing the CO emission to trace gas
mass throughout molecular clouds \citep[e.g.][]{Dickman1986}. The conversion factor, \aco, in the relation $\Mh=\aco\lcoa$ cannot be determined using standard optically thin line formation physics due to the high line optical depths. This created the need for a \aco\ calibration as soon as  the ubiquity of CO line emission in
H$_2$ clouds was established. Observational and theoretical investigation of \aco\ suggests it is sensitive to metallicity, molecular gas surface density and kinematic state in galaxies \citep[e.g.][]{Pelupessy2009,Narayanan2011,Papadopoulos2012,Bolatto2013}. 

Three distinct problems are now recognised regarding  the use of CO as a global tracer of H$_2$ mass
in galaxies:
\begin{enumerate}
\item{The \aco\ factor is sensitive -- in a highly non-linear fashion -- to
  the   ISM   metallicity   and  ambient   far-UV   radiation   fields
  \citep[e.g.][]{Israel1997, Pak1998, Bolatto2013}.}
 
\item{Non-self-gravitating molecular  clouds --  and/or very different average ISM states in terms of average temperature and gas density range from those found  in spiral galaxies where  \aco\ was
  first   calibrated --   can   yield  systematically   different   \aco\
  factors. For  example, \aco $\sim  1/5-1/4\times$ Galactic
  was  initially  reported   for  a   sample   of  four  ULIRGs
  by \citet{Downes1998}.}
\item{Elevated  cosmic ray  (CR)  energy densities can   destroy  CO below a certain gas density threshold, leaving  behind more C-rich  gas. This density threshold depends on the CR energy density in a highly non-linear fashion, as explored by \citet{Bisbas2015}, who found that regions of CO suppression may occur even in moderately enhanced CR conditions if the gas density is low, while the very high CR energy densities expected in ultraluminous infrared galaxies (ULIRGs) may be partly compensated by higher gas densities in such starbursts. Modelling [\CI/CO] ratios as a function of CR, turbulence, gas density and metallicity is an active area of theoretical research  \citep[e.g.][]{Bisbas2015,Bisbas2017,Bisbas2021,Glover2016, Clark2019ci,Papadopoulos2018,Gong2020}.}
\end{enumerate}

In the distant Universe, additional problems arise. High-redshift galaxies  are often observed  solely in high-$J$ CO  lines ($J=3$--2
and higher), due to the observational challenge of observing the two low-$J$ CO  lines\footnote{Prior to the commissioning of its bands 1 and 2, low-$J$ lines from high-redshift galaxies are inaccessible  to the Atacama Large Millimetre Array (ALMA). The Jansky Very Large Array (JVLA), the Australia Telescope Compact Array  (ATCA) and the Greenbank Telescope (GBT), have in some cases been able to access the faint low-$J$ ($J_{\rm u}\leq2$) CO
lines, but it requires huge amounts of observing time in the best available weather.}.  Using the high-$J$ lines means that global CO($J+1,J$)/(1--0) ratios must be assumed
before  an \aco\  factor can  be used; given the wide range of  CO spectral-line energy distributions (SLEDs)
found  for LIRGs  for  $J=3$--2  and higher  \citep{PPP2012xco,Greve2014,Kamenetzky2016}, these assumptions come with large uncertainties.  Finally,  at the  highest redshifts  ($\ga $4),
low-$J$  CO lines  (and dust  emission) can  be severely  suppressed for cold gas (and dust) reservoirs  due to their low contrast against
the ambient, rest-frame cosmic microwave background \citep{daCunha2013,Zhang2016}.

In  principle,  radiative  transfer  models  of  well-sampled  CO  (and
$^{13}$CO) SLEDs can  yield \aco\ values appropriate  for a particular
galaxy (or even galaxy class) \citep[e.g.][]{PPP2012xco,PPP6240, Harrington2021}.  Nevertheless, the size of the CO line datasets per galaxy required to do this make it impractical (in terms of telescope time) to obtain  $M$(H$_2$)  for  large  galaxy samples. Amassing a large sample typically means only one or two lines can be gathered per galaxy, and thus a calibration  of   \aco\  and its uncertainties  remains very valuable. The only practical way to achieve this is to  cross-calibrate against the other galaxy-scale  $\rm H_2$  mass tracers.

Large-area far-infrared (FIR) and submillimetre (submm) surveys
\citep[e.g.][]{Armus2009,Eales2010,Vieira2010,Kennicutt2011,Oliver2012,Hodge2013}
ushered in a new era in which submm continuum emission from dust has been used widely as an
alternative tracer of \Mh, although it has been clear that
submm-derived dust masses ($\propto \lsub$) and CO-derived molecular
gas masses ($\propto \lcoa$) are tightly correlated ever since the
first statistical submm survey of 100 local FIR-bright galaxies
\citep[SLUGS --][]{Dunne2000}. The first suggestions to use dust as an
alternative to CO at high redshift
\citep[e.g.][]{Santini2010,Magdis2012,Scoville2014} were followed quickly by work
demonstrating its potential
\citep[e.g][]{Scoville2016,Hughes2017,Orellana2017}.

An advantage  of using submm continuum  emission from dust as  an $\rm
H_2$ gas tracer is that it becomes easier to measure at high redshift,
because  of  the  negative  $K$-correction  \citep[e.g.][]{blain1993},
while recent  technological advances made  it possible to  image areas
large enough to  be free of cosmic variance, leading  to the FIR/submm
detection of  many thousands  of galaxies by  the {\it  Herschel Space
  Observatory}, for  example.  The  use of  dust as  a gas  mass proxy
requires  an estimate  of  metallicity, since  the dust-to-gas  ratio,
\gdr,      is      roughly       proportional      to      metallicity
\citep[e.g.][]{MM2009,Magdis2012,    Sandstrom2013,Draine2014}.    The
appropriate        \gdr\       can        then       be        applied
\citep[e.g.][]{Valentino2018}. Whilst this requirement is often raised
as a  problem regarding  the use  of dust  as a  gas mass  tracer, its
dependence on metallicity is in fact  weaker than that of CO\footnote{Moreover,
since $\rm H_2$ cannot be traced (in bulk) by any of its own lines, regardless of which other tracer  (X) is  used 
(dust emission, CO, or $^{13}$CO, or \CI\ line emission), it will always be necessary to assume a $\rm[X/H_2]$ abundance in
  order to proceed to a final $\rm H_2$ gas mass estimate.}.

For  galaxies selected  at  FIR/submm/mm wavelengths,  it  is safe  to
assume that  the metallicity  will be  high, such  that \gdr\  will be
broadly similar  to those found  for local metal-rich spirals  and the
Milky                       Way                      \citep{Dunne2001,
  Draine2009,Magdis2012,Sandstrom2013,Rowlands2014,
  Yang2017,Berta2021}.  A  detailed discussion  of the  advantages and disadvantages  of using  dust  as a  tracer  of gas  can  be found  in
\citet{Genzel2015}  and \citet{Scoville2017}\footnote{Continuum dust emission does not yield information on kinematics, unlike spectral lines.}.

A third method of tracing molecular gas -- the use of atomic carbon
lines -- has come to the fore since ALMA became operational.  Its promise was recognised by
\citet{PPP2004} and its first application as a tracer for molecular gas mass in galaxies gave good results
\citep{Weiss2003,PPP&Greve2004}, implying that: a) the
\CIfull\ lines are optically thin for the bulk of $\rm H_2$ gas
\citep{PerezB2015} and b) atomic carbon is present throughout CO-rich
molecular cloud volumes.

The  latter contradicts the earlier simple plane-parallel PDR model where  atomic carbon  (and its  line
emission) occupied  only a thin layer, sandwiched between $\rm
C^{+}$ in  the outer and CO  in the inner regions  of FUV-illuminated
molecular  clouds
\citep{Tielens1985}.   However  observations   have  repeatedly shown
excellent concomitance of \CI\ line  emission with CO
line emission, by area and by velocity, and \CI\ shows a tighter correlation with $^{13}$CO than with
$^{12}$CO. \CI\  is now thought to  arise from same volume  as the CO,
with             similar             excitation             conditions
\citep[e.g.][]{Plume1999,Ikeda2002,Schneider2003,Beuther2014,PerezB2015}.
Moreover, it  may be  that \CI\  lines can  also trace
CO-dark  molecular  gas,  should  such  phase  exist  in  galaxies  in
significant amounts, e.g. due to CR-induced dissociation of CO to C (and O) \citep{Bisbas2015}.

Despite being much  fainter than the $\rm C^{+}$ line  at 158\,$\mu$m
(the  prime ISM  cooling line),  atomic carbon  lines do  hold certain
advantages,  namely: a)  they solely trace  $\rm  H_2$ gas, whereas the $\rm C^{+}$ line also traces the \HI\ and H\,{\sc ii} gas reservoirs, which can
be  significant, especially  in metal-poor  systems
\citep[e.g.][]{Madden1997,Liszt2011,PPP&Geach2012,PerezB2015,Clark2019ci}; b) the \CI\ lines  can
remain excited  for cold  gas (e.g.\  for  [\CI](1--0):  $E_{10}/k_{\rm B}\sim
24$\,{\sc k}) unlike the  $\rm C^{+}$ line, where the $\Delta E/k_{\rm B}\sim 92$\,{\sc k} will keep it very faint for cold gas; c) the frequencies of the two
\CI\  lines,  at 492  and  809\,GHz, remain  accessible  for
galaxies over a much larger redshift range (and thus cosmic volume) than
the $\rm  C^{+}$ line.  In the  latter case, its rest-frame  frequency,
$\rm \nu (C^+)  \sim 1.9$\,THz, means the $\rm  C^{+}$ line is observable
by ALMA's most sensitive receivers only at $z\ga 4$.

Nevertheless, the high rest-frame frequencies of the \CI\ line made
early observations  (and thus  any calibration  efforts) in  the local
Universe very  difficult. Initially  there had been  relatively little
observational  work outside  of  the Milky  Way,  largely confined  to
extreme    systems   such    as   quasars    and   starburst    nuclei
\citep[e.g.][]{White1994,Weiss2005,   Walter2011}.     These   studies
advocated  a  higher  carbon  abundance  for  these  extreme  systems,
$\Xci=\rm  [C^0/H_2]   =  5$--$12\times  10^{-5}$,  compared   to  the
$\Xci=1$--$2.5\times    10^{-5}$    seen     in    the    Milky    Way
\citep{Frerking1989}.

More recently,  {\it Herschel}  observed many  local galaxies  in \CI,
although the  [\CI](1--0) line was at  the edge of the  observable range
for the {\it  Herschel} Fourier Transform Spectrometer (FTS),  such that the
sensitivity  was  somewhat compromised.   As  a  result, most  of  the
detected galaxies  were either  ULIRGs, starbursts  or low-metallicity
dwarfs \citep{Kamenetzky2014,  Rosenberg2015,Lu2017,Jiao2017}. A small
sample of normal  disk galaxies was mapped  in \CI\ \citep[][hereafter
  J19                   --                  see                   also
  \citealp{Crocker2019}]{Jiao2019}.  J19  studied  the  spatial
distribution of \lci\ and \lcoa\ at a $\sim 1$-kpc scale in 15 local
galaxies.  They concluded that \CI\ is a good tracer of molecular gas,
in  the  sense that  it  correlates  well with  CO  and  the ratio 
\lci/\lcoa\  is  distributed   smoothly  across  galaxies.   Comparing
against  CO(1--0) maps  and the  independent estimates  of \aco\  from
\citet{Sandstrom2013},     these     resolved    studies     suggested
$\Xci$=1.3--2.5$\times10^{-5}$, similar  to the  range in  the Galaxy,
and that found by the absorber study of \citet{Heintz2020}.

\begin{table*}
\caption{\label{SampleT} Samples used for our comparisons.}
\begin{adjustbox}{center}
\begin{tabular}{lccccclcc}
\toprule
Sample & Selection  & $z$ & $N_{\rm CO}$ & $N_{\CI}$ &
                                                                 $N_{\rm
                                                           sub}$ &
                                                                     Notes & SF mode & References\\
name & $\lambda_{\rm obs}$ (\mic) &&&&&&&(see below)\\
\midrule
high-$z$ SMG & 850--2000 & 2--6 & 89    & 42     & 114  & Corrected for lensing & Both & $a$\\
Local SF  &           & 0     & 35    & 19     & 35 & \CI\ from FTS & MS  &$b$\\
(U)LIRGs & 60 & 0      & 85   & 19     & 114    & \CI\ from FTS  & Both  & $c$\\
$z=1$       & 850 & 1        & 11     & 18      & 9 & CO(2--1) & MS & $d$\\
$z=0.35$    &  250 & 0.35    & 12    & 12     & 12     &  & MS & $e$\\
$0.04<z<0.3$ & 160 & 0-0.3    & 48    & 0      & 54     & VALES & MS & $f$\\
\bottomrule
\end{tabular}
\end{adjustbox}
\flushleft{In columns 4--6, $N$ refers to the number of detections in each of the tracers.\\
$a$: \citet{Chapman2005,Chapman2010,Weiss2005,Weiss2013,Coppin2006,Hainline2006,Kovacs2006,Daddi2009,Wu2009,Carilli2010,Carilli2011,Engel2010,Harris2010,Ivison2010,Ivison2011,Ivison2013,Frayer2011,Frayer2018,Riechers2011,Riechers2013,Riechers2020aspecs,Walter2011,Walter2012,Cox2011,Danielson2011,Lestrade2011,McKean2011,Magnelli2012,Thomson2012,AZ2013,Bothwell2013,Bothwell2017,Bussmann2013,Bussmann2015,Emonts2013,Sharon2013,Sharon2016,Cooray2014,Messias2014,Messias2019,Negrello2014,Negrello2017,Swinbank2014,Tan2014,Canameras2015,Dye2015,Aravena2016,Scoville2016,Spilker2016,Huynh2017,Oteo2017,OteoGRH,Popping2017,Falgarone2017,Wong2017,Yang2017,Yang2019,Bethermin2018,Enia2018,Pavesi2018coldz,Pavesi2018cosmos,Perna2018,Valentino2018,Valentino2020,Wang2018,Dannerbauer2018,GomezG2019,Jin2019,Kaasinen2019,Leung2019,Nesvadba2019,Bakx2020z,Boogaard2020,Berta2021,Ciesla2020,Drew2020,Neri2020,Harrington2021}.\\
$b$: \citet{Mirabel1990,Tinney1990,Young1995,Casoli1996,Zhu1999,Curran2000,Dunne2000,Dunne2001,Gao2004,Thomas2004,Stevens2005,Albrecht2007,Kuno2007,Ao2008,Baan2008,Young2008,Galametz2011,Koda2011,Iono2012,Pappalardo2012,Schruba2012,Alatalo2013,PS2013,Wong2013,Ueda2014,Liu2015,Rosenberg2015,Bolatto2017,Cao2017,Jiao2019,Jiao2021,Clark2018,Valentino2018,Valentino2020,Hunt2019,Lapham2019,Sorai2019};\\
$c$:\citet{Dunne2000,Yao2003,Gao2004,Wilson2008,Chung2009,PPP2010,Papadopoulos2012,GarciaB2012,Alatalo2016,Chu2017,Jiao2017,Lu2017,Yamashita2017,herI19,Michiyama2020,Izumi2020};\\
$d$: \citet{Valentino2018,Valentino2020,Bourne2019};\\
$e$: \citet{Dunne2021};\\
$f$: \citet{Villanueva2017,Hughes2017}.}
\end{table*}

With ALMA now in routine operations, studies of \CI\ have expanded to
a broader range of galaxies, with a greater variety of average
ISM conditions, over a wider range of redshift. These include
SMGs, which lie mainly at $z>1$ \citep[e.g.][]{AZ2013, Bothwell2017,
  Popping2017, OteoGRH,Nesvadba2019, Dannerbauer2018, GomezG2019}, and
main-sequence (MS) galaxies at $z=0.35$--1.2
\citep{Valentino2018,Bourne2019,Valentino2020,Dunne2021}. \CI\ has
even been detected in the intracluster medium of the Spiderweb galaxy
cluster at $z=2.16$, as well as in several of its individual galaxies
\citep{Emonts2018}. Routine use of \CI\ as a tracer of molecular gas
is currently limited by the lack of calibration studies to explore and
determine the values and behaviour of the parameters involved, i.e.\
\Xci\ and \aci.  $\Xci=3\times10^{-5}$ has been adopted by almost all
recent studies, taken from \citet{Weiss2003}, determined from a
comparison of analyses of CO and \CI\ in the centre of M\,82, which
has unusually high $\rm [C^0/CO]\sim 0.5$, whereas attempts to estimate
\Xci\ in other ways -- e.g.\ from absorption studies of Gamma-ray bursts and
quasar absorbers \citep{Heintz2020} -- have found lower values,
consistent with the range seen in the Milky Way.

This paper presents the first dedicated cross-calibration study of the
dust, \COa\ and \CIfull\ emission in  a sample of \Ntot\ galaxies from
the literature, including MS galaxies  and SMGs, such that we can
compare  their  properties  and tracer-($\rm H_2$ mass) conversion  factors. We include  the
250-$\mu$m-selected  galaxies at  $z=0.35$ observed  with ALMA  in all
three tracers by \citet{Dunne2021} where  our method was first briefly 
presented.

In \S\ref{obsS} we describe the samples used in this analysis, the
observables,  and the derived quantities. In \S\ref{optS} we describe
the Bayesian approach for producing optimised, self-consistent
tracer-($\rm H_2$-mass) conversion parameters between multiple tracers simultaneously. We
then examine correlations of the observables to look for trends in
\S\ref{correlationsS}.  In \S\ref{caltrendS} we investigate the trends
we have found in the conversion factors and provide refined calibration
recipes. Finally, in \S\ref{DiscS} we
discuss the results and highlight the open questions. Throughout, we
use a cosmology with $\Omega_{\rm m} = 0.27, \Omega_{\Lambda} =0.73$
and $H_0 = 71$\,km\,s$^{-1}$\,Mpc$^{-1}$.

\section{Deriving observational quantities}
\label{obsS}
\subsection{Sample}
\label{sampleS}

\begin{figure}
\includegraphics[width=0.48\textwidth,trim=0cm 0cm 0cm 0cm,
clip=true]{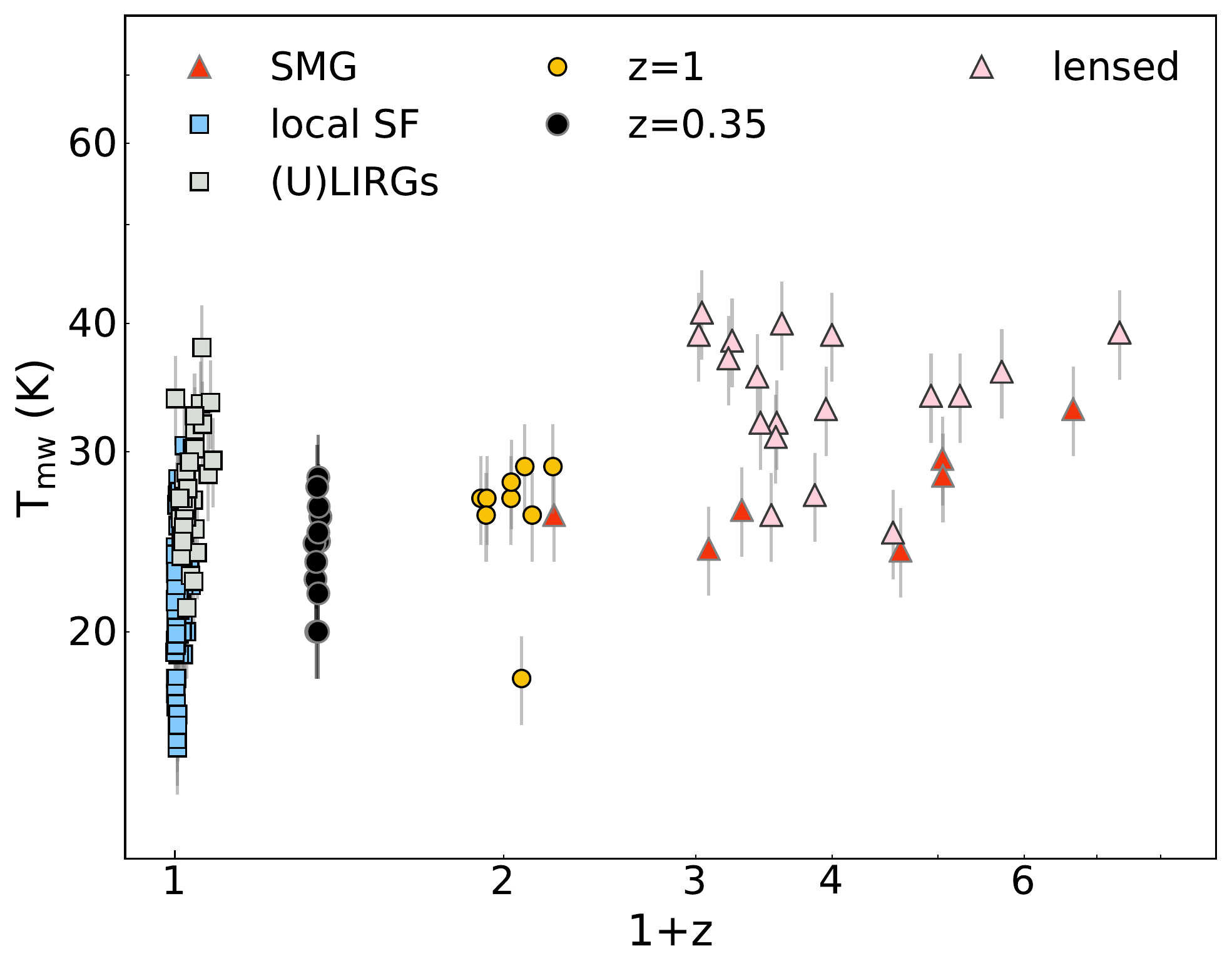}

\includegraphics[width=0.48\textwidth,trim=0cm 0cm 0cm 0cm,
clip=true]{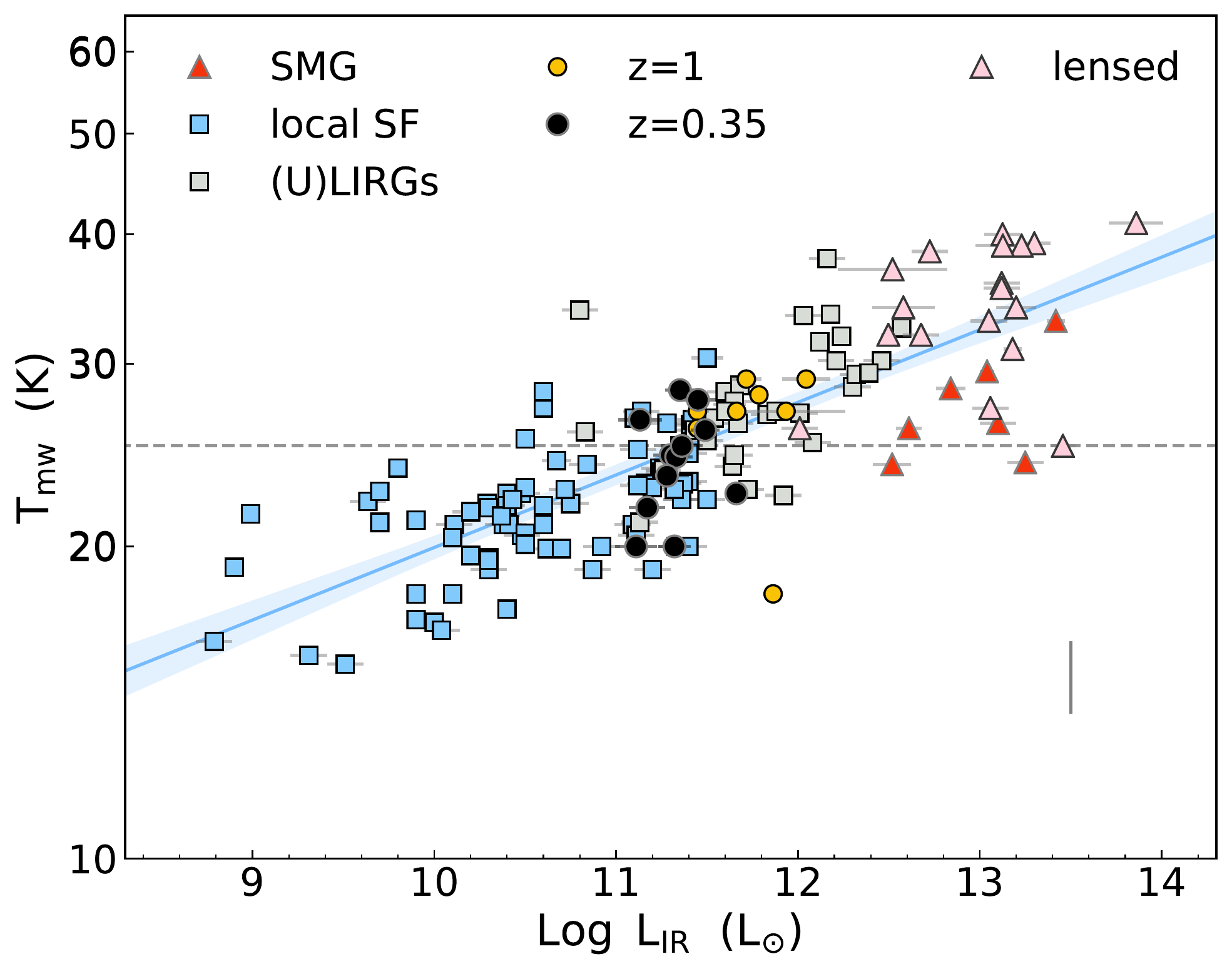}

\includegraphics[width=0.48\textwidth,trim=0cm 0cm 0cm 0cm,
clip=true]{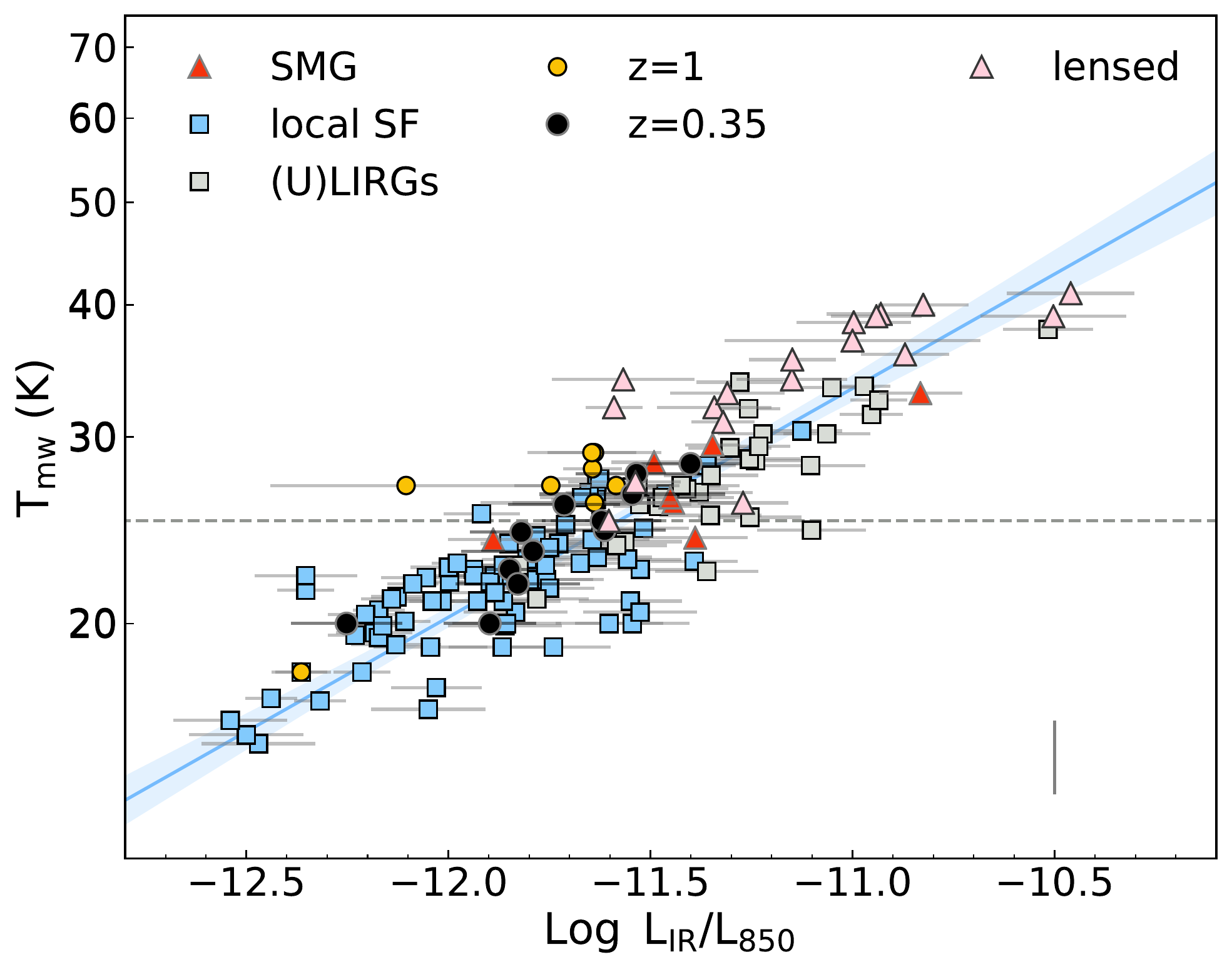}

\caption{\mwtd\ measurements from SED fitting as a function of
  redshift (top), \Lir\ (middle) and SED colour \Lir/\lsub
  (bottom). The best-fit line and 2$\sigma$ error are shown as a blue
  line and shaded region, the gray-dashed horizontal line shows the
  value of \mwtd=25\,{\sc k} used by Sco16. There is a significant
  correlation with all three observables, such that the average \mwtd\
  increases with redshift ($r=0.64$), \Lir ($r=0.74$) and SED colour
  ($r=0.80$). Fit parameters are listed in Table~\ref{fitsT}.}
\label{mwtdzF}
\end{figure}

The samples used in our study are those available in the
literature -- up-to-date as of early 2022 -- which have at least two
of the three tracers: submm dust continuum emission at
$\lambda_{\rm rest} >500$\mic, \COa\ or (2--1), and
\CIfull. Summarising: \Nad\ galaxies have both CO and submm continuum
detections; \NXd\ have both \CI\ and submm dust continuum detections;
\NXa\ have both \CI\ and CO detections; \NdaX\ have all three
tracers. The sample covers the redshift range $0<z<6$, and includes galaxies lying within 1\,dex of the MS as well as extreme starbursts such as local ULIRGs and most high-$z$ submm-selected galaxies. Full details and references
are listed in Table~\ref{SampleT}. Lensed galaxies are included only
where there is an estimate of the magnification, $\mu$, and all luminosities have been corrected by the magnification factor. Our sample includes the galaxies from one of the most comprehensive studies of dust as a tracer of molecular
gas across cosmic time - \citet{Scoville2016},
henceforth Sco16\footnote{Although lensed galaxies were included in their work, the luminosities
  were not de-magnified.}. The \citeauthor{Scoville2016} sample has been updated as described in
Appendix~\ref{notesS}.

In order to test for the effect of SF intensity or `SF-mode' on any later results,  we divide the sample into two groups, referred to hereafter as `MS galaxies' and `SMGs', the names of the groups are not meant to be accurate definitions but rather a reference to familiar categories. For this heterogeneous data-set, defining a simple criterion for two groups is not possible, and even if it were, a fuzzy boundary would still remain due to measurement errors and the inability to capture the complexity in a single parameter. The extreme starburst `SMG' group contains the high-redshift submillimeter selected galaxies which were discovered in the pre-ALMA era and as such are extreme star forming systems (else they could not have been detected), plus the local ULIRGs and some LIRGs which have evidence for very intense and obscured regions (e.g. NGC~4418, IC~860) where conditions are likely to be extreme \citep{DiazSantos2017,Falstad2021}. The `MS galaxy' group contains the lower luminosity local disk galaxies plus the LIRGS which are not extreme, the intermediate redshift sources selected at 250\mic\ from the {\em Herschel}-ATLAS -- $z=0.35$ galaxies from \citet{Dunne2021} and the $z<0.3$ VALES galaxies
\citep{Hughes2017}, the $z\sim 1$ galaxies
\citep{Valentino2018,Bourne2019} and the ASPECs sources denoted as `MS' in that survey \citep{Boogaard2020}.  (Full references are provided in
Table~\ref{SampleT}.) 

There are two situations where corrections to luminosities may be required:

\paragraph*{\HI-dominated galaxies at low \textit{L}$_{\textbf{IR}}$.}

For galaxies with a large fraction of \HI\ within their
optical disk, their dust tracing \HI\ rather than \mol\ makes a significant
contribution to the submm continuum emission. Since our intention is
to provide a calibration for \mol\ rather than total gas, we apply a
correction to \lsub\ for galaxies with $\fhi= \HI/\mol>1$, as described
in Appendix~\ref{HIS}. Galaxies corrected in this way are shown as
cyan diamonds in the plots.

\paragraph*{Local galaxies mapped in \CI\ by the \textbf{\textit{Herschel}} FTS.}

The local galaxies mapped using the {\it Herschel} FTS by J19 present
some complex issues. Some do not have \CI\ and dust continuum
measurements in matched apertures, and those same galaxies are often
only detected in \CI\ in the inner few kpc of the galaxy, where the
ratios of \lci/\lcoa\ may also be biased -- for example, by a lower
\aco\ in galaxy centres.  We discuss the issues in more detail in
\S\ref{J19S} and Appendix~\ref{J19A}. Galaxies requiring a significant
correction ($>0.1$ dex) to \lci\ are labelled as \CIcor; they are
shown in the plots as pink diamonds, but not included in the analysis
unless specified.

\subsection{Observables}
\label{calparamS}
\label{observablesS}

We will compare three tracers of molecular gas, where the observables
(the luminosities \lsub, \lcoa\ and \lci) are empirically related to
the molecular gas mass as

\begin{equation}
\label{MhE}
\Mmol=\lsub/\asub =\aco\lcoa =\aci\lci 
\end{equation}

\noindent The goal of this analysis is to determine self-consistent
conversion factors \asub, \aco\ and \aci\ and study the physical
properties they depend on, e.g.\ C abundance, gas-to-dust ratio (\gdr),
dust emissivity. Our definition of the `observables' is intended to
be independent of as many assumptions as possible. For CO and \CI, we
use $L^{\prime}$ as defined by \citet{Solomon2005}:
\begin{equation}
L^{\prime} = \frac{3.25\times10^7}{\nu_{\rm rest}^2} \left(\frac{D_{\rm
      L}^2}{1+z}\right) \left[\frac{\int _{\Delta V} S dv}{\rm Jy\, km\, s^{-1}}\right] \,\,\,\,{\rm K\,km\,s}^{-1}\,{\rm
  pc}^2 ,
\end{equation}
where $\int _{\Delta V}S dv$ is the velocity-integrated line flux density,
$D_{\rm L}$ is the luminosity distance (Mpc), and $\nu_{\rm rest}$ is
the rest frequency\footnote{Where we use $\nu_{\rm rest}$,
  \citet{Solomon2005} use $\nu_{\rm obs}$, hence the different
  exponent for $(1+z)$ cf.\ their equation (3).} of the transition in
GHz.

Most of the galaxies in our compilation have been observed in the
$^{12}$CO($J=1$--0) transition. However, some observations at high redshift target the
$^{12}$CO($J=2$--1) line. We convert \lcob\ to $L^{\prime}_{10}$ using
the line luminosity ratio $R_{21}=0.8$; if instead we were to set
$R_{21}$ to unity, this would not affect any of our conclusions. We do
not use $J\geq 3$ CO lines because the uncertainties in the global excitation
corrections become too large for a useful calibration study.

We use only the \CI\ $^3P_1$--$^3P_0$ line, as it is the least
sensitive to the average excitation conditions, and correlates better with the low-$J$ CO
emission \citep{Jiao2017,Jiao2019,Crocker2019}. Moreover, there is now evidence
of strongly sub-thermal excitation for \CI(2--1) \citep{Harrington2021,PPPDunne2022},
making it difficult to use this line as an $\rm H_2$ mass tracer since its excitation
is extremely uncertain.

For the dust continuum emission, we use \lsub, the luminosity at
rest-frame 850\,\mic:
\begin{equation}
\lsub= 4\pi S_{\rm{\nu(obs)}}\times K \left(\frac{D_{\rm
      L}^2}{1+z}\right) \,\,\,\,\, {\rm W\,Hz}^{-1} ,
\end{equation}
where $D_{\rm L}$ is the luminosity distance, $S_{\rm{\nu(obs)}}$ is the
observed flux density and $K$ is the $K$-correction to rest-frame 850\mic,
defined as
\begin{equation}
\label{KcorE}
K=\left(\rm{\frac{353\,GHz}{\nu_{\rm
        rest}}}\right)^{3+\beta}\,\left(\frac{e^{\rm{h\nu_{\rm
          rest}/k\td}}-1}{e^{16.956/\td}-1}\right) .
\end{equation}
Here, $\nu_{\rm rest}= \nu_{\rm obs}(1+z)$, \td\ is the
luminosity-weighted dust temperature, from an isothermal fit to the
spectral energy distribution (SED) with the dust emissivity, $\beta$,
allowed to vary between 1.8--2.0.

Sco16 assumed a \td=25\,{\sc k} and $\beta=1.8$, respectively, to extrapolate (or $K$ correct) their observed
submm luminosities to rest-frame 850\,\mic. We make full use of the available data to refine this procedure as follows: 1) With sufficient data, we fit the SED ourselves with $\beta=1.8$ and estimate the rest-frame 850\,\mic\ luminosity directly from the SED fit. 2) Failing that, we use the reported \td\ in the literature to make the extrapolation from the longest wavelength measurement available. 3) For SMGs with insufficient data points to have had their SED fitted, we adopt their observed average, \td=38\,{\sc k} \citep{daCunha2015}. The bulk of the high-$z$ samples now have observations between 2-3~mm with ALMA, as such the extrapolation to rest-frame 850\mic\ is small, even at the highest redshifts. The shortest rest-frame wavelengths we deal with are $\lambda_{r}\sim 250$\mic\ for sources at $z\sim 2-3$ observed at 850\mic, which require K-corrections in the range 50--140. However, the important consideration is the potential uncertainty in that K-correction, not its absolute value. We tried alternatively using the Sco16 method of assuming \td=25\,{\sc K} to extrapolate to rest-frame \lsub\ and found a maximum difference of a factor 1.6, with the average being 1.15 times. The true uncertainty due to the SED sampling and K-correction will be smaller than this, as we know from our work in Section~\ref{dustS} and Figure~\ref{tdhistF} that the dust temperatures in SMG are much higher than 25\,{\sc K}. We thus do not consider the extrapolation to rest-frame 850\mic\ to be a significant source of uncertainty or bias in this analysis.

\subsection{Physical dependencies of gas mass tracers}
\label{physparamS}

\subsubsection{Dust--$\rm{H_2}$ calibration}
\label{dustS}
Large dust grains ($a\sim 0.1$\mic) in thermal equilibrium with their
incident radiation field emit as a modified black body (MBB), where
the emission is related to the mass of hydrogen as:
\begin{equation}
\Mh = \frac{L_{\nu}}{4\pi B(\nu,\mwtd)}\,\,\,\kh(\nu) .  \label{MdE}
\end{equation}
\noindent The two physical quantities needed to calibrate dust
continuum emission as a tracer of gas are therefore \mwtd\ and
\kh. Expressing Eqn~\ref{MdE} in astronomical units for
$\lambda= 850$\,\mic, we can write:
\begin{equation}
\frac{\Mh}{[\msun]} = 6.14\times10^{-14}\frac{\kh}{[\rm{kg\,m^{-2}}]}\frac{\lsub}{[\rm{W\,Hz^{-1}}]}\left(\frac{24.5}{\mwtd}\right)^{-1.4} , \label{MdaE}
\end{equation}
\noindent
where we have simplified the exponential term in the Planck function
as $\sim (24.5/\mwtd)^{1.4}$ for $17<\mwtd<30$\,{\sc k}. 

The mass-weighted dust temperature, \mwtd, is often lower than the
luminosity-weighted dust temperature, \td, as derived from an
isothermal MBB fit to the dust SED, because warm dust outshines cold
dust per unit mass. There is an excellent discussion of this in the
Appendix to Sco16 which we will not repeat here \citep[see
also][]{Dunne2001}.

To determine \mwtd, we require a multi-component MBB fit to a
well-sampled dust SED \citep[e.g.][]{Dunne2001}, or an SED fit using a
model that allows a range of radiation field strengths, leading to a
range of dust temperatures \citep[e.g.][]{Draine2007}. These methods
give broadly consistent results.  As a rule of thumb, the range of
\mwtd\ in local star-forming galaxies is 15--25\,{\sc k}
\citep{Dunne2001,Draine2007,Hunt2015,Dale2002,daCunha2008,Bendo2014,Clark2015},
increasing to 25--30\,{\sc k} in luminous starbursts at higher
redshifts \citep{Rowlands2014,daCunha2015}.

As the dust  (\asub) factor is only weakly dependent on the
assumed temperature at rest-frame  850\,\mic, Sco16 and others assumed
a constant \mwtd=25\,{\sc k}. If instead the true \mwtd\ were to be 15
[30]\,{\sc k}, the  dust and gas mass  would be under-[over-]estimated
by a factor $\sim 2\times$ [$1.3\times$], which is overshadowed by the
other uncertainties.   On the other  hand, failure to account  for any
{\it systematic} trend  of \mwtd\ with another  physical parameter can
introduce  or mask  correlations of the conversion factors with
that physical parameter.

We therefore explore the validity of assuming constant \mwtd\ by
collating measurements of \mwtd\ from the literature
\citep{Dunne2001,Hunt2019}, and additionally making our own fits where possible. In
Fig.~\ref{mwtdzF} we show that there are indeed strong correlations of
\mwtd\ with the observables, namely $z$, \Lir\ and the SED colour,
\Lir/\lsub. There is a clear difference between samples with low and
high SFRs, with $\langle\mwtd(\rm MS)\rangle=23.0\pm0.4$\,{\sc k}
while $\langle\mwtd(\rm SMG)\rangle=30.1\pm0.7$\,{\sc k}. We fit
empirical relations for the correlations in Fig.~\ref{mwtdzF} (see
Table~\ref{fitsT}).  Where there is no direct estimate of \mwtd\ from
an SED fit, which is the case for two-thirds of galaxies, we use these
empirical relations\footnote{We restrict the predicted \mwtd\ such
  that $\mwtd\leq\td$.} to derive \mwtd\ for use in our subsequent
analysis. Appendix~\ref{QTS} compares our approach, where we use
individual estimates of \mwtd, to the adoption of a constant
\mwtd=25\,{\sc k}, where we will discuss those findings in
\S\ref{caltrendS}.

In their work on strongly lensed SMGs at high redshift, which includes
dust continuum emission as a constraint in a large-velocity gradient
(LVG) model, \citet{Harrington2021} find that \td $\sim$ \mwtd\ for
most SMGs, with both measures of temperature higher than the
\mwtd=25\,{\sc k} commonly used in the literature. To ensure
consistency with our other estimates of \td, we fitted the
\citet{Harrington2021} photometry with three simple models: 1) an
isothermal optically thin MBB; 2) an MBB with variable optical depth
and -- where there were enough data -- 3) a two-component MBB. In
agreement with \citeauthor{Harrington2021}, we find that a single dust
temperature adequately describes the SED of these galaxies, in contrast
to lower redshift (U)LIRGs and normal galaxies which are better fit
with multiple dust components (\td\ $>$ \mwtd) and/or fits with higher
FIR optical depths. The temperatures from the \citeauthor{Harrington2021}
turbulence model correlate best with our isothermal \td\ meansurements
for these galaxies, and the temperatures returned when allowing
variable optical depth, $\td(\tau)$, are always significantly higher
than those from the \citeauthor{Harrington2021} model. We therefore do
not use optically thick fits to yield \mwtd\ for our high-redshift
SMGs.  We instead use two-component SED fits to the lensed {\it Planck} sources
and the handful of SMGs with sufficient data for the
empirical relations shown in Fig.~\ref{tdcorrF}.

The other key physical parameter in the dust--\mol\ conversion is \kh,
which is a combination of the dust mass absorption coefficient (\kd)
and \gdr, such that\footnote{Literature studies generally present the
  dust-\mol\ conversion in terms of \gdr\ for a fixed emissivity,
  \kd. Given the mounting evidence that \kd\ varies within our own
  \citep{Remy2017,Ysard2015,Ysard2018,Kohler2015} and other galaxies
  \citep[e.g.][but see also \citealt{Priestly2020}]{Clark2019}, we
  prefer to work with \kh\ to avoid projecting all the variation in the
  $\rm H_2$-dust conversion factor onto \gdr.}  \kh = \gdr/\kd. Briefly, \kd\ is sensitive
to the grain composition and structure (amorphous; crystalline;
coagulated; mantled), while \gdr\ is roughly proportional to
metallicity and, for galaxies with metallicity within a factor 2 of $Z_{\odot}$, as expected for those in our
samples, can be taken to be roughly constant, at \gdr $= 100$--150
\citep{Sodroski1997,Dunne2000,Dunne2001,
  Draine2007,MM2009,Leroy2011,Sandstrom2013,Planck2011,Jones2017,deVis2021}. 

Fortunately, observational measures of \kh\ are available, both for
the Milky Way and for external galaxies, with values of \kh\
$\sim 2000$\,\khunit\ in the Milky Way's diffuse interstellar medium
(ISM) and \kh\ $\sim 800$\,\khunit\ in dense
clouds. Appendix~\ref{kappaA} discusses in more detail how it is
measured, and Table~\ref{kappalitT} provides a comprehensive set of
observational and theoretical values for \kh\ from the literature.

It is impossible to disentangle the effect of changing dust properties
(\kd) from changes in \gdr\ in observational determinations of
\kh. While the decrease in \kh\ towards denser sightlines in the Milky
Way is thought to be due to the dust grains coagulating in denser
environments -- a process expected to increase their emissivity
\citep[e.g.][]{Kohler2015} -- there may also be some decrease in \gdr\
if the gas is accreted into dust mantles or ices (i.e.\ grain
growth). Both effects are to be expected
\citep[e.g.][]{Jones2017,Jones2018} and both act to decrease
\kh. Counter to that, the higher estimates of \kh\ in the diffuse
atomic phase (lowest $N_{\rm H}$ sightlines at high latitudes) in the
Milky Way may be due in part to a lower dust emissivity for grains
without ice mantles, where only the refractory cores remain, subjected
to harsher ultraviolet (UV) irradiation. Additionally, there is likely
a metallicity gradient at high latitudes, leading to a higher \gdr,
further increasing \kh. There is thus a qualitative expectation that
denser regions with higher metallicity will have higher dust
emissivity, \kd, and lower \gdr, producing a lower \kh. More diffuse
regions with lower metallicity will move in the opposite direction. In
\S\ref{caltrendS}, we find that we can constrain the {\it range} of
\kh, at least, and therefore the combination of \gdr/\kd.

\subsubsection{$\rm C\,I$--$\rm{H_2}$ calibration}
\label{CIS}

Here, we introduce the two physical parameters the  \Xci=$\rm [C^0/H_2]$ average
abundance ratio and the average excitation factor \Q$=\rm N_1/N_{tot}$, pertinent to the use of \CI\ as a tracer of \mol.
The relationship between \Mh\ and the `observable' -- [\CI](1--0) line emission -- is (in astronomical units):

\begin{equation}
\Mh ({\rm M}_\odot) = \frac{0.0127}{X_{\rm
    C\,I}\,\,Q_{10}}\left(\frac{D_{\rm L}^2}{1+z}\right)\,\,\left[\frac{\int _{\Delta V}S_{\rm [CI](1-0)} dv}{\rm Jy\, km\, s^{-1}}\right]
\end{equation}
with $D_{\rm L}$ in Mpc and $\int_{\Delta V}S_{\rm [CI](1-0)}\Delta v$ in Jy\,\kms. Expressed in
units of line luminosity, this becomes:
\begin{equation}
\Mh (\msun) = \frac{9.51\times 10^{-5}}{\Xci\, Q_{10}}\,\lci  .\label{MhciE}
\end{equation}
 
\noindent
The excitation term, \Q, describes the relative fraction of carbon atoms in the
$J=1$ state. Under general non-LTE conditions it  is a function of both gas density, $n$, and \tk\, and is
 derived analytically in the Appendix to
\citet*{PPP2004}. A recent study of the [\CI](2--1)/(1--0) line ratio
\citep*{PPPDunne2022} finds that the \CI\ lines are both sub-thermally excited in the ISM of galaxies, with the [\CI](2--1) especially 
so \citep[see also][]{Harrington2021}. Thus the LTE expressions for \Q\ 
should not be used, nor will the \CI\ line ratio produce an estimate of \tk\ (both methods having been widely used in the literature to date). Details for \Q\ are in the Appendix~\ref{QA}, but in summary we find:

\begin{enumerate}
\item The [\CI](1--0) excitation term, \Q, is a non-trivial function of
  density and temperature, but for the range $\tk \geq 20$~K and log $n\geq 2.5$ -- which is where the bulk of \mol\ in star forming galaxies is thought to reside -- $\langle\Q\rangle=0.48\pm 0.08$ where the 99
  per cent confidence range is quoted (see \citealt{PPPDunne2022} and Figure~\ref{QulF} for details).
\item Due to a slight super-thermal behaviour, higher density, higher \tk\
  conditions can produce similar or even lower \Q\ than lower
  density, lower \tk\ conditions. This breaks any intuitive link
  between \Q\ and the ISM conditions, i.e.\ we do not necessarily
  expect a higher \Q\ in SMGs compared to MS galaxies (see
  Fig.~\ref{QulF}).
\item As the [\CI](2--1) line is even more strongly sub-thermally excited, its $Q_{21}=N_2/N_{\rm tot}$ factor varies strongly\footnote{The $Q_{21}(n, T_k)$ that enters the estimates of molecular gas
  mass when the [\CI](2--1) line is used can vary almost by a factor of $\sim5$, depending on $(n, T_k)$.}. This is the main reason why our current
  study is restricted to the [\CI](1--0) line. 
\end{enumerate}

As the \CIfull\ line is optically thin for most conditions expected in
spiral disks \citep{Weiss2005,PerezB2015,Harrington2021}, the
relationship between \lci\ and \Mh\ is proportional to \Xci\ -- the abundance
of carbon atoms relative to H$_2$. This dependence on abundance is as expected for any method  that employs tracers of  $\rm H_2$ gas mass  other than
the  $\rm  H_2$  lines themselves\footnote{Even  for  optically  thick
  tracers of  $\rm H_2$  gas, such as CO(1--0) line emission,  a $\rm
  [CO/H_2]$ abundance still  enters the method via  the CO--H$_2$ cloud
  volume-filling factor, $f_{\rm CO}$, albeit not in a sensitive fashion
  unless   a  combination   of   strong  FUV   radiation  and/or   low
  metallicities selectively dissociate CO in the outer cloud layers while
  leaving the largely self-shielding $\rm H_2$ intact (then $f_{\rm CO}$ can be $\ll 1$,
see Pak et al. 1998 for details).}.
  
With the  excitation factor \Q\ varying  no more than 16 per cent over
the typical range of \mol\ conditions  in   galaxies ($\tk\geq 20$~K, log $n\geq 2.5$), the major source  of  uncertainty  in  \CI-based
molecular  gas mass  estimates (and thus the major source of scatter in the \aci\ conversion factor) is the neutral carbon abundance, \Xci. The relatively recent introduction of the [\CI](1--0) line as a gas tracer means that \Xci\ has not  been widely explored -- constraining it and investigating any potential trends is  a  key  outcome  of  our
cross-calibration work.

In the Milky Way, \Xci\ is found to vary only modestly, from
0.8--$2.2\times10^{-5}$ \citep[e.g.][]{Zmuidzinas1988,
  Frerking1989,Tauber1995,Ikeda2002}, while a much higher value
($\Xci=5\times10^{-5}$) has been inferred for the nearby starburst nucleus of M\,82 \citep{Schilke1993,White1994,
  Stutzki1997}\footnote{The measurement is in fact the $\rm [C^0/CO]$ abundance, and a value for [CO/\mol] has then to be assumed to infer \Xci.}. Thanks to ALMA, very high localised ratios of \lci/\lcoa\ (translating to high \Xci=5-7$\times 10^{-5}$) have also been measured in extreme regions, such as the Circum-Nuclear Disk (CND) of NGC~7469 which is believed to host an X-ray Dominated Region (XDR) \citep{Izumi2020} and the outflow region in NGC~6240 \citep{Cicone2018}. More modestly elevated \lci/\lcoa\ ratios tend to be found in the central nuclear regions of starburst galaxies \citep{Jiao2019,Salak2019,Saito2020}. However, when averaged over larger kpc scale regions -- the ratios become consistent with the average global ratios measured for this sample (see Figure~\ref{tdcorrF}).  Independent measurements of $\Xci=1.6^{+1.3}_{-0.7} \times 10^{-5}$ (for solar
metallicity)  were made by
\citet{Heintz2020} using UV absorption measures for a range of absorber systems across cosmic time. Cosmic rays (and X-rays) are
expected to dissociate CO in favour of atomic carbon, increasing $[{\rm C}^0/{\rm CO}]$, a
hypothesis supported by both simulations and observations
\citep[e.g.][]{Bisbas2015,Clark2019ci,Israel2020,Izumi2020}.

\subsubsection{\rm CO--$\rm{H_2}$ calibration}
\label{COS}

The $^{12}$CO(1--0) line  is optically thick in most (but  not all see
\citealp[e.g.][]{Aalto1995})  ISM conditions expected in galaxies.  Unlike
dust  continuum  emission where  optical  depths  build up over  large
columns of  dust, the  entire CO line  optical depth  builds up within
very small gas `cells'  ($<$0.1\,pc) due to the very turbulent nature of the velocity fields,
and the  small thermal line widths \citep{Tauber1991,Falgarone1998}. This localised nature  of CO line optical  depths and
the  macro-turbulent  CO  line   formation  mechanism  allows  a  great
simplification of  the radiative transfer  models of such  lines, i.e.\
the use of the so-called Large Velocity Gradient (LVG) approximation. However, it also complicates the relationship  between the CO line luminosity and
the underlying $\rm H_2$ gas mass, making the corresponding conversion
factor, \aco,  dependent on the  thermal state  of the gas,  its average
density, as well as its dynamic state.

 Following \citet{PPP2012xco} the \aco\ factor in an LVG setting is given by:

\begin{equation}
    \aco = 2.65\frac{\sqrt{n_{\rm H2}}}{T_{\rm b}}\,K_{\rm vir}^{-1}\,\,\,\, [\aunit] \label{acoE}
\end{equation}

\noindent
where $n_{\rm  H2}$ and $T_{\rm b}$  are the average density  (in \cc)
and  the  CO(1--0)   brightness  temperature\footnote{Here  the  cloud
  CO-H$_2$  volume filling  factor  is set  $f_{\rm CO}=1$.} for  the
molecular  cloud ensemble  while $K_{\rm  vir}$ describes  the average
dynamic state  of the gas (self-gravitating  clouds $K_{\rm vir}\sim 1$,  unbound clouds  $K_{\rm vir}>1$). In  principle, multi-phase  LVG
models of CO  (and $^{13}$CO) SLEDs can be used  to constrain \aco, but in
practice this demands   large  line   datasets  per   galaxy  \citep[e.g.][]{PPP6240, Harrington2021},  making  it
impractical for use in large galaxy  samples. This is why in our study
\aco\ remains an empirical conversion factor to be (cross)-calibrated.

\subsubsection{Conversion factors and physical parameters}

The two optically thin tracers  -- thermal dust continuum emission and
\CI\  --  have a  simple  relation  between the  empirical `mass-to-light' conversion
parameter ($\alpha_{\rm  X}$) and the  physical conditions in  the ISM
(e.g.\ abundance, emissivity, temperature). We can write the empirical
factors  (Eqn.~\ref{MhE}) in  terms  of these  physical parameters  as
follows\footnote{Hereafter  we  omit the  units  for  \aco, \aci\  and
  \asub.}:

\begin{equation}
	\asub =
        \frac{1.628\times10^{16}}{1.36\,\kh}\left(\frac{24.5}{\mwtd}\right)^{-1.4}
        \,\,\,{\rm W\,Hz}^{-1}{\rm M}_{\rm mol}^{-1} \,\, , \label{asubE}
\end{equation}
where the factor 1.36 corrects to total molecular mass, including
He. 

\begin{equation}
	\aci=16.8\,\left[\frac{\Xci}{1.6\times10^{-5}}\right]^{-1}\left[\frac{Q_{10}}{0.48}\right]^{-1} \,\,\,\rm{\msun\,(K\,km\,s^{-1}\,pc^2)^{-1}} \label{aciE}
\end{equation}
Eqn.~\ref{aciE} also includes the factor 1.36 for He.

\section{Deriving self-consistent calibration of conversion factors}
\label{optS}

We next describe how we combine the measurements of multiple gas
tracers in the most efficient way, in order to determine their
cross-calibrations. Our goal is to find the empirical conversion factors (\asub, \aci, \aco) or physical parameters (\kh, \Xci), which
produce a consistent estimate for \Mh\ in a given galaxy.

Our dataset provides nested samples, each with a different set of
available gas tracers.  The daX sample has all three tracers
available -- dust continuum, CO and \CI, and contains \NdaX\ galaxies
($N_{\rm daX}$ = \NdaX). The names and statistics for the other
samples are as follows: ad -- CO and dust, $N_{\rm ad}$ = \Nad; Xd
-- \CI\ and dust, $N_{\rm Xd}$ = \NXd; Xa -- \CI\ and CO, $N_{\rm Xa}$
= \NXa. The properties of these samples are  in
Table~\ref{namesT}. 

The best constraints at log$_{10}$ \Lir\ $> 11$\footnote{Hereafter we will refer to $\log_{10}$ as simply log.} come from the daX
sample because it has three independent tracers of gas mass, but
it lacks coverage of luminosities below $l_{\rm IR}  = 11$. The ad sample is the
largest and spans the widest range in \Lir, reflecting the longer time
for which CO observations have been possible for nearby galaxies.

We begin with the daX sample, to illustrate the method of
optimisation for the estimates of all three conversion factors
simultaneously.\footnote{This method was first presented in brief in
  \citet{Dunne2021}, where it was applied to the sample of $z=0.35$
  galaxies.} There are four unknowns namely: $m={\rm log} (\Mh)$, $X={\rm log} (\Xci)$,
$\kappa = {\rm log} (\kh)$, and $\alpha= {\rm log} (\aco)$, and 
 three observables: \lcoa, \lci\ and \lsub.

With an independent measure of the true \Mh, the observables would
provide direct estimates of the three conversion factors; however,
the value of \Mh\ is not known {\it a priori}, so we must use a
probabilistic argument based on the fact the observations do provide
constraints on the {\it relative} values of the conversion factors
for each galaxy. There is thus a set of self-consistent conversion
factors which link the observables to the true \Mh, with an unknown
common constant factor.

The Bayesian approach we use is described in detail in
Appendix~\ref{bayesS} and requires an estimate of the intrinsic
scatter for the logarithms of each of the factors:
$s_{\rm X}$, $s_{\kappa}$ and $s_{\alpha}$. The observable luminosities relate
to these factors as follows, where the coefficients of proportionality are listed in Table~\ref{methodT}:

\begin{equation}
	\begin{aligned}
	\label{ratioE}
\frac{\lcoa}{\lci}\propto\aco\Xci,\,\, &
\frac{\lsub}{\lci}\propto\kh\Xci,\, &
\frac{\lcoa}{\lsub}\propto\frac{\aco}{\kh} .
    \end{aligned}
\end{equation}
We begin by measuring the intrinsic scatter between the three pairs of
observables using an orthogonal distance regression (ODR) fitting
method, which includes the intrinsic scatter, $\lambda$, as a third
parameter in the analysis\footnote{We need to multiply $\lambda$ from
  the ODR fitting routine by $\sqrt{2}$ because we need to know the
  intrinsic scatter of $X-Y$ in our dataset in order to determine the
  intrinsic scatter of each conversion factor in turn.}  (see
Appendix~\ref{ODRS} for full details). The three pair variances
derived from the data are then used to estimate the intrinsic variance
of the three individual conversion factors (the derivation can be
found in Appendix~\ref{pairwiseS}). The values of the intrinsic
scatter for the parameters are given in Table~\ref{methodT}, with
\Xci\ having the smallest scatter between galaxies. This finding is
purely empirical, requiring no assumptions about the values or trends
of the conversion factors, and as such is very interesting.

\begin{table}
	\caption{Samples used in cross-calibration analysis.}
	\label{namesT} 
	\begin{adjustbox}{center}
		\begin{tabular}{cccc}
			\toprule
			Sample & Tracers present  & $N$ & median log \Lir \\
			\midrule
			daX  & Dust, CO and \CI & \NdaX\  (\NdaXhi) & 11.65 (11.77) \\ 
			Xa   & CO and \CI       & \NXa\   (\NXahi)  & 11.66 (11.88) \\
			Xd   & Dust and \CI      & \NXd\  (\NXdhi) & 11.88 (12.06)\\
			ad   & CO and dust      & \Nad\   (\Nadhi)  & 11.54 (12.07) \\
			\bottomrule
		\end{tabular}
	\end{adjustbox}
        \flushleft{$N$ is the size of the sample upon which the analysis has been performed, excluding those with uncertain and potentially large corrections -- see \S\ref{J19S}. Values in parenthesis are the number of galaxies in the samples with \Lhi\ and their median log \Lir.}
\end{table}

\begin{table}
	\caption{\label{methodT} Summary of the parameters required to reproduce this analysis.}
	\begin{adjustbox}{center}
	    \begin{tabular}{ccccc}
	        \toprule
	        Quantity & Set & \CI & CO & Dust\\
	        \midrule
	        physical &     & \Xci & \aco & \kh\\
	        empirical &    & \aci & \aco & \asub\\
	        \midrule
	        $s_{X,\alpha,\kappa}$ & \Lhi & 0.082  & 0.1646 & 0.1339\\ 
			& BL & 0.1125 & 0.1436 & 0.1294\\
	        \bottomrule
	    \end{tabular}
	\end{adjustbox}
	\flushleft{BL = baseline (excludes \CIcor\ and lo-VALES galaxies).}
            \vspace*{1cm}
  \begin{adjustbox}{center}
		\begin{tabular}{lccc}
			\toprule
			\multicolumn{1}{c}{Pair}&\multicolumn{1}{c}{Set}&\multicolumn{2}{c}{Mean log pair}\\
			\cmidrule{3-4}
			\multicolumn{2}{c}{}&\multicolumn{1}{c}{BL}&\multicolumn{1}{c}{\Lhi}\\
			\midrule
			\aco\Xci   & Xa   & $-4.400\pm0.020$ & $-4.383\pm0.021$ \\
			&  daX & $-4.393\pm0.021$ & $-4.373\pm0.022$ \\
			
			\aco/\kh    & ad & $-2.769\pm0.015$ & $-2.798\pm0.018$ \\
			& daX & $-2.867\pm0.025$ &  $-2.875\pm0.027$ \\
			\kh\Xci    & Xd & $-1.529\pm0.021$ & $-1.509\pm0.021$ \\
			& daX & $-1.526\pm0.024$ & $-1.498\pm0.024$\\             	\bottomrule
		\end{tabular}
	\end{adjustbox}
	\flushleft{Notes: Values here can be used to reproduce our method
          and should be applicable to other metal-rich
          samples. $s_{X,\alpha,\kappa}$ are the intrinsic scatter on
          the log conversion factors, $X$, $\alpha$ and
          $\kappa$. `Mean log pair' are the means of the log
          combinations of calibration factors listed in the `Pair'
          column, quoted with the standard error on the mean. We list
          in the second column the sample used to derive these
          means, both the sample with the largest number of pairs
          and also for daX, which provides our reference set. We
          provide numbers both for the BL galaxies (excluding those
          discussed in \S\ref{J19S}) and also those with
          \Lhi. The differences are not significant.
          }
\end{table}

\begin{figure*}
	\centering
	\includegraphics[width=0.32\textwidth,trim=0cm 0cm 0cm 0cm, clip=true]{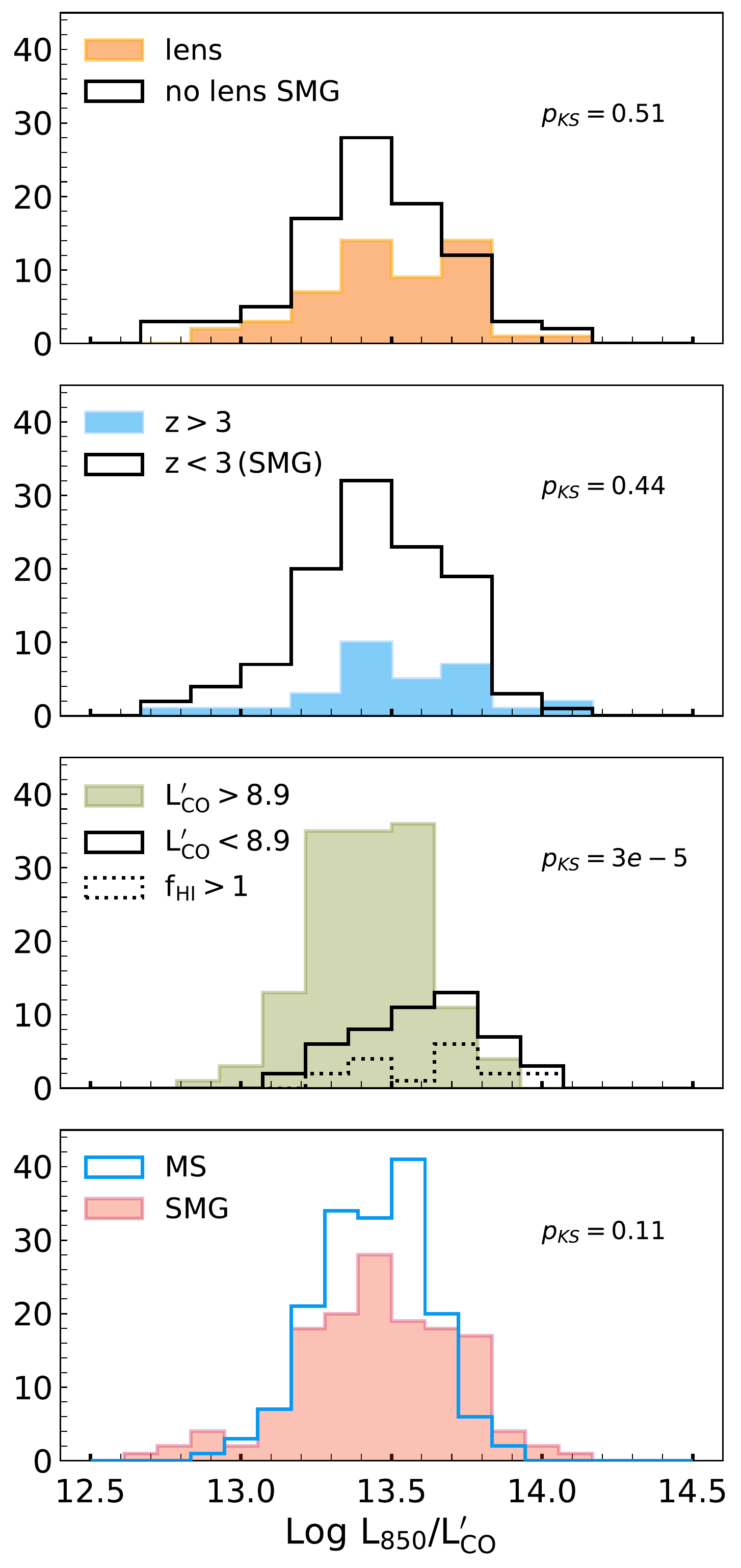}
	\includegraphics[width=0.32\textwidth,trim=0cm 0cm 0cm 0cm, clip=true]{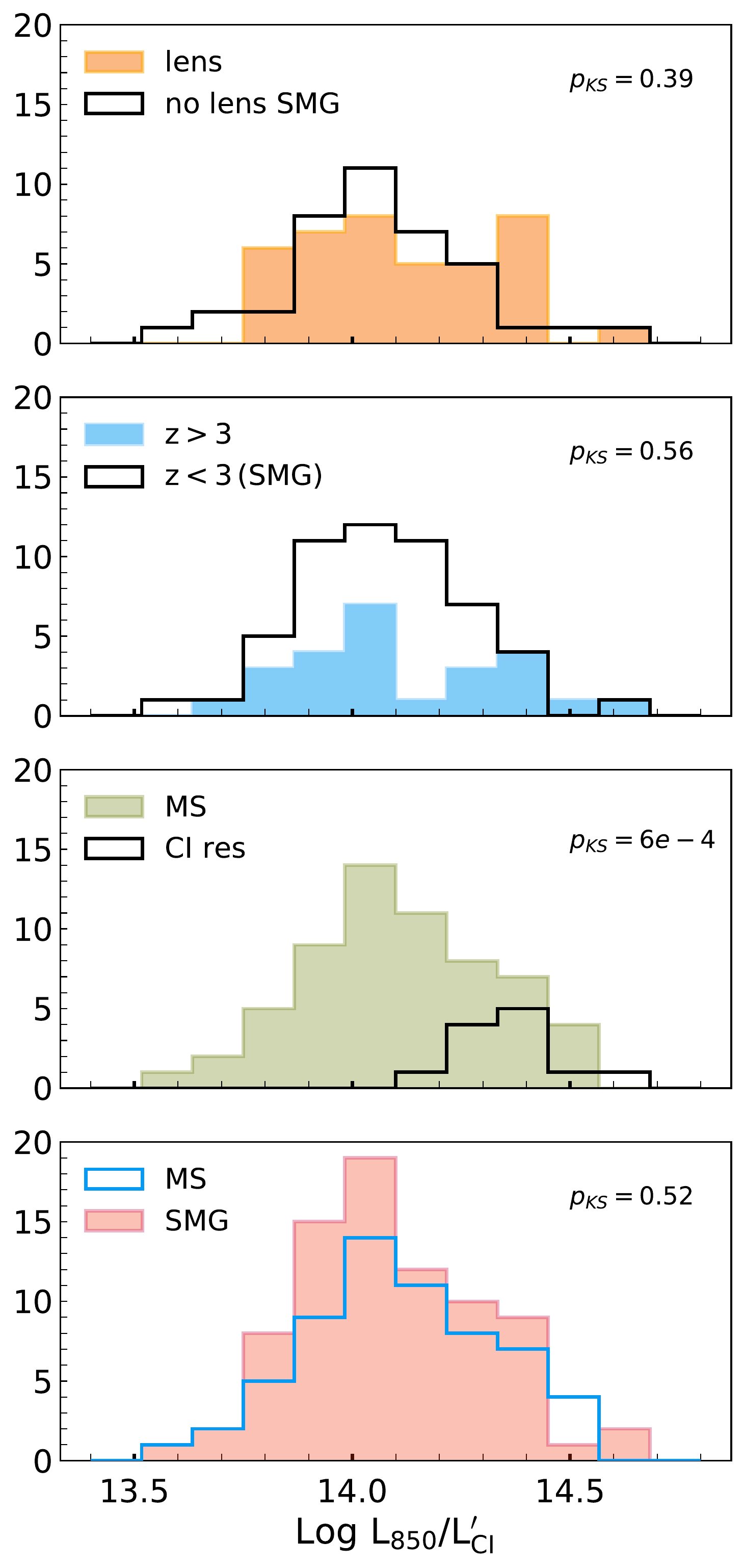}
	\includegraphics[width=0.32\textwidth,trim=0cm 0cm 0cm 0cm, clip=true]{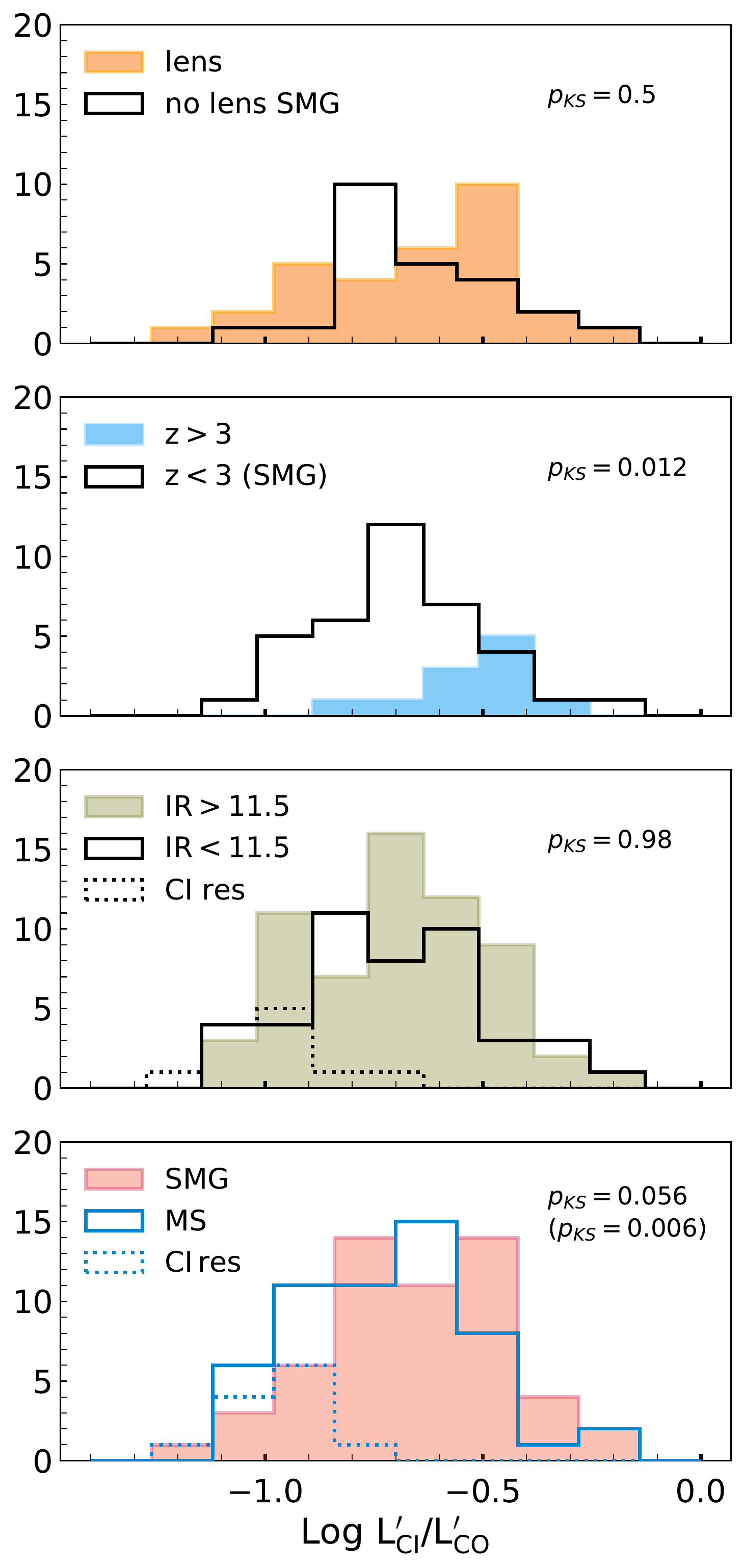}
	\caption{\label{LhistsF} Tracer luminosity ratios as a
          function of lensing, redshift and SFR. $p_{\rm KS}$ is the
          probability of no difference from a two-sided KS test. {\bf
            Left column:} \lsub/\lcoa\ -- there are no indications of any bias
          due to lensing or redshift. The third panel shows
          the effect of high \fhi\ in MS galaxies (we have not applied
          the H\,{\sc i} correction factor to show the difference in the
          raw ratio). The green-shaded region consists of MS galaxies
          with log $\lcoa>8.9$ and $\fhi<1$. The black line is for all galaxies
          with $log \lcoa<8.9$, including the lo-VALES galaxies suspected
          of having $\fhi>1$. Galaxies known to have $\fhi>1$ are shown
          as the dotted line to illustrate the similarity. The MS
          sub-group in the lower panel excludes the $\fhi>1$ and
          lo-VALES galaxies. {\bf Middle column:} \lsub/\lci\ -- there are no
          indications of bias due to lensing or redshift. The third panel shows the \CIcor\ galaxies from J19
          compared to the rest of the MS sub-group; even with the
          aperture correction applied to the \CI\ flux
          (Appendix~\ref{J19S}), they have a significantly different
          distribution of observed ratios. In the lower panel, we
          exclude \CIcor\ galaxies and find no difference between
          galaxies with high and low
          SFRs. {\bf Right column:} \lci/\lcoa\ -- the highest redshift
          SMGs ($z>3$) have higher \lci/\lcoa\ at marginal significance
          ($p=0.012$), but there are only 11 galaxies at $z>3$. The
          third panel shows once more the difference between the
          \CIcor\ and other galaxies in both high and low luminosity
          bins. Even though the \CI\ and CO are measured in the same
          apertures, the \CIcor\ galaxies have much lower
          \lci/\lcoa. The bottom panel shows a low signficance
          difference between high and low SFR galaxies ($p=0.056$), becoming
          significant when including the \CIcor\ galaxies. Means and KS
          results are given in Table~\ref{hypT}.}
      \end{figure*}

As we do not have any independent measure of the gas mass with which
to normalise our cross-calibration (four unknowns but only three
measurements), we must make an assumption about the sample average of
one of the physical or empirical conversion factors.  However, with that assumption made
transparent, the individual values can always be scaled to whichever
normalisation a reader wishes to adopt. {\it The relative values, however,
are always the optimal solution.}

We choose to use the  dust  parameter \kh =
\gdr/\kd\ for this normalisation, because there are no trends of
\lsub/\lcoa\ or \lsub/\lci\ with \Lir\ (see
Figs~\ref{LhistsF},~\ref{tdcorrAF}) and \kh\ also has the best
observational constraints.

For the Xa sample, where there are no dust continuum measurements,
we normalise to $\Xci^{\rm N}=1.6\times10^{-5}$, which is the
mid-range of the values suggested by the independent study of
absorption lines by \citet{Heintz2020}.

All the galaxies in our sample are metal rich ($0.5 < Z/Z_{\odot} < 2$) and so we assume that
\kh\ (\gdr) should be similar to that in the Milky Way and other local
disks. Throughout the rest of this work, we will use as our reference
point the mid-range of extragalactic determinations, \kh =
1500--2200\,\khunit, which are consistent with measurements of the
diffuse ISM in the Milky Way. Our chosen normalisation value, then, is
\kh$^{\rm N}$ = 1884\,\khunit, which is a good match to current
theoretical dust models (THEMIS: \citealp{Jones2017}, and the updated
\citealp{Draine2007} modified by \citealp{Hensley2021}).

For the standard Milky Way value of \gdr\ \citep[=
135,][]{Jones2017,Magdis2012}, \kh\ = 1884\,\khunit\ implies that \kd\ =
0.071\,\kunit, similar to that used in many extragalactic studies
\citep{Dunne2000,James2002,daCunha2013}. For a \gdr\ fixed to 135, the
range of \kh\ in extragalactic studies implies a range in \kd\ of
0.06--0.09\,\kunit. Table~\ref{kappalitT} lists \kh\ values from
extragalactic and Galactic observations, as well as from theoretical
dust models.

Note that the choice of normalisation does not affect any of the
trends, nor the ratio of the conversion parameters in the pairings;
it merely sets the average value of the reference calibration
parameter, to which the others are relative.

The sample mean expectation values for the other two 
parameters, $\langle\aco\rangle$ and $\langle\Xci\rangle$, are next derived
from our assumed value of $\langle \kh \rangle$, together with the mean
ratios of the observables listed in Table~\ref{methodT}. The {\it effective standard
  deviation} is also calculated -- the intrinsic scatter of each parameter added in
quadrature to the measurement error for that gas tracer. For example,
for CO:
\[
\sigma_{\rm eff} = \sqrt{s_\alpha^2 + \sigma_{\rm CO}^2},
\]
where $s_{\alpha}$ is the intrinsic scatter in log(\aco), and
$\sigma_{\rm CO}$ is the measurement error on log(\lcoa).
 
We can now estimate the probability of finding a particular set of
conversion factors for any given galaxy. We use $a_i,\, i=1,2,3$ to
denote the logarithms of the three conversion factors\footnote{For
  ease of representation, $a_{850}=-\log(\asub)$.}, and write the mean
expectation values and effective standard deviations as
$\langle a_i \rangle$, and $\sigma_{i,\rm eff}$ respectively.
Assuming that these follow Gaussian distributions, the probability of
finding the factors, $a_i$, for any galaxy is:

\begin{multline}
\label{eqn:chi2}
P \propto \displaystyle\prod_{i=1}^N \exp\left(-\frac{(a_i-\langle a_i \rangle )^2}{2\sigma_{\rm i, eff}^2}\right)\\
= \exp \left(- \displaystyle\sum_{i=1}^N \frac{(a_i-\langle a_i\rangle
  )^2}{2\sigma_{\rm i, eff}^2} \right) .
\end{multline} 

\noindent
Thus, the ratios of observable luminosities for any given galaxy can
be used to determine the ratios of conversion factors
(Eqn.~\ref{ratioE}), and the common scaling factor that maximises the
probability in Equation~\ref{eqn:chi2} is the best estimate of
\Mh. The derivation in Appendix~\ref{bayesS} shows that this reduces
analytically to a simple inverse variance weighted mean, such that:

\begin{equation}
\log M_{\rm H_2}^{\rm opt} = \frac{\sum^N_{i=1}(m_i\times
  w_i)}{\sum^N_{i=1} w_i}   \label{MoptE} ,
\end{equation}
\noindent
where $w_i = 1/\sigma_{i, \rm eff}^2$, and $m_i$
is the log mass estimate for each tracer. 
\[
m_i = l_i + \langle a_i \rangle ,
\]
\noindent where $l_i$ is the measured observable (log luminosity) and
$\langle a_i \rangle$ is the sample mean expectation value for the
conversion factor. Once the optimal mass is determined this way, we
can then estimate the corresponding optimal conversion factor on a
per-galaxy basis, as:
\begin{equation}
a_i = m^{\rm opt} - l_i  .
\end{equation}
\noindent The error on the optimal mass is simply the error on the
inverse variance weighted mean:
\begin{equation}
\sigma_m^{\rm opt} = \left(\sum^N_{i=1} w_i\right)^{-1/2} ,
\end{equation}
\noindent and the error on each of the conversion factors, accounting
for co-variance is:
\begin{equation}
\sigma_{ai} = \sqrt{\sigma_{\rm m^{opt}}^2 + \sigma_{li}^2 \left(1 -
    \frac{2 w_i}{\sum^N_{j=1} w_j}\right)} ,
\end{equation}
\noindent where $\sigma_{li}$ is the logarithmic measurement error on
the observable quantity, e.g.\ \lcoa, \lci, \lsub.

By design, each tracer for a given galaxy, together with its optimised
conversion factors, will produce the same gas mass, such that
$\Mh^{\rm CO}=\Mh^{\rm C\,I}=\Mh^{\rm dust}$.

\section{Trends in the luminosity ratios}
\label{correlationsS}

As  our  cross-calibration  process  relies  on  measurements  of  the
luminosity ratios,  it is first instructive  to look at the  trends in
these observables  to better understand  any subsequent trends  in the
derived conversion factors.

\begin{figure*}
\includegraphics[width=0.48\textwidth,trim=0cm 0cm 0cm 0cm, clip=true]{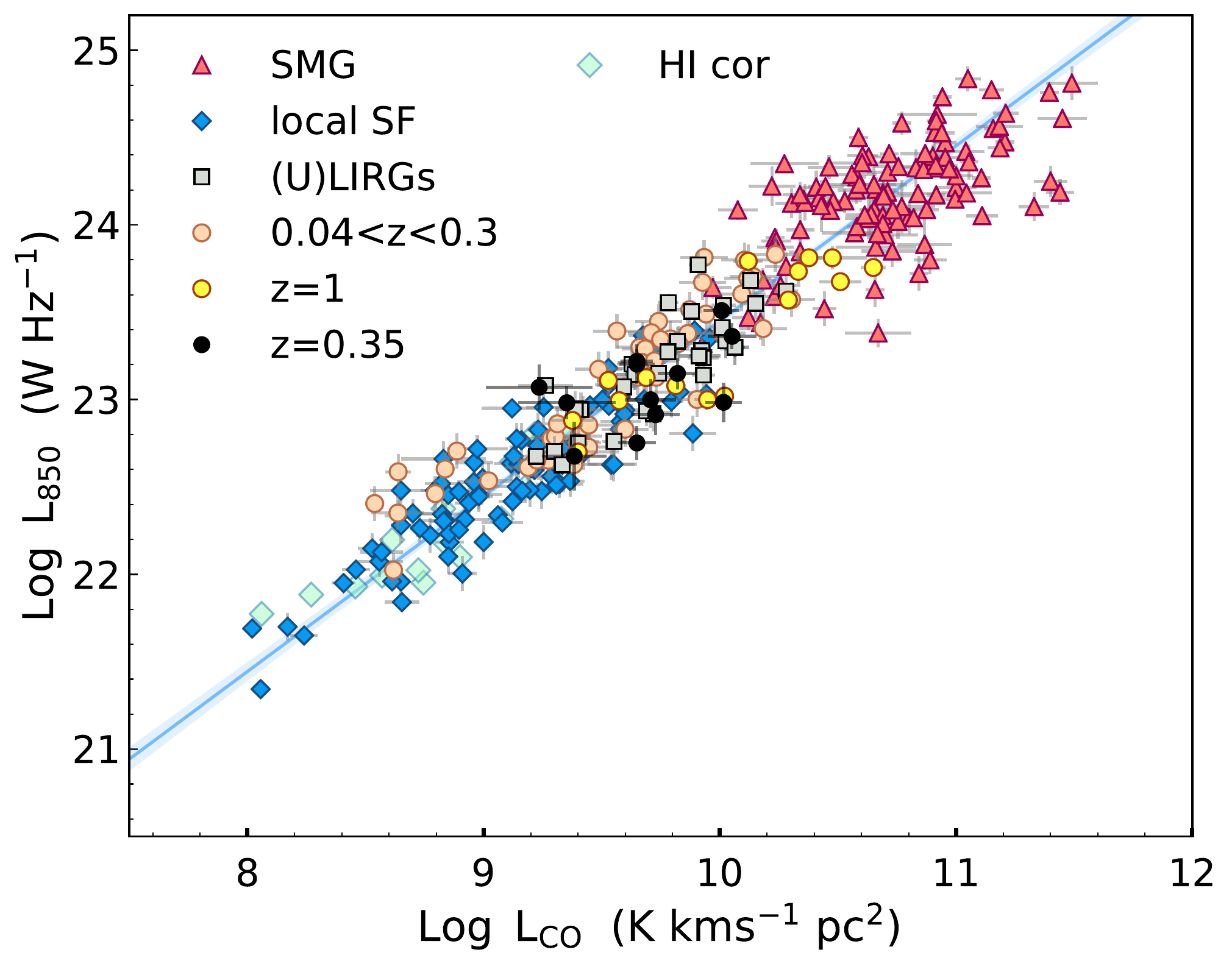}
\hspace*{0.2cm}
\includegraphics[width=0.48\textwidth,trim=0cm 0cm 0cm 0cm, clip=true]{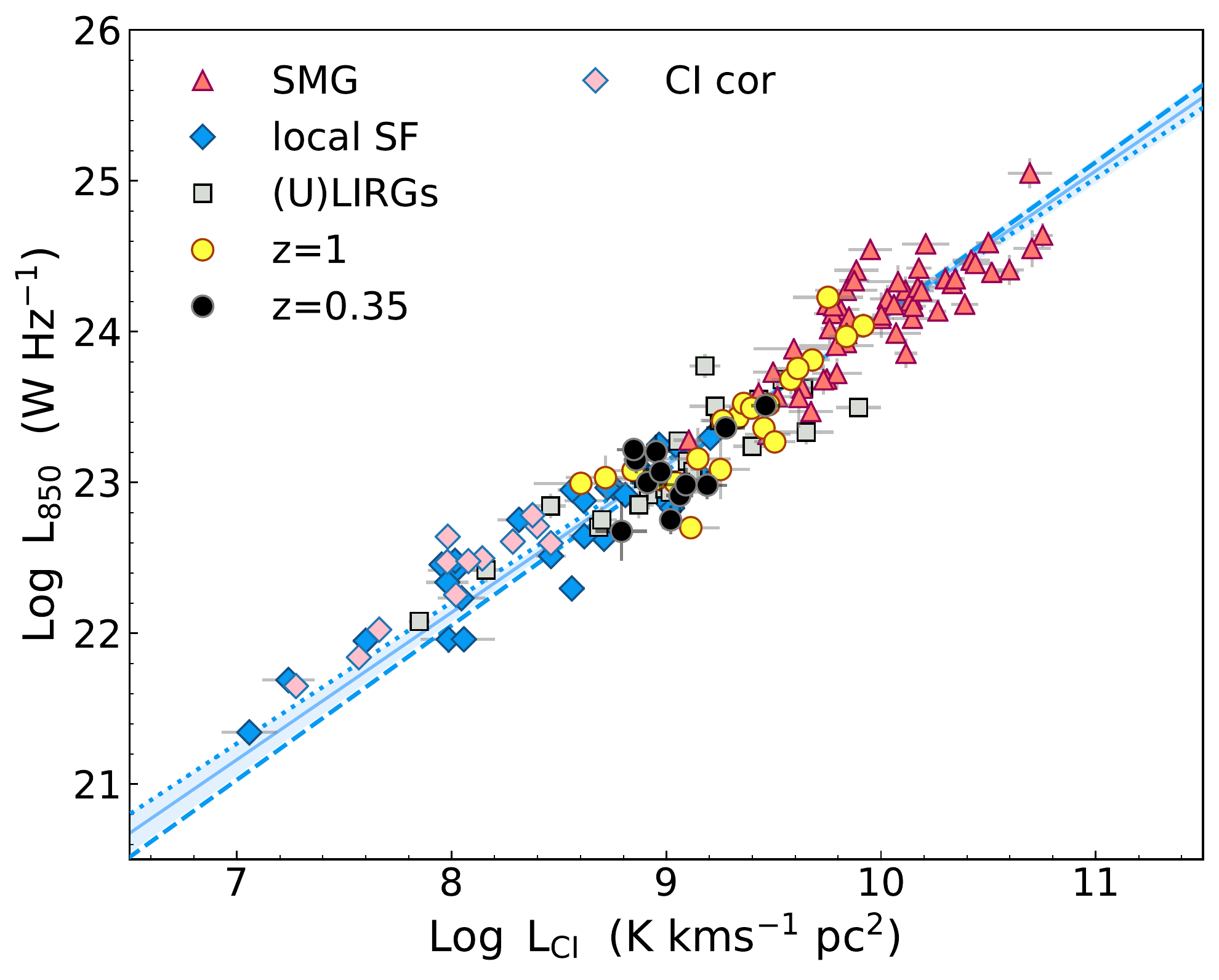}\\
\includegraphics[width=0.48\textwidth,trim=0cm 0cm 0cm 0cm, clip=true]{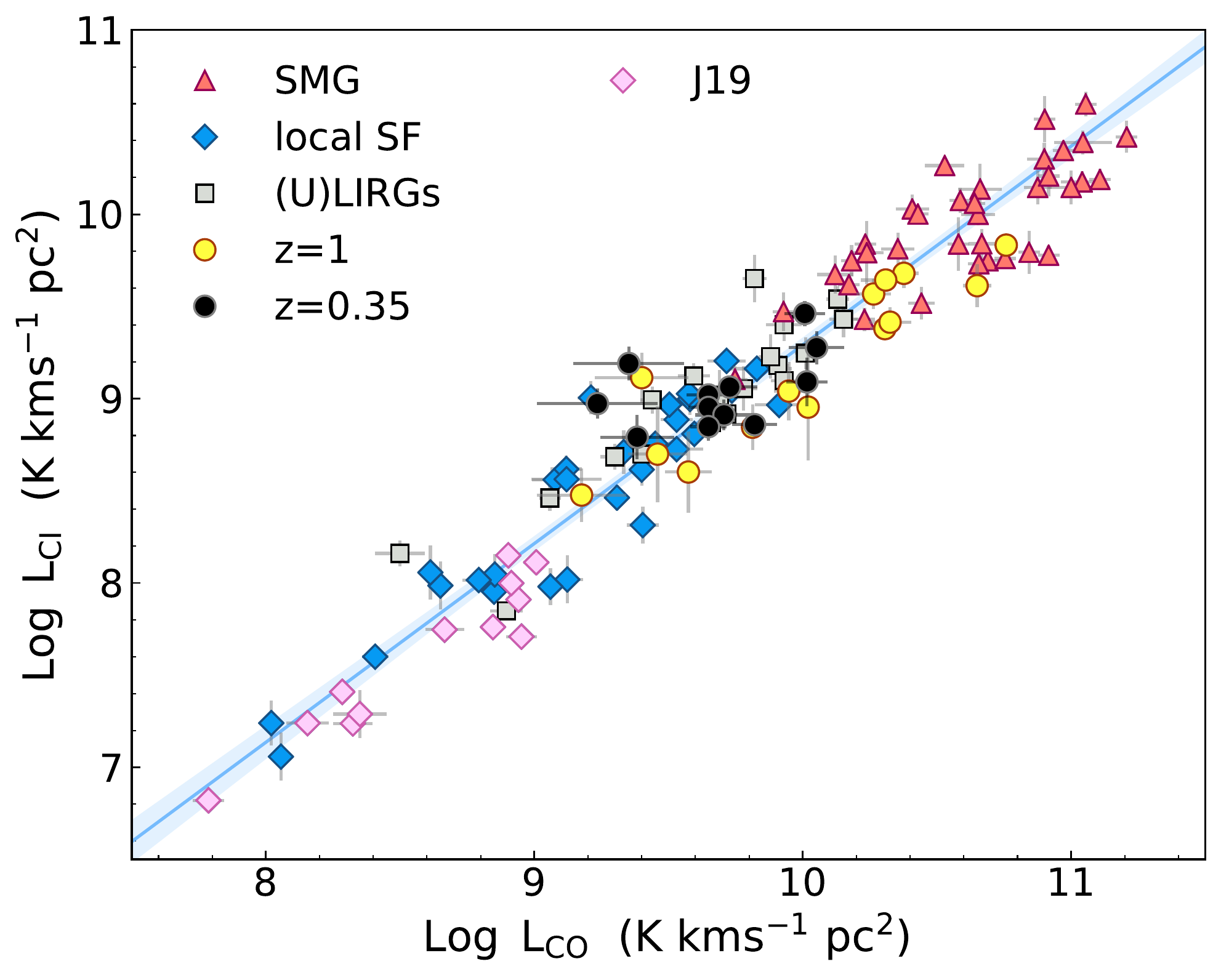}
\hspace*{0.2cm}
\includegraphics[width=0.48\textwidth,trim=0cm 0cm 0cm 0cm, clip=true]{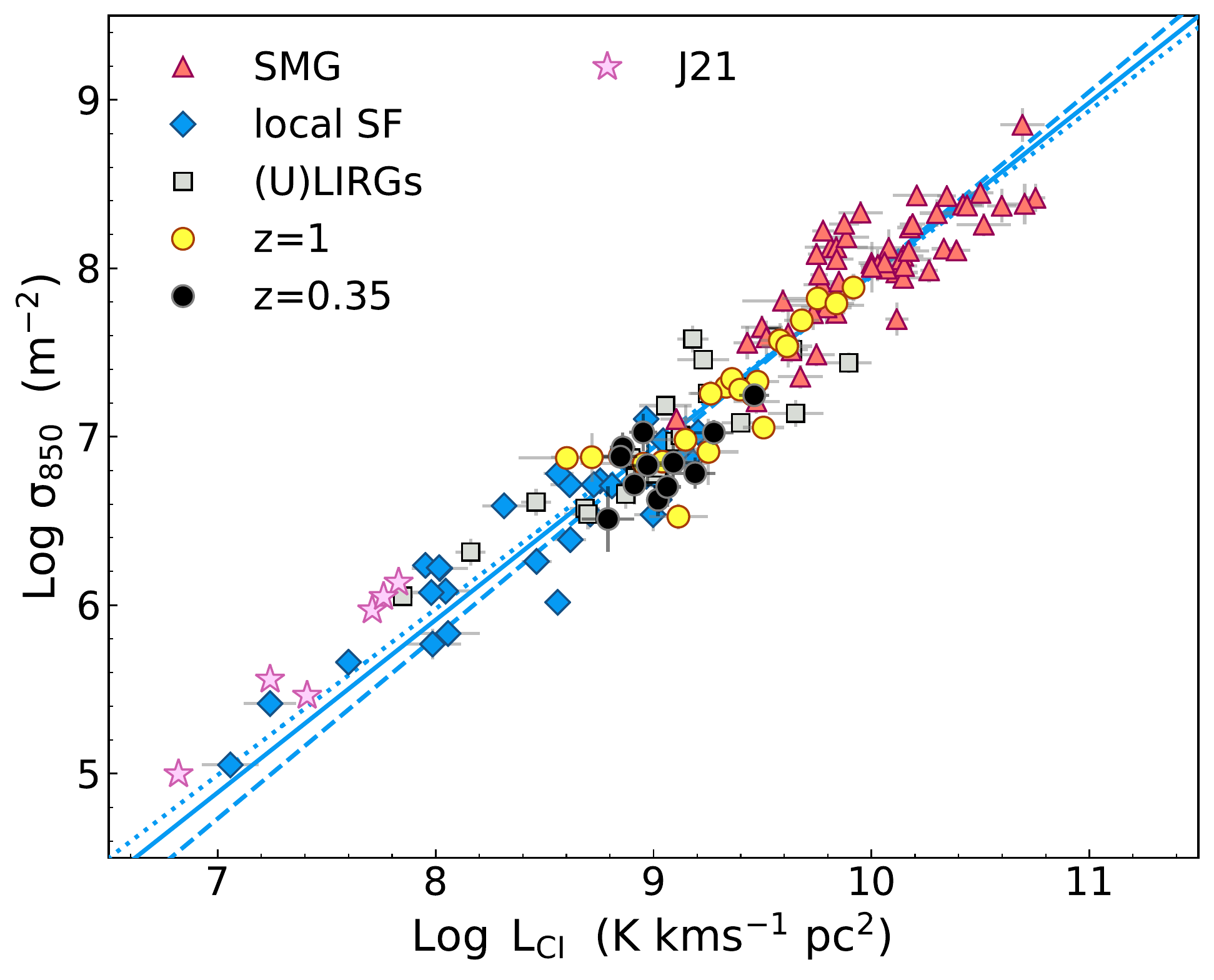}\\
\caption{Relationships between the observed luminosities. The blue
  line and shaded regions in each panel show the best fit and the
  $2\sigma$ uncertainty (see Table~\ref{fitsT} for parameters). {\bf
    Top left:} \lsub--\lcoa. Cyan diamonds indicate those local
  galaxies which have \fhi\ $>1$ and have corrected values of
  \lsub. The best fit is shown, including the H\,{\sc i}-corrected set, but
  excluding the lo-VALES galaxies (peach circles with log \lcoa\ $<8.9$; see
  Appendix~\ref{HIS}). Choosing only galaxies with \fhi\ $<1$ does not
  change the results; the fit is consistent with a linear slope with a
  high degree of confidence. {\bf Top right:} \lsub--\lci. The
  resolved \CIcor\ galaxies from J19 that have an aperture correction
  (Appendix~\ref{notesS}) applied to \lci\ in order to compare to
  \lsub\ are shown as pink diamonds. The dotted blue line shows the
  fit when these galaxies are included. The solid blue line is the fit that
  excludes these galaxies, while the dashed blue line is the fit to
  the \Lhi\ galaxies only. {\bf Bottom left:} \lci--\lcoa. The pink
  diamonds are the \CIcor\ galaxies, where now the \CI\ and CO
    are from matched regions, however, they are still offset to lower
  \lci\ for a given \lcoa. The fit including the pink diamonds is
  steeper than linear, while excluding them (solid line) returns a
  linear slope.  {\bf Bottom right:} \cs--\lci. We compare the
  850-\mic\ emissivity ($\cs=\Md\times\kd$) to \lci. This allows us to
  show dust data in the resolved galaxies from \citet{Jiao2021} (pink
  stars) measured in the same region as [\CI](1--0). The dotted blue
  line shows the fit including the pink stars, the solid line shows
  the fit excluding the \CIcor\ galaxies and the dashed line shows the
  fit to \Lhi\ galaxies only. The local resolved (\CIcor) galaxies
  from J19/J21 appear to have less \CI\ emission for a given dust or
  CO luminosity, an effect which persists even when comparing matched
  regions (bottom row).}
\label{LcorF}
\end{figure*}

Histograms of the tracer ratios are shown in Fig.~\ref{LhistsF}, split
by factors such as lensing, redshift, SFR, and other notable
quantities. The correlations of the three tracer luminosities are
shown in Fig.~\ref{LcorF}, where the various samples are colour coded
and labelled and in each panel the blue line and shaded region
represent the best fit and $2\sigma$ error interval. Fitting was
performed using our own Orthogonal Distance Regression (ODR) method, which includes $x$ and $y$ errors,
intrinsic scatter as a third parameter, and covariance in errors where
required. The method is described in detail in
Appendix~\ref{ODRS}. Fit parameters, slope $m$, intercept $c$, and scatter ln $\lambda$
, are listed in Table~\ref{fitsT}
and statistics for the various subsets from Fig.~\ref{LhistsF} are given
in Table~\ref{hypT}. It is instructive to look at these two plots
together for the same luminosity pairs.

\subsection{\lsub\ vs.\ \lcoa}
\label{L850LCOS}

The left-hand column of Fig.~\ref{LhistsF} and Fig.~\ref{LcorF}(a)
show the quantities \lsub\ vs.\ \lcoa. There are no significant differences in the distribution of
\lsub/\lcoa\ with lensing, redshift or SF-mode, but the observed
uncorrected ratio is significantly higher for galaxies with log \lcoa\
$<8.9$ (Fig.~\ref{LhistsF}; left green histogram). These log \lcoa\ $<8.9$
galaxies tend to have optical disks dominated by atomic hydrogen
(\fhi\ $>1$) and the likely increased contribution to \lsub\ from dust
associated with H\,{\sc i} rather than \mol\ results in the offset to
higher \lsub/\lcoa\ ratios. For local galaxies with \fhi\ $>1$ we
apply a correction (Appendix~\ref{HIS}) which appears to remove this
offset (cyan diamonds in Fig.~\ref{LcorF}(a)). Galaxies with log \lcoa\
$<8.9$ from the VALES sample at $0.04<z<0.3$
\citep{Villanueva2017,Hughes2017} have noticeably higher \lsub/\lcoa\
ratios than VALES galaxies with log \lcoa\ $>8.9$ (peach circles in
Fig.~\ref{LcorF}(a) and Table~\ref{hypT}). There are no published
H\,{\sc i} measurements for VALES, but we suspect that the lo-VALES
sample with log \lcoa\ $<8.9$ is likely to be H\,{\sc i}-rich, based on
the similarity between the log \lcoa\ $<8.9$ and \fhi\ $>1$ categories in
the third panel. We therefore exclude lo-VALES from our averages, as
we suspect they are in need of a correction for H\,{\sc i} but we have
no means to apply one. We also recommend that a low \lcoa\ requires
careful consideration of H\,{\sc i}-related dust. In
Fig.~\ref{LcorF}(a), the tracers show a linear dependence, regardless
of exactly which galaxies are included (Table~\ref{fitsT}). The
lo-VALES galaxies are excluded from all the fits.

\subsection{\lsub\ vs.\ \lci}

The central column of Fig.~\ref{LhistsF} and Fig.~\ref{LcorF}(b) show
\lsub\ vs.\ \lci, the pair with the least scatter
($\ln \lambda$ in Table~\ref{fitsT}). There are no significant differences in the distributions of \lsub/\lci\ as a function of lensing, redshift or SF-mode. The \lsub--\lci\ sample has
12 galaxies with large angular sizes from J19 that have only part of
their disk detected in \CI, denoted \CIcor. To compare \lci\ and
\lsub\ for these galaxies, we need to apply an aperture correction to
their \CI\ fluxes, thereby assuming that their \lci/\lcoa\ ratios
remain roughly constant across the disk (see
Appendix~\ref{J19S}). Some of the correction factors are very large
(up to 0.7 dex); even after correction, their \lsub/\lci\ ratios are
significantly offset from the rest of the \ms\ sample (green
histogram). Fig.~\ref{LcorF}(b) shows \lsub\ vs.\ \lci, with the
\CIcor\ galaxies as pink diamonds. The fit to all galaxies including
the \CIcor\ subset is shown as a dotted blue line, and has a
sub-linear slope $m$ ($3\sigma$), although the difference is very small
($m=0.937\pm0.020$). Excluding the \CIcor\ galaxies from the fit
leaves a linear relationship, shown by the blue solid line.  At first
glance, this and the green histogram in Fig.~\ref{LhistsF} suggest
that our CO-based corrections are insufficient; however,
Fig.~\ref{LcorF}(d) shows that the problem is not simply the
assumption used to correct \lci\ to match the global \lsub. This plot
shows the dust emissivity cross-section, \cs, equivalent to \lsub\ but
with the temperature sensitivity removed\footnote{\cs\ is derived from
  the data provided in \citet{Jiao2021} by multiplying the dust mass
  in the \CI\ aperture by the \kd\ used in their method, \kd\ =
  0.034\,\kunit\ \citep{draine2003,Draine2007}.}.  Importantly, this quantity is
measured in the same aperture as \lci. The \CIcor\ galaxies are shown
as pink stars, and they -- along with many other low-luminosity
galaxies in J21 -- still appear to have less \CI\ emission for a given
amount of dust.

\begin{table*}
\caption{Parameters of robust ODR fits between variables using
  MCMC, co-variant errors and including intrinsic scatter, $\ln \lambda$.}
\begin{adjustbox}{center}
\begin{tabular}{llcrrrrlc}
\toprule
\multicolumn{1}{l}{Log $x$}&\multicolumn{1}{l}{Log
                             $y$}&\multicolumn{1}{c}{Group}&\multicolumn{1}{c}{$m$}&\multicolumn{1}{c}{$c$}&\multicolumn{1}{c}{ln
                                                                                                             $\lambda$}&\multicolumn{1}{c}{$r_{\rm
                                                                                                                         s}$}&\multicolumn{1}{c}{$p$}&\multicolumn{1}{c}{$N$}
  \\
\midrule
\lcoa & \lci & BL & 1.023 (0.029) & $-0.93$ (0.29) & $-2.09$ (0.09) & 0.94& & 109\\
\lcoa & \lci & BL+\CIcor & 1.078 (0.026) & $-1.49$ (0.25) & $-2.03$ (0.09) & 0.96 & & 121\\

\lcoa & \lci & \Lhi & 0.950 (0.035) & $-0.18$ (0.37) & $-2.07$ (0.10) & 0.92 & &97\\

\midrule

\lci & \lsub & BL & 0.976 (0.024) & 14.33 (0.23) & $-2.11$ (0.09) & 0.95 &&  140\\
\lci & \lsub & BL+\CIcor & 0.937 (0.020) & 14.71 (0.19) & $-2.08$ (0.08) & 0.96 &  &152\\
\lci & \lsub & \Lhi & 1.024 (0.030) & 13.86 (0.30) & $-2.17$ (0.09) & 0.94 &  &128\\

\lci & \cs & BL & 1.024 (0.025) & $-2.28$ (0.24) & $-2.05$ (0.08) & 0.95 & & 140\\
\lci & \cs & BL+\CIcor & 0.997 (0.021) & $-2.01$ (0.20) & $-2.07$ (0.08) & 0.96 & & 152\\

\midrule
\lcoa & \lsub & BL & 1.003 (0.015) & 13.42 (0.15) & $-2.01$ (0.05) & 0.96 & & 326\\
\lcoa & \lsub & \fhi<1 & 1.002 (0.017) & 13.43 (0.16) & $-1.99$ (0.05) & 0.96 & & 310\\

\lcoa & \lsub  & \Lhi & 0.983 (0.026) & 13.63 (0.26) & $-1.91$ (0.06) & 0.93 & & 226\\

\midrule

\Lir & \lci/\lcoa & BL+\CIcor & 0.071 (0.02) & $-1.57$ (0.23) & $-1.70$ (0.09) & 0.26 & 0.005 & 121\\
\Lir & \lci/\lcoa & BL & 0.034 (0.02) & $-1.11$ (0.25) & $-1.70$ (0.09) & 0.09 & 0.34 & 109\\

\midrule
\td & \lci/\lcoa & BL+\CIcor & 1.23 (0.23) & $-2.60$ (0.35) & $-$2.20 (0.12) & 0.31 &   & 115\\
\td & \lci/\lcoa & BL & 0.83 (0.28) & $-2.0$ (0.4) & $-$2.00 (0.15) & 0.15 & 0.12   & 103\\

\midrule
\Lir/\lsub & \mwtd &  & 0.216 (0.010) & 3.90 (0.12) & $-3.76$ (0.15) & 0.82 && 152\\

\Lir  & \mwtd &  & 0.070 (0.004) & 0.60 (0.05) & $-3.10$ (0.10) & 0.80 && 152\\
\midrule

\Lir & \asub & BL & 0.045 (0.007) & 12.271 (0.082) & $-3.36$ (0.30) & 0.46 && 230\\
\Lir & \aco & BL & 0.59 (0.09) & $-0.91$ (1.10) & $-1.00$ (0.50) & 0.46 && 230\\
\Lir & \aci & BL & $-0.052$ (0.010) & 1.896 (0.124) & $-3.90$ (0.20) & $-0.48$ && 82\\
\Lir & \Xci & BL & $0.028$ (0.011) & $-5.136$ (0.133) & $-3.70$ (0.23) & 0.29 & 0.008 & 82\\
\bottomrule
\end{tabular}
\end{adjustbox}
\flushleft{Notes: $y=mx+c$ fit parameters are given with 1$\sigma$
  errors in parentheses. Parameters are calculated accounting for the
  errors in both $x$ and $y$ using the robust orthogonal distance
  regression described in Appendix~\ref{ODRS}. Errors are sampled
  using the {\sc emcee} MCMC sampler. Intrinsic scatter ($\lambda$) is
  fitted as a third parameter. $r_{\rm s}$ is the Spearman rank
  correlation coefficient, and $p$ is the probability, shown when
  $p>0.005$.  $N$ is the number of galaxies in that
  regression. `Group' defines the galaxies on which the regression is
  performed: BL = baseline (excludes \CIcor\ and lo-VALES galaxies),
  while galaxies with \fhi>1 are corrected as described in
  Appendix~\ref{HIS}.}
\label{fitsT}
\end{table*}

\begin{table*}
\caption{Two-sample KS-test result and Z-test statistic for the following parameter pairs shown in Figs~\ref{LhistsF} and \ref{tdhistF}.}
\begin{adjustbox}{center}
\begin{tabular}{@l^l^c^l^c^c^c^c^c}
\toprule
Quantity & $A$ & $N_{\rm A}$ & $B$  & $N_{\rm B}$ & $\bar{A}$ & $\bar{B}$ & $Z(\sigma)$ & $P_{\rm KS}$\\
\midrule
\lsub/\lcoa$^\dag$ & log \lcoa<8.9 & 50 & \lcoa>8.9 (MS) & 138 & $13.600\pm 0.030$ & $13.420\pm 0.016$ & 5.3 & 3e-5\\
\lsub/\lcoa$^\dag$ & \fhi<1 (MS) & 168 & \fhi>1 (MS) & 24 & $13.456\pm 0.014$  &  $13.685\pm 0.042$ &  5.2  & 3e-6\\
\lsub/\lcoa$^\dag$ &  lo-VALES & 7 & \fhi>1 & 17 & $13.741\pm 0.061$  &  $13.663\pm 0.053$ &   & 0.57\\
\lsub/\lcoa &  \fhi<1 (MS) & 168 & \fhi>1 (MS$^{\ast}$) & 17 & $13.456\pm 0.014$  &  $13.448\pm 0.039$ &    & 1.0\\

\lsub/\lci  & MS  & 61  & \CIcor & 12 & $14.108\pm 0.027$  & $14.36\pm 0.036$ & 5.6  & 6e-4\\

\lci/\lcoa  & $z<3$ (SMGs) & 37 & $z>3$ & 11 & $-0.686\pm 0.031$ & $-0.512\pm 0.044$ & 3.2 & 0.012\\

\lcoa/\lci & MS & 55 & SMGs & 54 & $-0.743\pm 0.028  $ & $-0.651\pm0.028$ & 2.3 & 0.056\\
\lcoa/\lci & MS+\CIcor & 66 & SMGs & 54  & $-0.786\pm0.026$ & $-0.651\pm0.028$ & 3.5 & 0.006 \\

\midrule

\td         & MS & 174 &  SMGs  & 160 & $31.1\pm0.4$    & $38.3\pm0.7$    & 8.8   &  2e-12\\ 

\mwtd     & MS & 82 & SMGs  & 52 & $23.0\pm0.4$ & $30.1\pm0.7$ & 8.8 & 7e-13\\
\bottomrule
\end{tabular}
\end{adjustbox}
\flushleft{$^{\dag}$Using \lsub\ without correction for \fhi>1 (as
  this is the driver of the difference).\\
  $^{\ast}$Not including the lo-VALES galaxies and with the \HI\ correction applied. } 
\label{hypT}
\end{table*}

\subsection{\lci\ vs.\ \lcoa}

The right-hand column of Fig.~\ref{LhistsF} and Fig.~\ref{LcorF}(c)
show \lci\ vs.\ \lcoa. There are no significant differences in \lci/\lcoa\ for strongly lensed vs. unlensed sources,
but the highest redshift, $z>3$, galaxies have higher \lci/\lcoa\
ratios at marginal significance ($p=0.012$). There are only 11
galaxies at $z>3$ and a larger sample is needed to determine if this
is a genuinely significant trend. Fig.~\ref{LcorF}(c) shows the
\CIcor\ galaxies as pink diamonds, where the \CI\ and CO fluxes are
measured in the same apertures by J19. The solid line shows the fit to
all galaxies, which is non-linear at 3$\sigma$ significance
($m=1.078\pm0.026$). The slope becomes linear once the \CIcor\
galaxies are removed. The green histogram in Fig.~\ref{LhistsF}
(right) shows more clearly why we see this: the \CIcor\ galaxies have
significantly lower \lci/\lcoa\ ratios compared to other galaxies at
similar or higher luminosity. The bottom histogram shows a marginal
difference between galaxies with different SFRs ($p=0.056$) when
excluding the \CIcor\ galaxies, which becomes significant when they
are included ($p=0.006$).
 
\subsection{Resolved \CI\ fluxes from Herschel FTS mapping}
\label{J19S}

The local resolved galaxies observed with {\it Herschel} FTS \citep{Jiao2019} lie off the global trends seen in Fig.~\ref{LcorF}. There are possible physical explanations why lower luminosity, and more quiescently star-forming galaxies might have lower \CI/CO line ratios (for example, different ISM environments in terms of their position in the CR energy density vs average molecular gas density diagram: see Figure 1 in \citealp{Bisbas2015}). Low ratios of \lci/\lcoa\ have been found in other studies, most intriguingly in the case of the interacting LIRG NGC~6052 using ALMA \citep{Michiyama2020}, and some high-$z$ strongly lensed sources \citep{Harrington2021}. Such ratios tend to be unusual in higher luminosity samples, however, whereas the resolved FTS sample has a very low {\em average} for the \CI\ line ratios with both CO and dust.

We recommend caution in the interpretation of the data for these resolved FTS sources because another team subsequently presented the same data but drew different conclusions \citep{Crocker2019}. We can therefore only note 
that the \CI\ fluxes from {\it Herschel} FTS mapping datasets are not always consistent when analysed by different teams.\footnote{\citeauthor{Crocker2019}
  did not  provide integrated fluxes,  nor a method to  determine them from their published measurements; hence,  we cannot use their work directly in our analysis. Q.~Jiao has provided us with the maps used in J19, enabling us to check the measurements independently and extend our analysis, but we have had no responses to our requests for integrated fluxes or the details of the method used from the Crocker team.} As these resolved FTS measurements are 
essentially the only source of \CI\ data at \Llo, and carry a lot of weight in  \Lir\ and SFR correlations, we chose not to include the \CIcor\ galaxies in the statistical analysis. If, instead, we take the J19 measurements at face value -- they signpost a fundamental physical change in \CI\ properties, a finding which clearly warrants further study with ground-based facilities. We discuss possible
physical mechanisms for changes in the \lci/\lcoa and \lci/\lsub\ ratios in Appendix~\ref{J19A}.

\subsection{Trends with global indicators of star-formation.}

\begin{figure}
\includegraphics[width=0.45\textwidth,trim=0cm 0cm 0cm 0cm, clip=true]{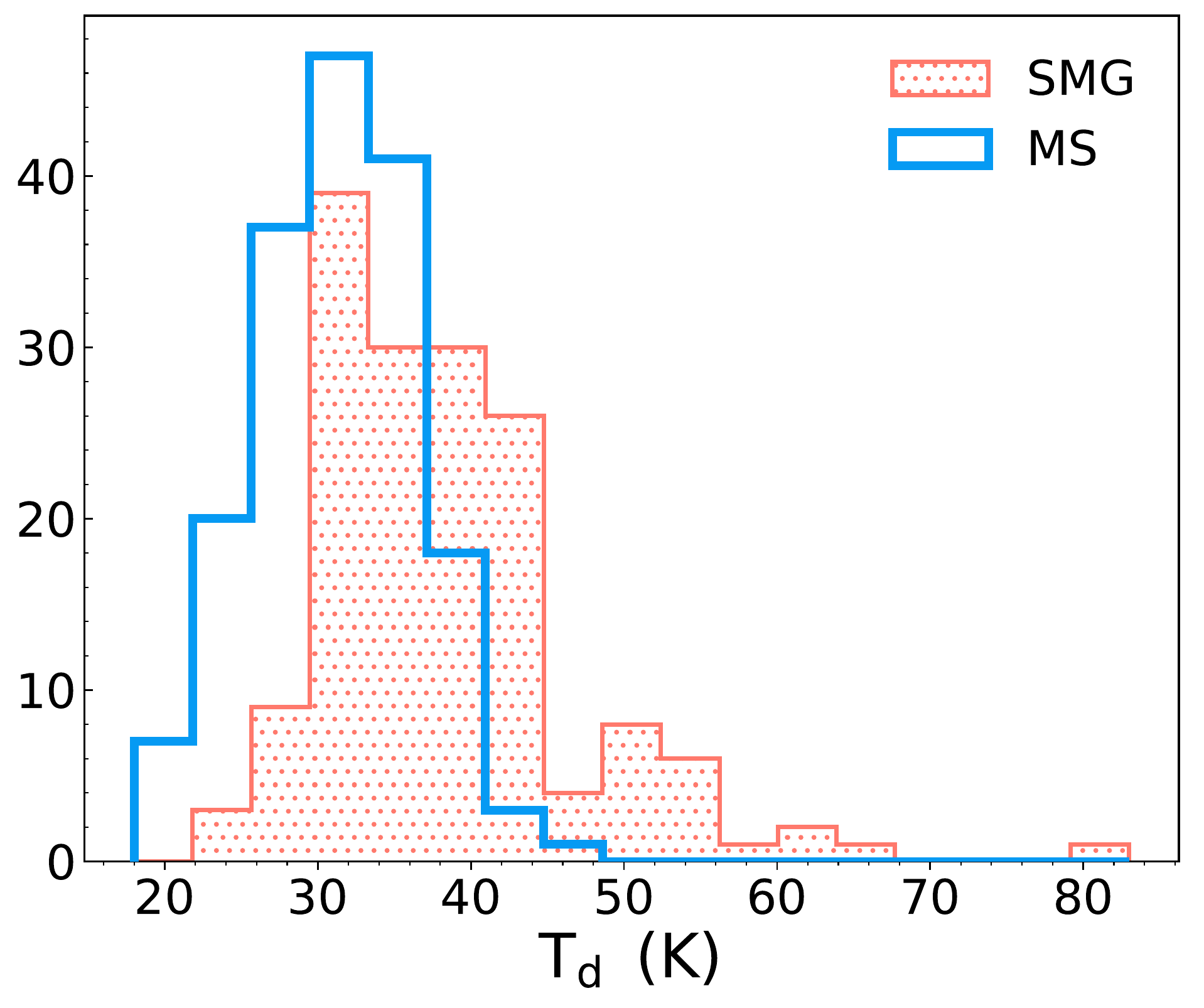} 
\includegraphics[width=0.45\textwidth,trim=0cm 0cm 0cm 0cm, clip=true]{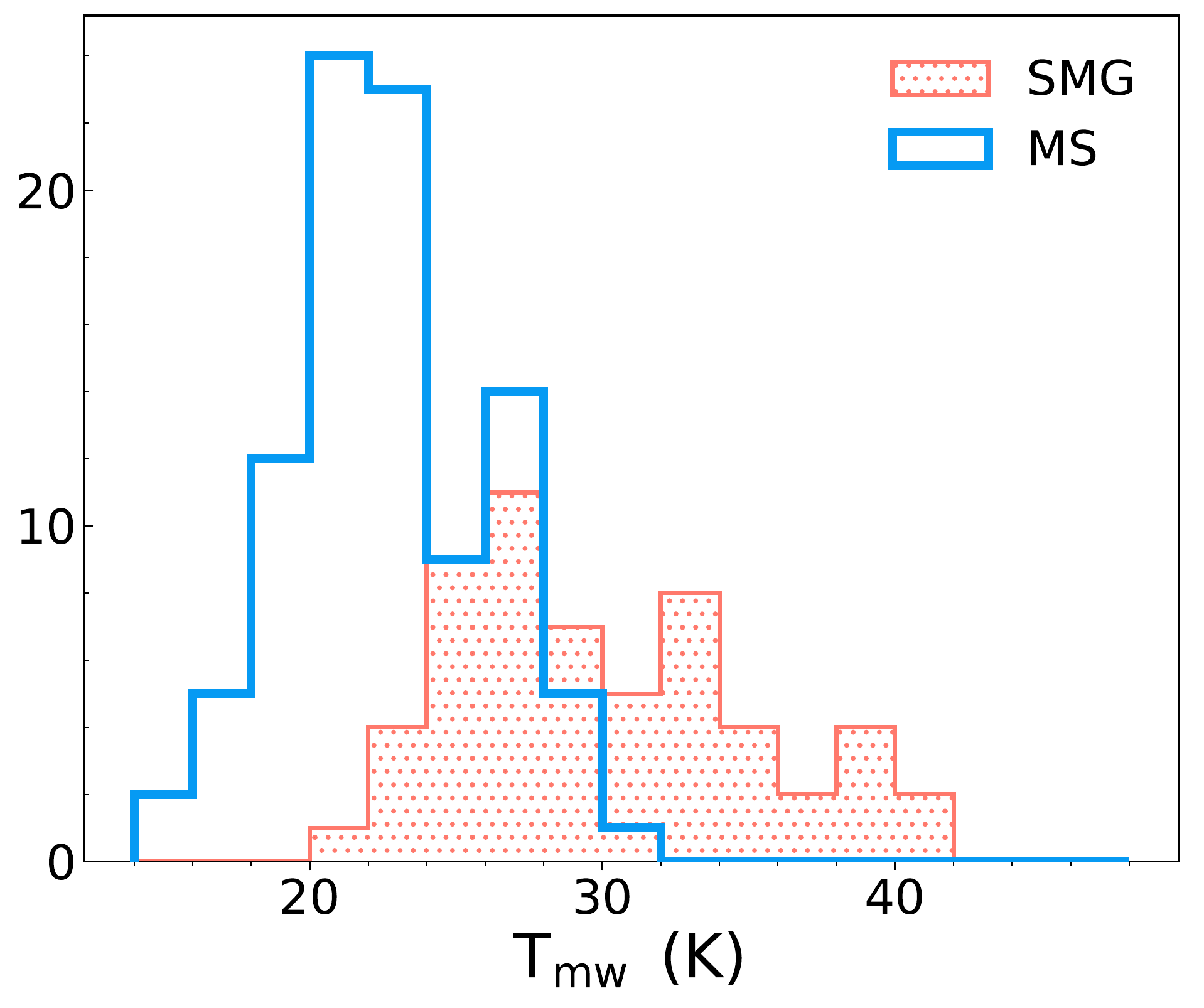}  
\caption{Histograms of {\bf (top)} luminosity-weighted dust
  temperature (\td) from MBB fits {\bf (bottom)} mass-weighted dust
  temperature (\mwtd) from fits allowing for multiple dust
  temperatures. Means and KS test results are given in Table~\ref{hypT}.}
\label{tdhistF}
\end{figure}

\begin{figure*}
\includegraphics[width=0.48\textwidth,trim=0cm 0cm 0cm 0cm, clip=true]{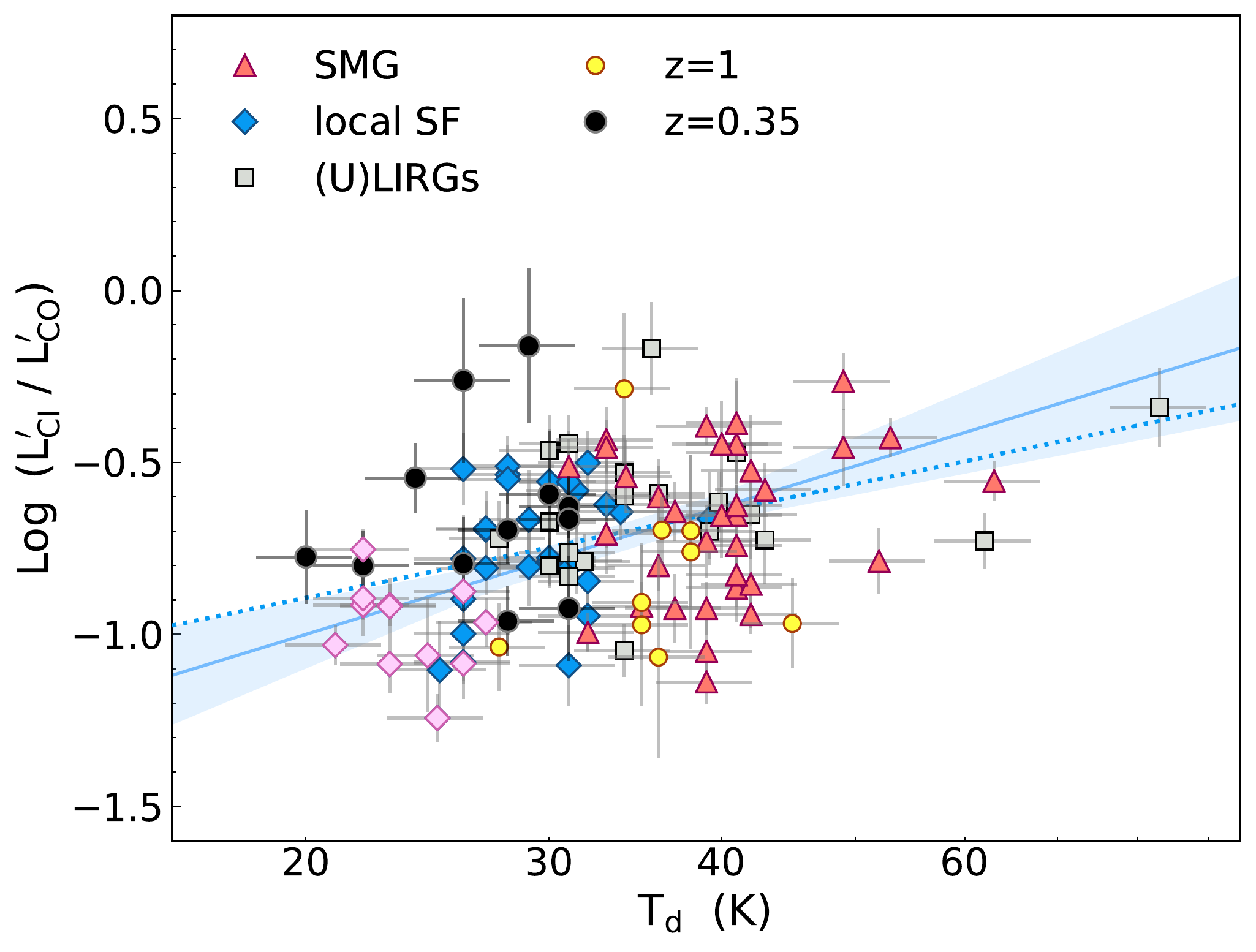}
\includegraphics[width=0.48\textwidth,trim=0cm 0cm 0cm 0cm, clip=true]{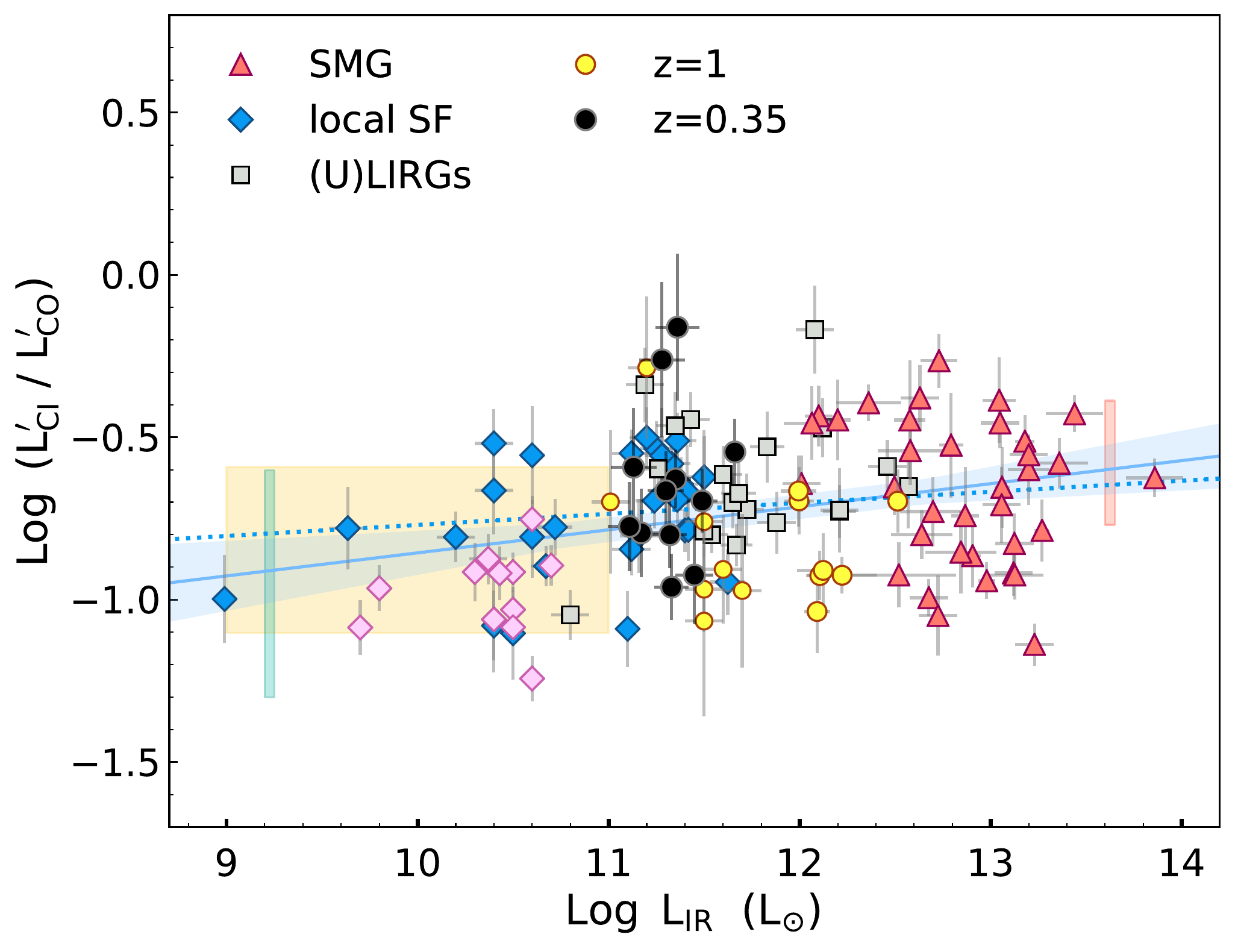}\\
\caption{\lci/\lcoa\ luminosity ratio (equivalent to \aco/\aci) as a
  function of luminosity-weighted (peak SED) dust temperature (left)
  and log \Lir\ (right). The yellow shaded regions in the right panel
  represent the 1$\sigma$ range of values found by \citet{Crocker2019}
  for resolved galaxies mapped with the {\it Herschel} FTS. The
  vertical red and blue bars show the 1$\sigma$ range of values for
  QSOs/SMGs and the Milky Way, respectively
  \citep{Walter2011,Frerking1989}. The best fit to all galaxies and
  2$\sigma$ confidence region are shown as the blue line and shaded
  area (parameters are given in Table~\ref{fitsT}). The \lci/\lcoa\
  ratio has a significant correlation with both \td\ and log \Lir\ when
  all galaxies are included, but when excluding the \CIcor\ galaxies,
  the correlation all but disappears (blue dotted line).}
\label{tdcorrF}
\end{figure*}

Finally, we check to see if any of the tracer ratios are sensitive to
SFR indicators. In our data-set the dust observables \Lir\ and \td\ are indicators of the intensity and magnitude
of star formation in galaxies
\citep[e.g.][]{Kennicutt1998,Foyle2012,Liu2021}. As expected, the
distribution of \td\ is very different for the MS galaxies and SMGs
(Fig.~\ref{tdhistF} and Table~\ref{hypT}), reflecting the increase in
the intensity of star formation in the SMGs. The only tracer ratio
sensitive to these SF indicators is \lci/\lcoa, which in Figure~\ref{tdcorrF} is seen to increase with
\Lir\ and \td\ when all galaxies are considered. Such a trend was not
reported for smaller samples over a more limited range of luminosity
\citep[e.g.][]{Jiao2017,Jiao2019}, presumably because of limited
statistics. However, if the \CIcor\ galaxies (pink diamonds) are
excluded, the correlation all but disappears (blue dotted line).

Naively, we might expect \lci/\lcoa\ to rise with increasing SFR intensity, due to the expected destruction of CO by cosmic rays (CR) in high-SFR environments \citep[e.g.][]{Bisbas2015}. For a given range of \mol\ densities in the typically hierarchical molecular clouds, any increased CR-induced ionisation rate, \zcr\ (due to a rising average CR energy density, $U_{\rm CR}$) will destroy CO in the lower-density, more extended areas, while leaving CO still tracing \mol\ in the more compact, denser regions (see also Figure 1 of \citealp{Papadopoulos2018} for a visual effect of this). Intriguingly, the gas density, $n$(\mol), and CR ionization rate, \zcr, will compete against each other in ULIRG/SMG environments, with the higher <n> expected in their highly turbulent ISM
tending to keep the ordinary CO/\CI\ chemistry in place, even when 
exposed to the higher \zcr\  values.\footnote{We here assume that CR energy density $U_{\rm CR}\propto \rm \rho_{SFR}$ and CR ionisation rate $\zcr \propto U_{\rm CR}$.}
Guessing which one will win this highly non-linear competition
(see Fig 1, 8 in \citealp{Bisbas2015}) is  dangerous in the absence of CO and \CI\ line data. These effects have been probed with a variety of simulations \citep[e.g.][]{Bisbas2015,Bisbas2021,Clark2019ci,Gong2020} and while showing similar trends, they are not easily parameterisable in terms of $n$(\mol) and \zcr; one reason why such cross-calibration efforts of the available gas mass tracers are so important.\footnote{On an individual galaxy basis one could assemble well-sampled CO, $^{13}$CO and \CI\ line SLEDs and overcome these problems with detailed analysis \citep[e.g.][]{PPP6240}. However, even in the ALMA era this remains very expensive in terms of telescope time making it prohibitive for large samples of galaxies.}

The other two tracer ratios show no trends
with either \Lir, \td\ (our proxies for SFR) -- we present the relevant plots in
Appendix~\ref{tdcorrAF} for completeness.

\section{Results}
\label{caltrendS}

In this section we present the results of the optimisation method,
firstly for the daX sample, for which we have all three gas
tracers, and later for the other three samples, for which we have
pairs of tracers.  We investigate trends of the conversion factors
with \Lir\ and SFR. Mean values for the conversion factors are
listed in Table~\ref{caloptT}.

\begin{figure*}
\includegraphics[width=0.98\textwidth,trim=0cm 0cm 0cm 0cm, clip=true]{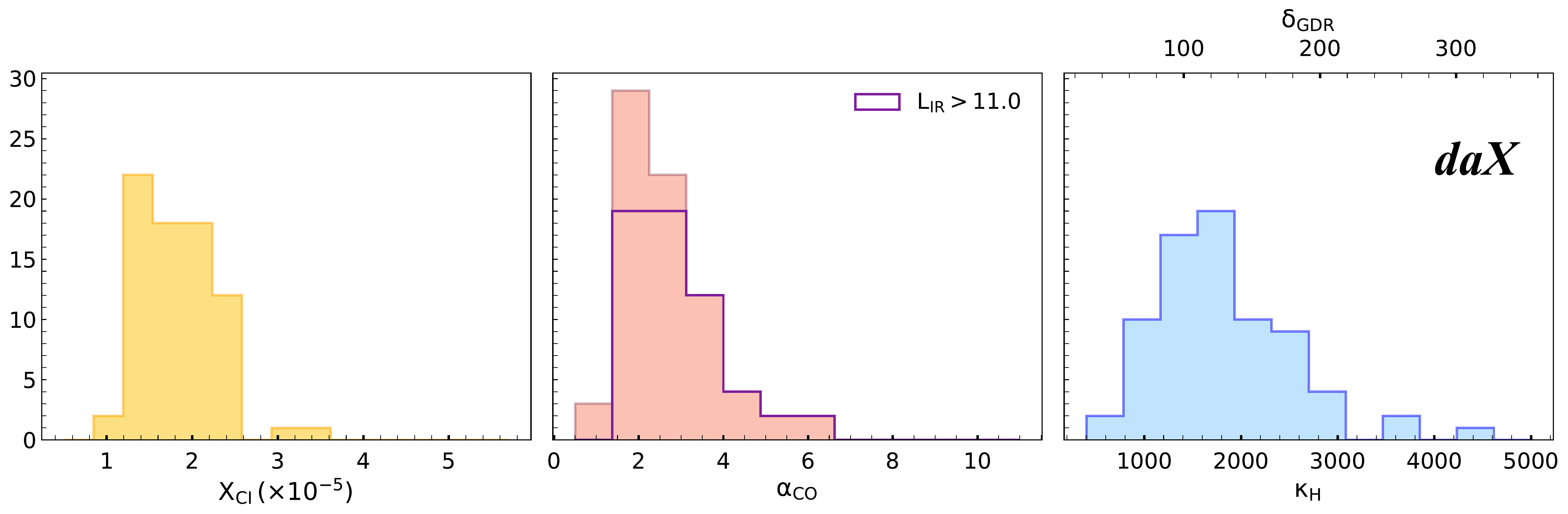}
\includegraphics[width=0.51\textwidth,trim=0.4cm 0cm 0cm 0cm, clip=true]{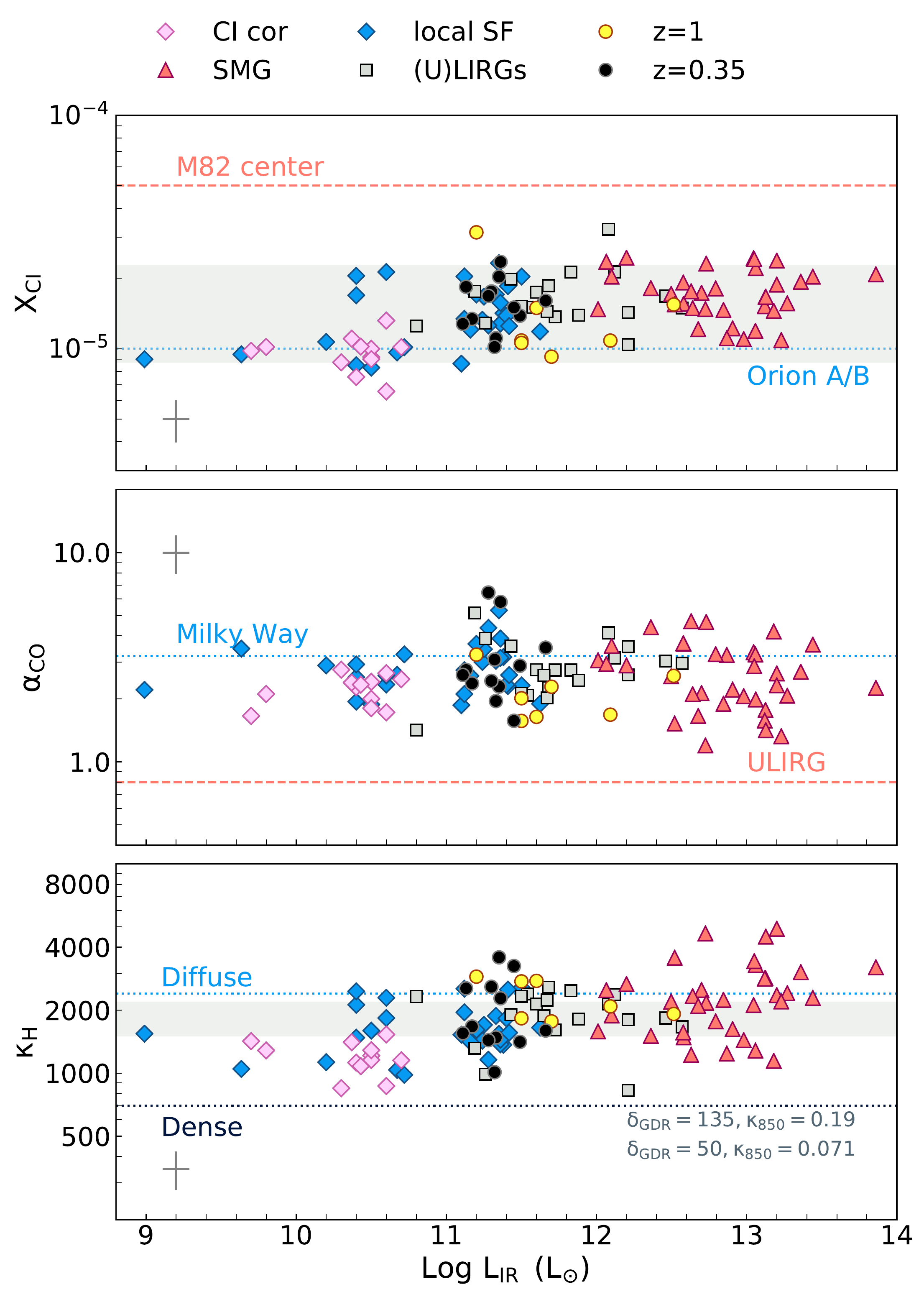}
\includegraphics[width=0.48\textwidth,trim=0cm 0cm 0cm 0cm, clip=true]{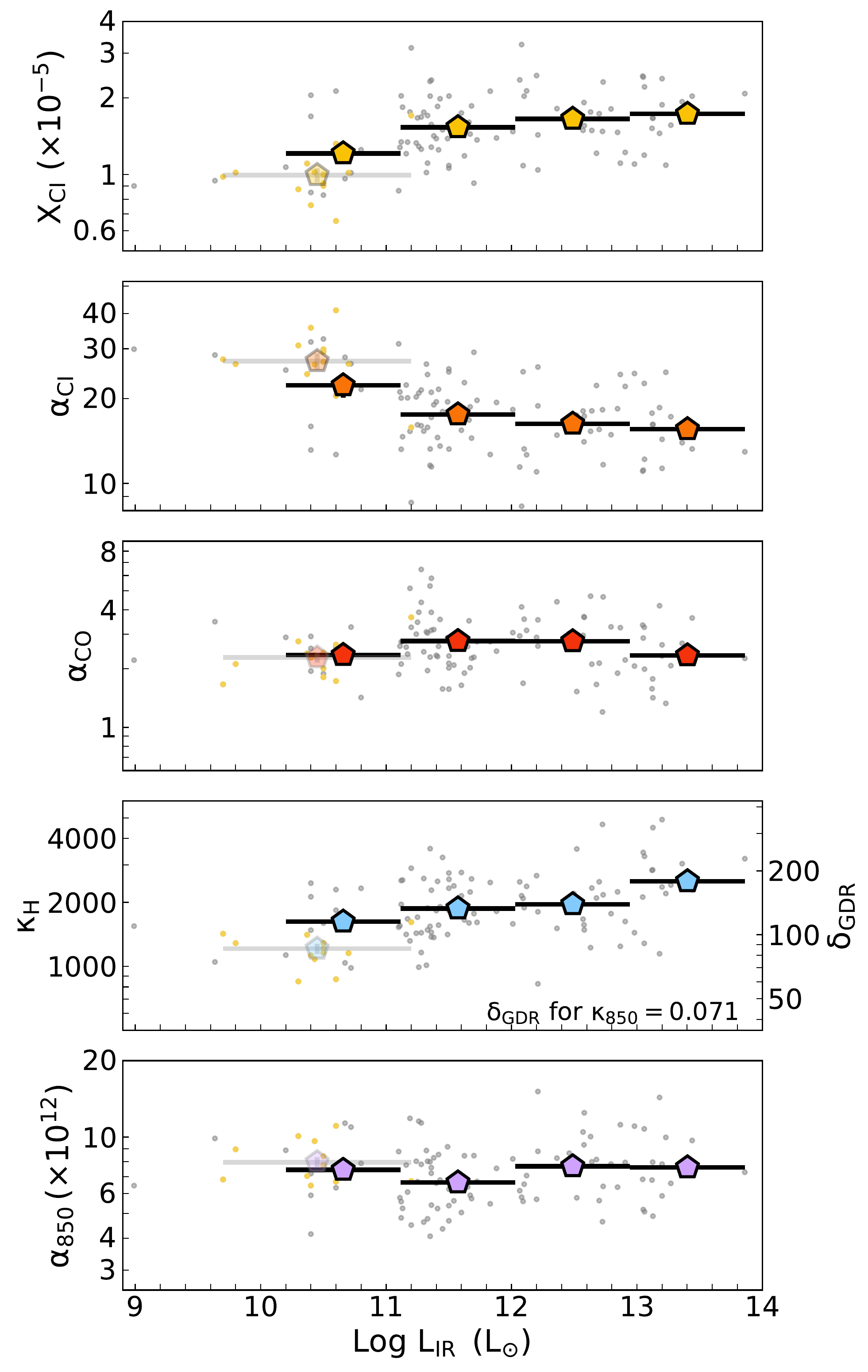}
\caption{{\bf Top row:} Distributions of the three conversion
  factors, \Xci, \aco\ and \kh\ for the daX sample. The overall
  normalisation has been set as $\kh^{\rm N}=1884$\,\khunit, which produces the
  $\gdr=135$ of the Milky Way for $\kd=0.071$\,\kunit
  \citep{Jones2018}. The intrinsic pairwise scatter and conversion factors
  were estimated excluding \CIcor\ and lo-VALES galaxies. {\bf Lower
    left:} The optimised conversion factors as a function of
  \Lir. The grey shaded band represents \Xci\ at solar metallicity
  using the relationship found in GRB and QSO absorber hosts by
  \citet{Heintz2020}, a measure which is independent of assumptions
  about \aco\ or \gdr. {\bf Lower right:} The running means of the
  conversion factors, with error bars of $\sigma/\sqrt{N_{\rm
      bin}}$. The solid shaded bins are the means for the grey points
  which are those used to determine the calibration, the yellow points
  are \CIcor\ and the semi-transparent pentagon is the mean of those
  -- see \S\ref{J19S} for more details.}
\label{caldaXF}
\end{figure*}

Fig.~\ref{caldaXF} shows the results for the daX sample
(Figs~\ref{calXdF}--\ref{calaodF} present the same results for each of
the samples in turn). The top row of each plot shows the
distribution of the relevant physical parameter for \CI\ and dust, and the conversion factor for CO: \Xci, \kh\ and \aco.
The lower left panels show the same quantities for the individual
galaxies as a function of \Lir; each panel indicates a reference
measure to give context. For \Xci, the horizontal lines indicate the
measured extremes found in the local Universe: Orion A/B clouds in the
Milky Way \citep{Ikeda2002} and the starburst centre of M\,82
\citep{White1994}, while the grey shaded region shows the range of
values inferred from observations of GRB hosts and QSO absorbers for
solar metallicity by \citet{Heintz2020}. They use a method which does
not rely on emission measures of dust, CO or \CI\ and so can be
considered independent. For \aco, the horizontal lines indicate the
typical \aco\ for the Milky Way \citep{Bolatto2013} and that commonly
adopted\footnote{In this panel, \aco\ does not include the factor 1.36
  for He.}  for ULIRGs and SMGs \citep{Downes1998}. For \kh, we show a
shaded band indicating the range derived for local galaxies (see
Table~\ref{kappalitT}), along with lines showing the value for the
most diffuse and dense sight lines in the Milky Way
\citep{Remy2017}. The right lower panels show the running log-means as
a function of \Lir\ to make it easier to see any trends, and
additionally includes\footnote{As elsewhere, the empirical parameters
  \aci\ and \asub\ include the factor 1.36 for He.} the empirical
parameters, \aci\ and \asub. The solid shaded bins are the means for
the grey points, which are those used to determine the calibration;
the yellow points are \CIcor\ and the semi-transparent pentagon is the
mean of those -- see \S\ref{J19S} for more details.

Fig.~\ref{caldaXF} shows that for galaxies with \Lhi\ there are only
weak trends of the conversion factors with \Lir. While the
normalisation ($\kh^{\rm N}=1884$\,\khunit) was chosen to produce average dust
properties consistent with the Milky Way and other nearby spirals, the
CO and \CI\ conversion factors derived from the luminosity ratios
also lie within the ranges expected from independent studies. The
averages at \Llo\ are based on only a small number of points (12) and
more \CI\ studies are required to probe quiescent local galaxies.

\begin{table*}
\caption{Mean optimised conversion factors for our various samples.}
\begin{adjustbox}{center}
\begin{tabular}{clcccccccc}
\toprule 
Sample & Selection & $N$ & \Xci & $\sigma_{\bar{X}}$ & \aco &
$\sigma_{\bar{\alpha}}$ & \kh & $\sigma_{\bar{\kappa}}$ & \gdr \\ 
\cmidrule(lr){4-5}
\cmidrule(lr){6-7}
\cmidrule(lr){8-9}
& & & \multicolumn{2}{c}{/$\times 10^{-5}$} & \multicolumn{2}{c}{\aunit} & \multicolumn{2}{c}{\kunit} & \\ 
\midrule

daX &  \Lhi & 90 & $1.59^{+0.45}_{-0.38}$  & 0.04 & $2.66^{+0.96}_{-0.70}$  & 0.10 & $1990^{+738}_{-607}$ & 86 & 141\\ \addlinespace[1pt]

daX &  \Llo  & 12 & $1.18^{+0.60}_{-0.29}$  & 0.13 & $2.44^{+0.56}_{-0.51}$ & 0.17 & $1571^{+732}_{-525}$ & 163 & 112\\ \addlinespace[1pt]

Xd & \Lhi   & 128 & $1.59^{+0.47}_{-0.35}$  &  0.04   &   & & $1946^{+654}_{-464}$  & 53 & 138\\
\addlinespace[1pt]
Xd &  \Llo    & 12 & $1.24^{+0.57}_{-0.27}$ &  0.13   &    & & $1503^{+604}_{-369}$ & 145 & 107\\
\addlinespace[1pt]

ad &  \Lhi   & 240 &     &  & $3.08^{+1.32}_{-0.81}$  & 0.07  & $1936^{+658}_{-504}$  & 45 & 137\\
\addlinespace[1pt]

ad & \Llo   & 88 &    &   & $3.52^{+0.95}_{-0.84}$  & 0.10   & $1718^{+502}_{-339}$ & 44 & 122\\
\midrule
Xa &  \Lhi & 97 & $1.61^{+0.39}_{-0.31}$ &   0.04 & $2.57^{+0.71}_{-0.62}$ & 0.08 & &  &\\
\addlinespace[1pt]
Xa & \Llo$+$\CIcor & 24 & $1.30^{+0.2}_{-0.23}$ &   0.05 & $1.88^{+0.41}_{-0.34}$ & 0.10  & &  &\\
\addlinespace[1pt]
Xa &  \Llo & 12 & $1.37^{+0.34}_{-0.31}$ &  0.09 & $2.11^{+0.18}_{-0.57}$ & 0.15  & &  &\\
\bottomrule
\end{tabular}
\end{adjustbox}
\flushleft{Means of the optimal conversion parameters (\Xci, \aco,
  \kh) and the error on the mean ($\sigma_{\bar{X}}$,
  $\sigma_{\bar{\alpha}}$, $\sigma_{\bar{\kappa}}$) for each
  subset. We calculate the log-mean and express here in the linear
  form. $^\dag$ We also report the gas-to-dust ratio, \gdr, for a
  fiducial \kd=0.071\,\kunit. We use two normalisations: where dust is
  one of the tracers, we use $\kappa^{\rm N}=1884$\,\khunit (equivalent to Milky Way
  $\gdr=135$ for $\kd=0.071$\,\kunit); otherwise, for the Xa
  sample we use $\Xci^{\rm N}=1.6\times10^{-5}$ -- the mid-range of the
  values found by \citet{Heintz2020} for solar metallicity. The errors
  are the 16th and 84th percentiles of the distribution. The \CIcor\
  and lo-VALES galaxies are removed for analysis and the variances are
  derived from the same set.}
\label{caloptT}
\end{table*}

\subsection{A calibration for the gas masses}
\label{prescS}

We next give a prescription for estimating gas mass, tailored to how
many tracers are available and -- where appropriate -- the type of
galaxy being investigated.

\subsubsection{Dual-band}

\begin{figure*}
\includegraphics[width=0.8\textwidth,trim=0cm 0cm 0cm 0cm, clip=true]{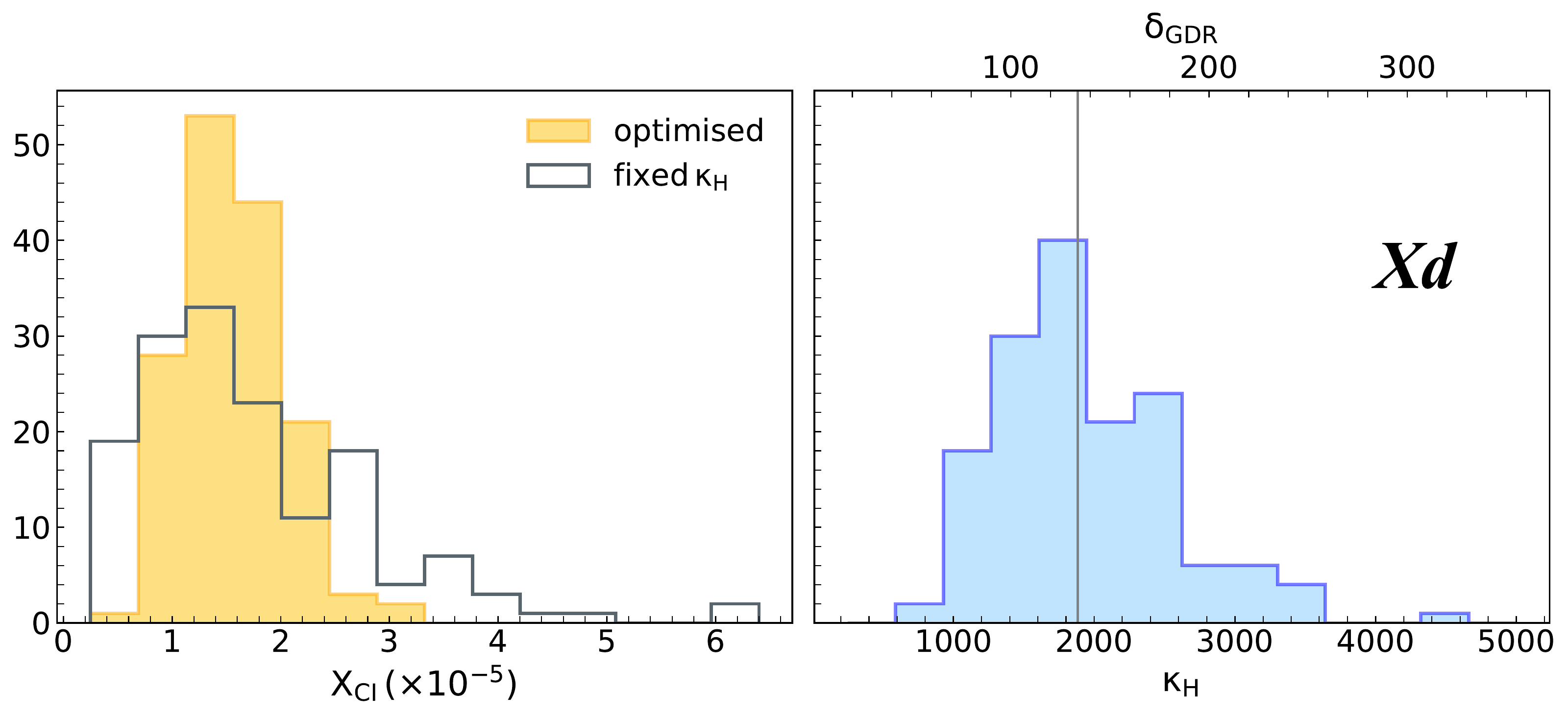}
\includegraphics[width=0.52\textwidth,trim=0cm 0cm 0cm 0cm, clip=true]{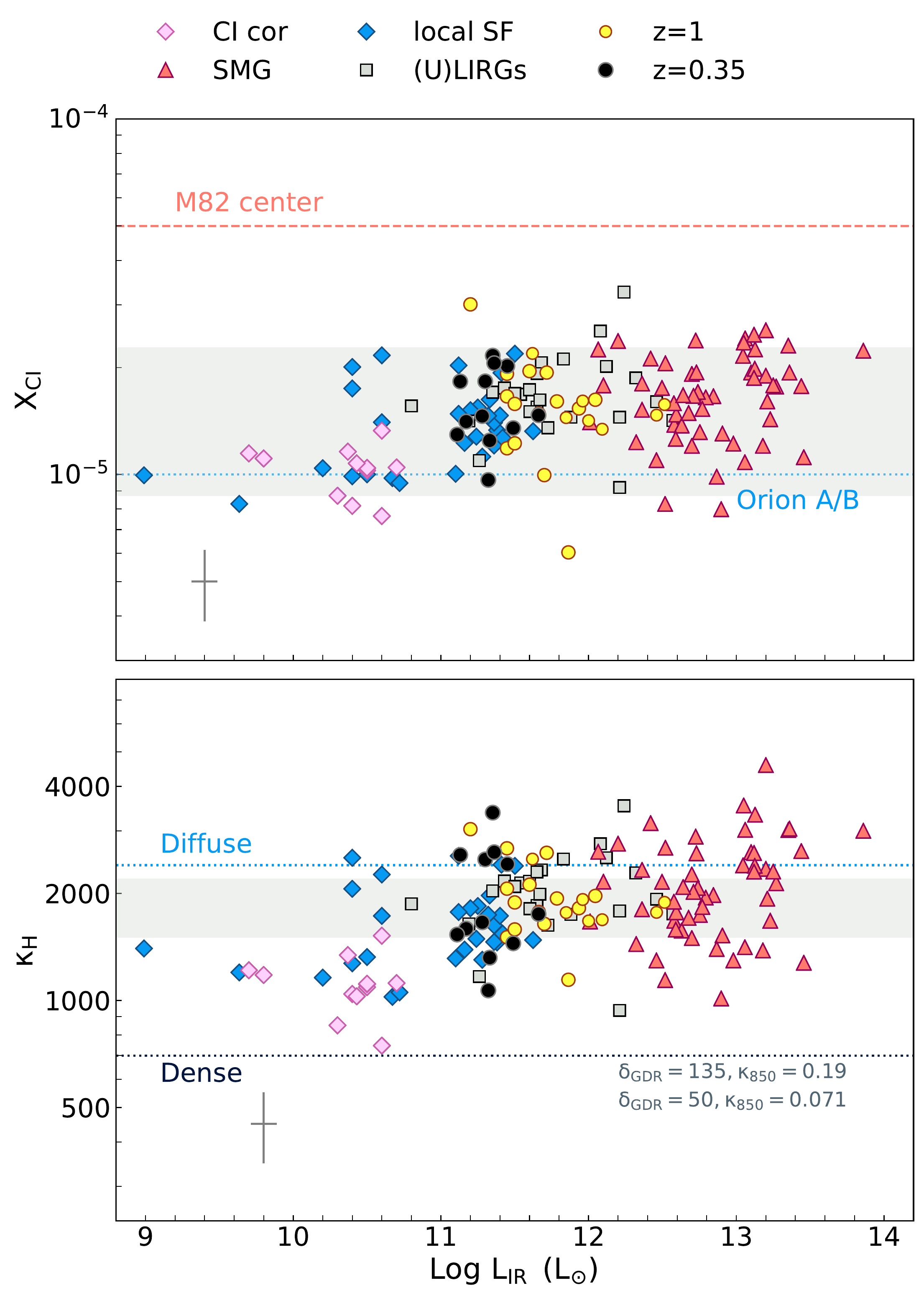}
\includegraphics[width=0.46\textwidth,trim=0cm 0cm 0cm 0cm, clip=true]{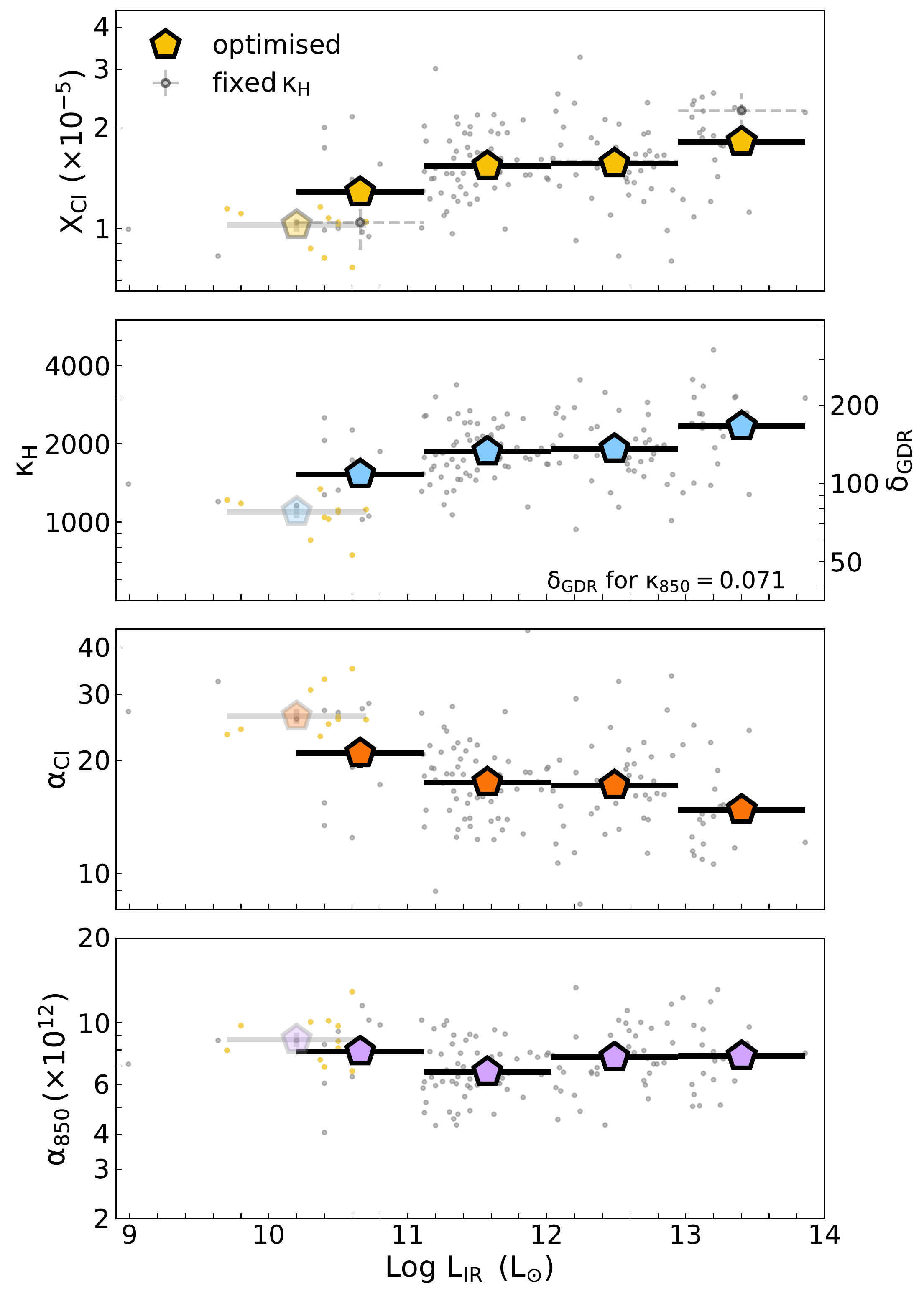}
\caption{{\bf Top:} Distributions of \Xci\ and \kh\ for the Xd sample,
  where the normalisation used is $\kh^{\rm N}=1884$\,\khunit. {\bf Left:} Optimised
  conversion factors as a function of \Lir\ for the galaxies in the
  Xd sample. The grey shaded bands represent the range of \Xci\
  ($Z=Z_{\odot}$) from a study of GRB and QSO absorber hosts by
  \citet{Heintz2020}, a measure which is independent of assumptions
  about \aco\ or \gdr, and also the range of \kh\ in nearby
  galaxies. {\bf Right:} Running means of the conversion factors with
  error bars of $\sigma/\sqrt{N_{\rm bin}}$. The solid coloured
  pentagons are the means for the grey points, which are those used to
  determine the conversion; the yellow points are the \CIcor\
  galaxies and the semi-transparent pentagon is the mean of those (see
  \S\ref{J19S} and Appendix~\ref{J19A}). The grey dashed error bars in
  the \Xci\ running mean are for the so-called simple method, where
  $\kh=1884$\,\khunit\ is fixed for all galaxies.}
\label{calXdF}
\end{figure*}

\begin{figure*}
\includegraphics[width=0.8\textwidth,trim=0cm 0cm 0cm 0cm, clip=true]{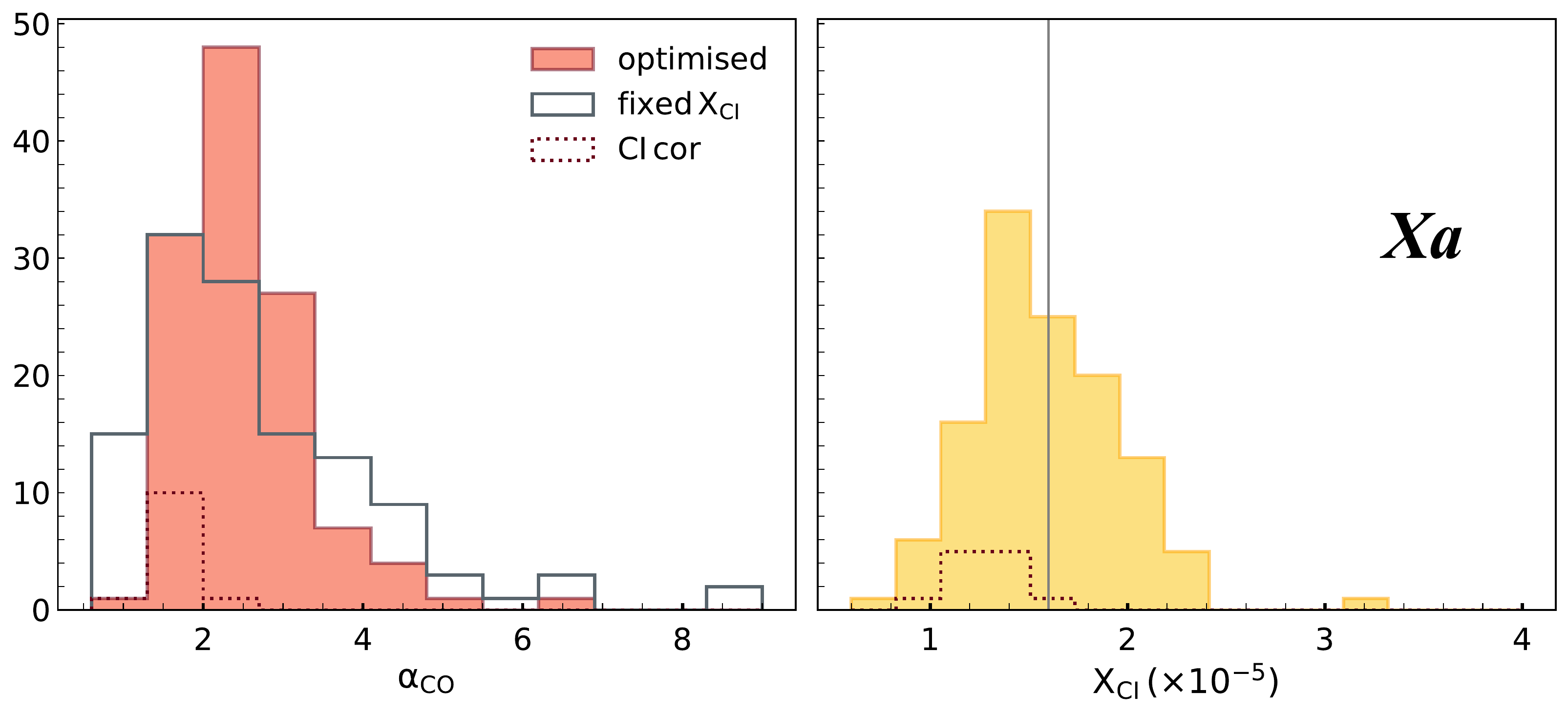}
\includegraphics[width=0.52\textwidth,trim=0cm 0cm 0cm 0cm, clip=true]{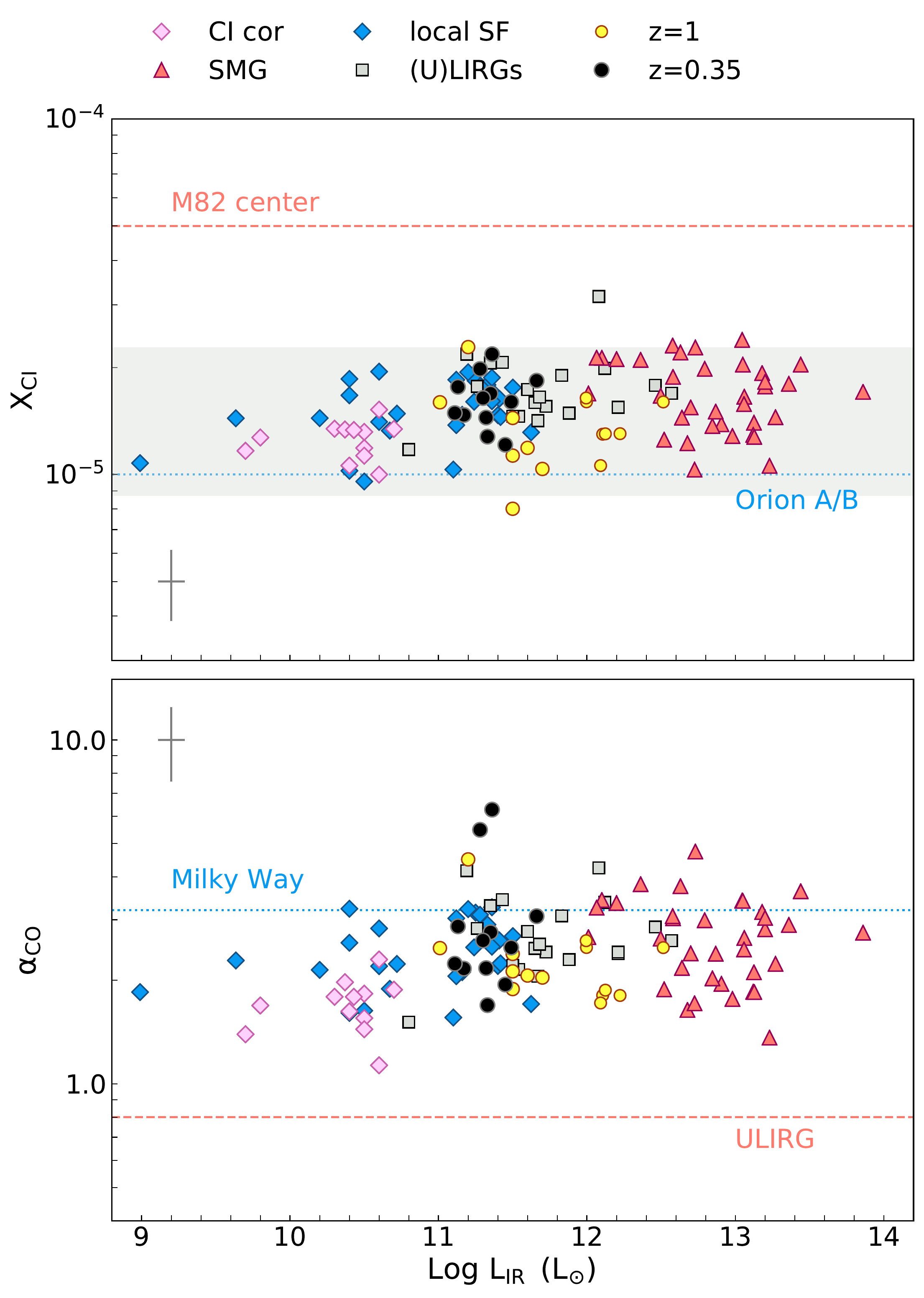}
\includegraphics[width=0.46\textwidth,trim=0cm 0cm 0cm 0cm, clip=true]{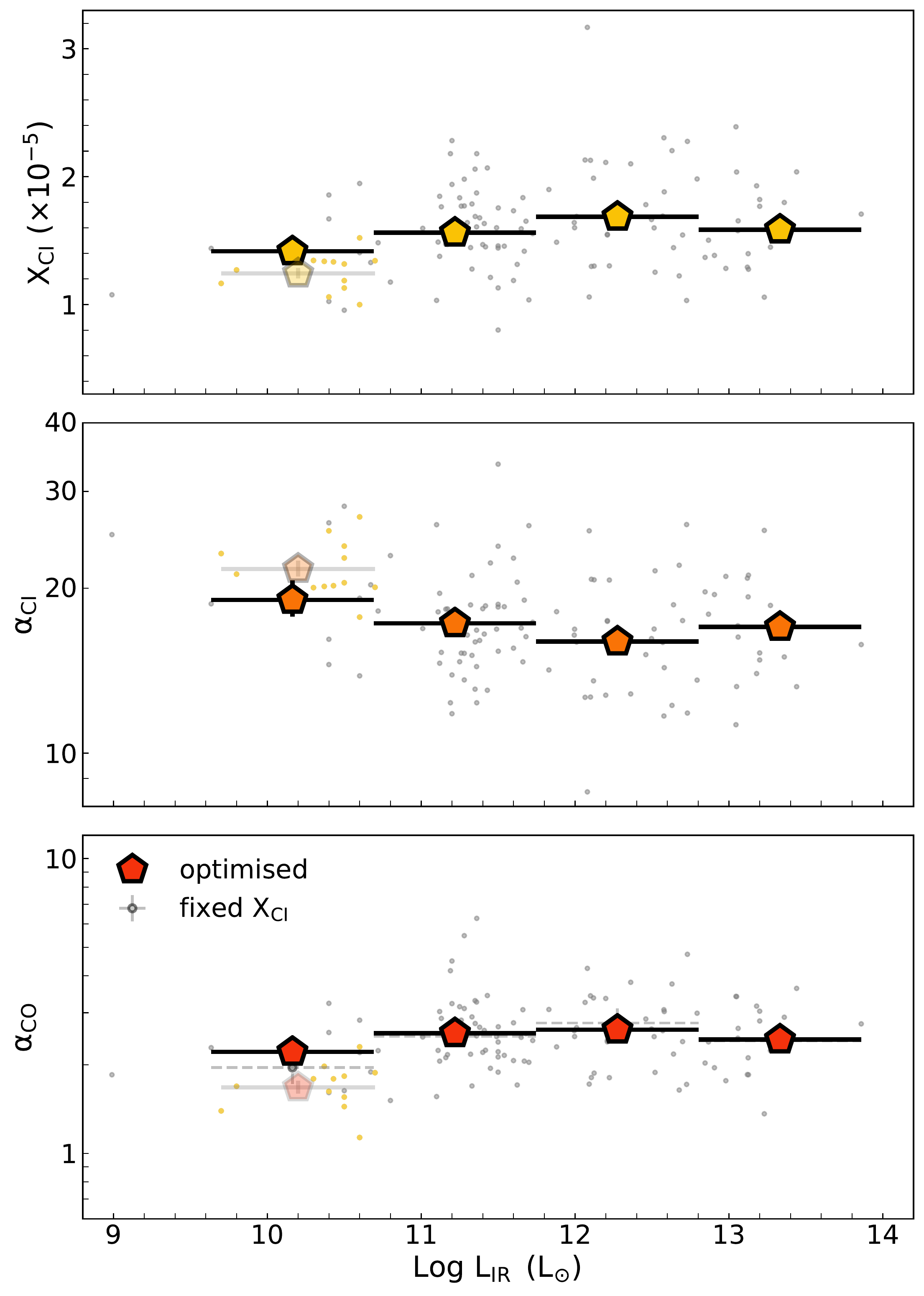}
\caption{{\bf Top:} Distributions of \Xci\ and \aco\ for the Xa
  sample, where the normalisation has been set to
  $\Xci^{\rm N}=1.6\times 10^{-5}$ -- the middle of the range for
  $Z_{\odot}$ found in GRB hosts and QSO absorbers by
  \citet{Heintz2020}. {\bf Lower left:} Optimised conversion factors
  as a function of \Lir\ for the galaxies in the Xa sample. {\bf
    Right:} Running means of the conversion factors with error bars
  of $\sigma/\sqrt{N_{\rm bin}}$. The solid coloured pentagons are the
  means for the grey points which are those used to determine the
  conversion; the yellow points represent the \CIcor\ galaxies and
  the semi-transparent pentagon is the mean of those -- see
  \S\ref{J19S} and Appendix~\ref{J19A}. The grey dashed error bars in
  the \aco\ running mean are for the so-called simple method, where
  $\Xci=1.6\times 10^{-5}$ is fixed for all galaxies. }
\label{calXaF} 
\end{figure*}

\begin{figure*}
\includegraphics[width=0.8\textwidth,trim=0cm 0cm 0cm 0cm, clip=true]{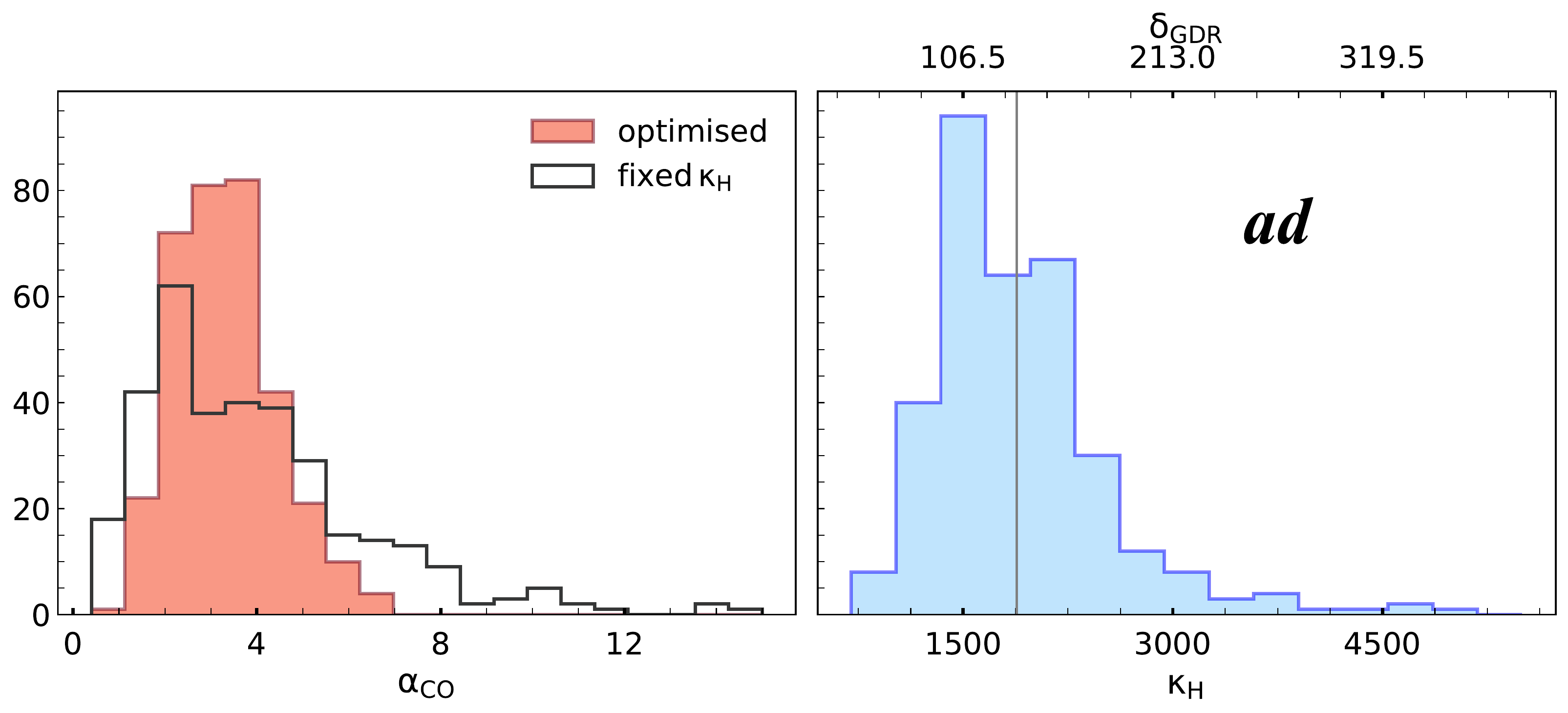}
\includegraphics[width=0.52\textwidth,trim=0cm 0cm 0cm 0cm, clip=true]{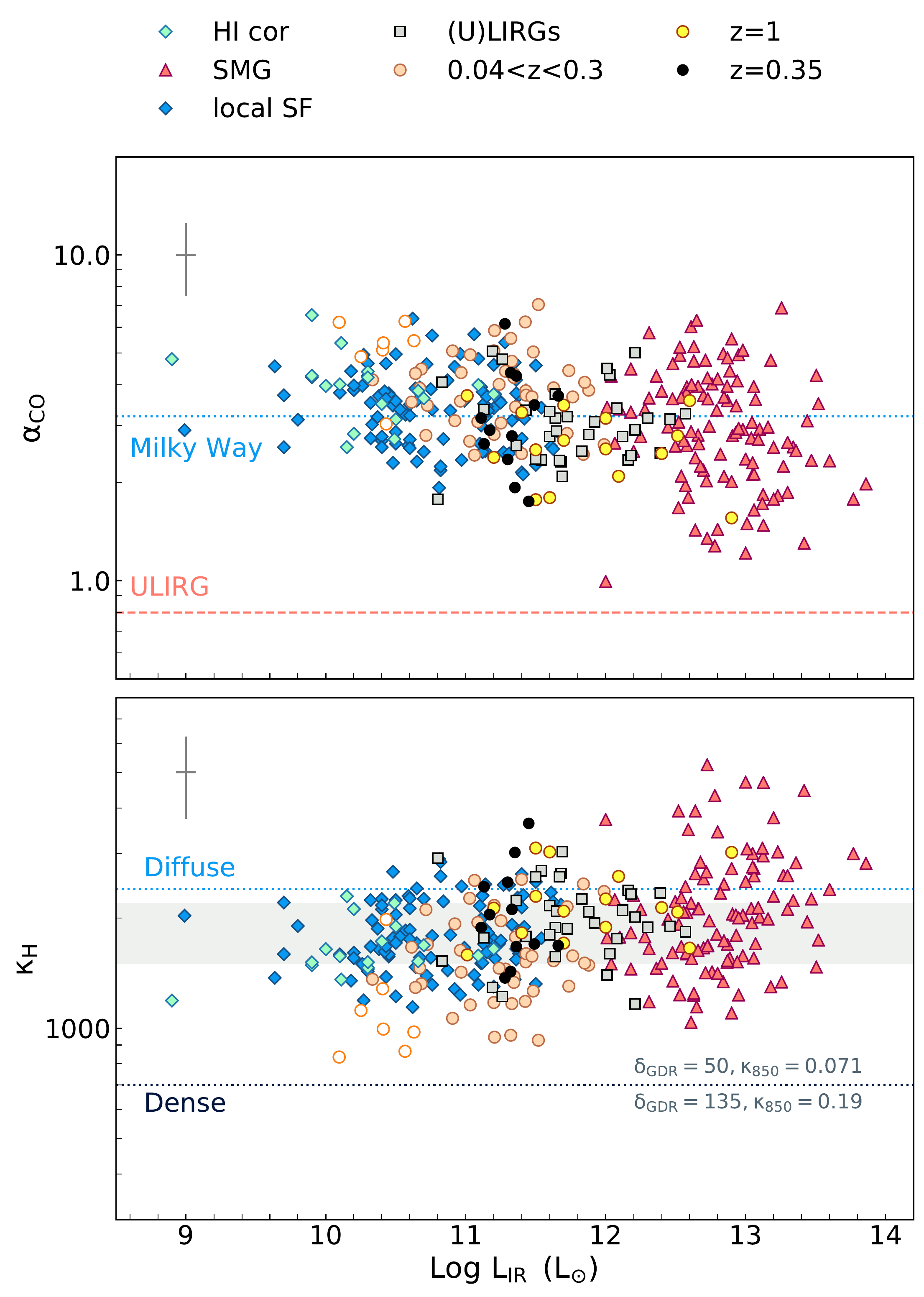}
\includegraphics[width=0.46\textwidth,trim=0cm 0cm 0cm 0cm, clip=true]{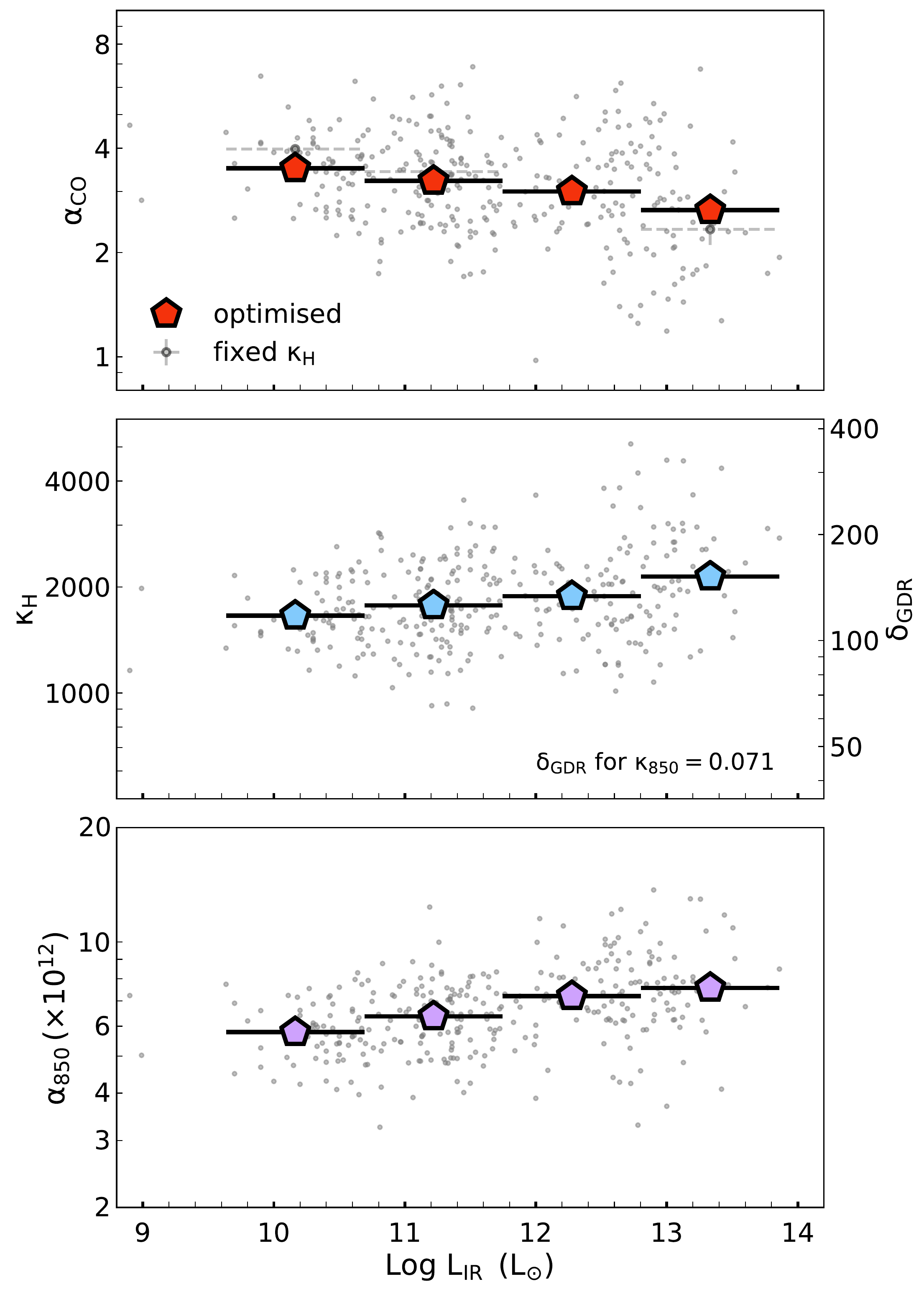}
\caption{{\bf Top:} Distributions of \aco\ and \gdr\ for the ad
  sample, where the normalisation has been set to $\kh^{\rm N}=1884$\,\khunit. {\bf
    Lower left:} Optimised conversion factors as a function of \Lir\
  for the galaxies in the ad sample. The cyan diamonds indicate
  galaxies with \fhi>1 which have been corrected for the contribution
  of dust mixed with H\,{\sc i}, following the procedure outlined in
  Appendix~\ref{HIS}. The open peach circles are the lo-VALES galaxies
  which we suspect have \fhi>1 -- these are not included in the
  analysis. {\bf Right:} Running means of the conversion factors with
  error bars of $\sigma/\sqrt{N_{\rm bin}}$. The coloured points
  reflect the optimised results and the grey dashed error bars for the
  \aco\ running mean are for the so-called simple method, where
  $\kh=1884$\,\khunit\ is fixed for all galaxies.}
\label{calaodF}
\end{figure*}

While the information content is greatest for the daX sample, which
has three tracer pairs to optimise, the method presented in
\S\ref{optS} still improves the cross-calibration for samples which
have two tracer measurements, i.e.\ one pair. The results for the Xd,
Xa and ad samples are shown in Figs~\ref{calXdF}--\ref{calaodF} and
behave similarly to the daX sample, as one would hope given that the
daX galaxies are a subset of the others. The pink diamonds in the
lower-left panels in Figs~\ref{calXdF} and \ref{calXaF} denote the
\CIcor\ galaxies. The cyan diamonds in the lower-left panel of
Fig.~\ref{calaodF} are galaxies with \fhi>1 which have been corrected
for the contribution of dust mixed with the \HI\ gas, as described in
Appendix~\ref{notesS}. The open peach circles are the lo-VALES
galaxies, which we suspect to have \fhi>1 (see \S\ref{L850LCOS}) but
which we cannot correct. We do not include these in any averages or
histograms.

The method previously used in the literature
\citep[e.g.][]{AZ2013,Scoville2016,Orellana2017,Hughes2017,Valentino2018}
has been to assume one tracer in a pair (e.g.\ \lcoa) has a known
conversion (\aco), then to fix that factor for all
galaxies in order to estimate the second (i.e.\ the one of
interest). We show this simple method alongside our optimal method as
grey lines and dashed grey error bars in the relevant panels of
Figs~\ref{calXdF}--\ref{calaodF}. The scatter in the conversion
factors for the optimised estimates are governed by the intrinsic
scatter we inferred in our analysis of the data in
Appendix~\ref{pairwiseS}, free of assumptions. In contrast, the simple method proscribes that there is no
scatter in the known conversion factor, and therefore all of the
intrinsic scatter in the luminosity ratio is attributed to the second
conversion factor of interest. The optimised method presented here
does not assume an {\it ad hoc} preference for any particular
conversion factor: as it is based on empirical variance analysis, it
uses more of the available information to improve the accuracy of the
estimated conversion factors. The histograms in
Figs~(\ref{calXdF}--\ref{calaodF}) show that the scatter in the factor
of interest is larger when using the simple method, and the trends in
the running medians are also more exaggerated.

Comparing the parameter estimates using three tracers to those using
two tracers for the same galaxies allows us to test the accuracy of
these two-tracer estimates. The details are in Appendix~\ref{testsA},
but in summary there is a reasonable correlation between the
three-tracer and two-tracer estimates, without bias
(Fig.~\ref{acocompF}) and an average scatter of 0.06--0.08 dex.

Thus, when multiple \mol\ tracers are available, we recommend
 the procedure outlined in the example below.

\paragraph*{Example:}

Take the example of a galaxy with observations of both dust and CO. We
take the mean value for the appropriate pair combination from
Table~\ref{methodT}: \aco/\kh=0.00133. For our adopted sample mean
normalisation of $\kappa^{\rm N} = 1884$\,\khunit, we now infer the sample mean
expectation value of $\langle\alpha\rangle= 0.00133 \times 1884 = 2.5$
(excluding He). The sample means, $\kappa^{\rm N}$ (assumed) and
$\langle\alpha\rangle$ (derived), are next used to estimate an initial
gas mass for our galaxy in each of the two tracers, \lsub\ and \lcoa.

\begin{multline}
	M_{\kappa} = \kappa^{\rm N} \lsub/4\pi B(\nu_{850},\mwtd) \\
	M_{\alpha} = \langle\alpha\rangle \lcoa \\ 	
\end{multline}

\noindent
Next, we calculate the effective standard deviation by adding the
observational error on the tracer luminosities in quadrature to the
intrinsic scatter for $\alpha$ and $\kappa$.
\begin{multline}
	\sigma^{\kappa}_{\rm eff} = \sqrt{s_\kappa^2 + \sigma_{850}^2}\\
	\sigma^{\alpha}_{\rm eff} = \sqrt{s_\alpha^2 + \sigma_{\rm CO}^2}\\
\end{multline}
where $\sigma_{850}$, $\sigma_{\rm CO}$ are the errors on $\log (\lsub)$ and
$\log (\lcoa)$, and $s_{\kappa}$ and $s_{\alpha}$ are the intrinsic
scatter on $\log (\kh)$ and $\log (\aco)$ listed in Table~\ref{methodT}. 
\noindent
The optimal \mol\ mass estimate is then calculated thus:
\begin{equation}
	M^{\rm opt} = \frac{M_{\kappa}/\sigma^2_{\rm \kappa, eff} + M_{\alpha}/\sigma^2_{\rm \alpha, eff}}{1/\sigma^2_{\rm \kappa, eff} + 1/\sigma^2_{\rm \alpha, eff}}
\end{equation}
We now work back to find the optimal conversion parameters for this galaxy:
\begin{multline}
	\kh^{\rm opt} = \kappa^{\rm N} M^{\rm opt}/M_{\kappa}\\
	\aco^{\rm opt} = \langle\alpha\rangle M^{\rm opt}/M_{\alpha}.\\
\end{multline}

\begin{figure}
	\includegraphics[width=0.49\textwidth,trim=0cm 0cm 0cm 0cm, clip=true]{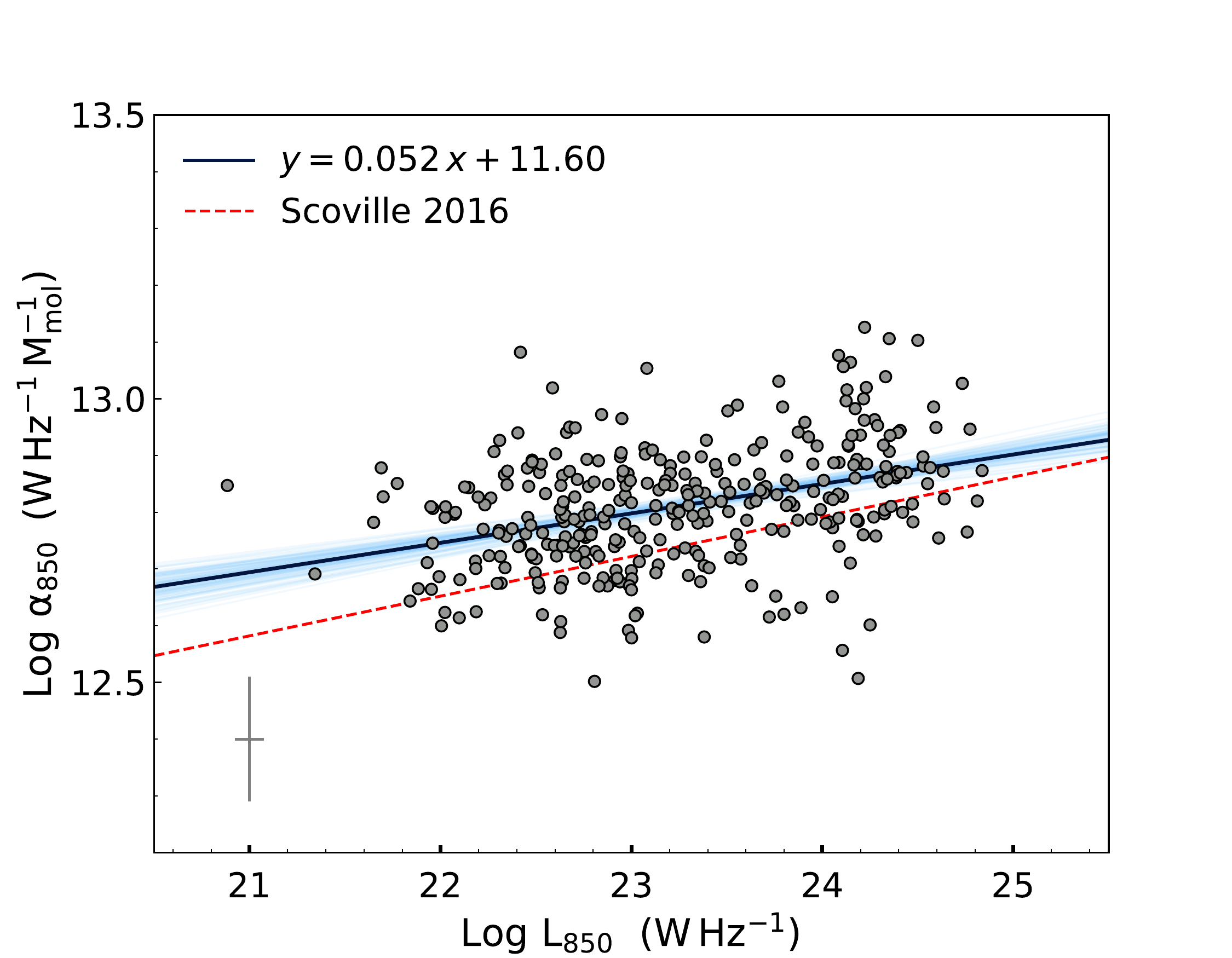}
	\includegraphics[width=0.49\textwidth,trim=0cm 0cm 0cm 0cm, clip=true]{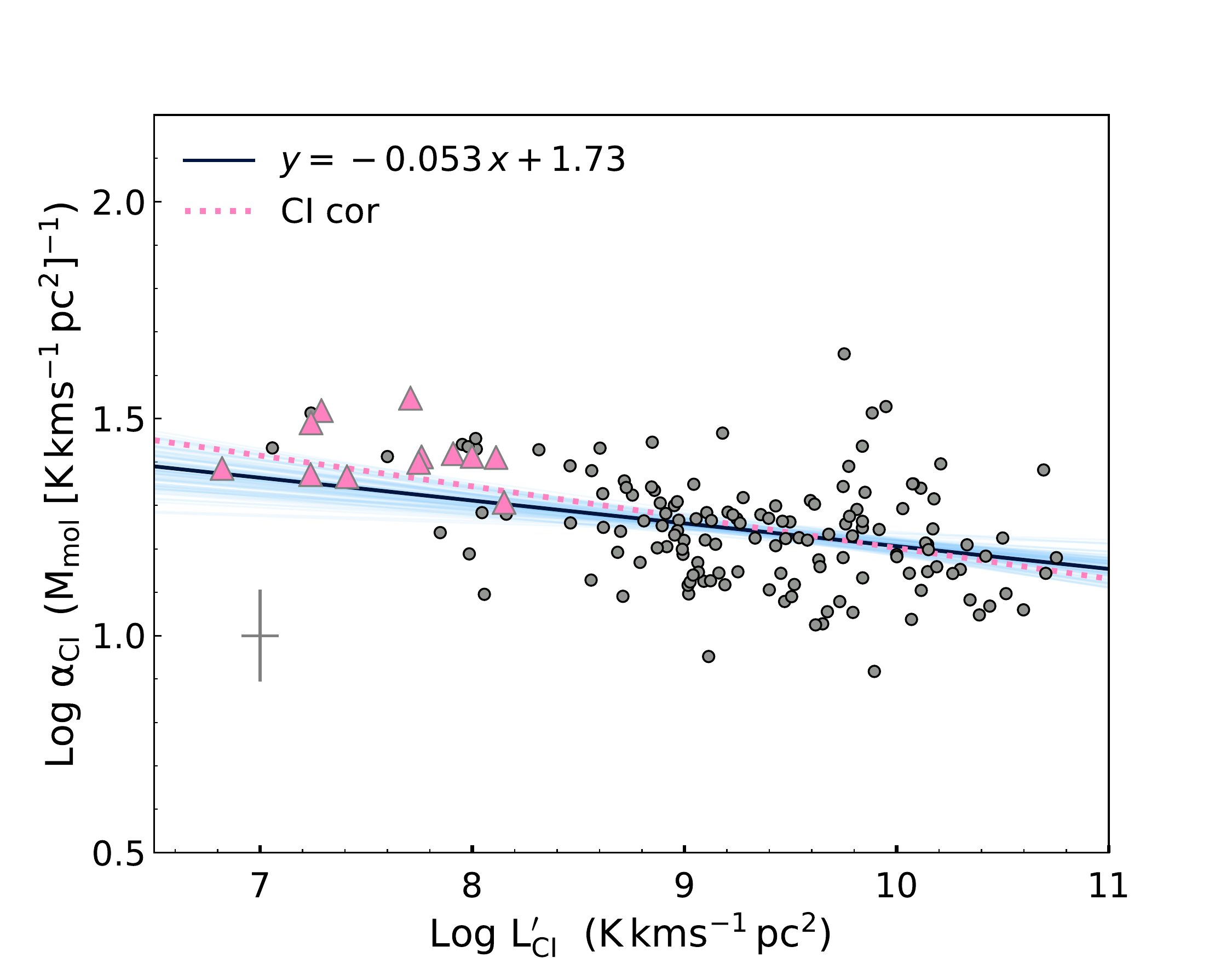}
	\includegraphics[width=0.49\textwidth,trim=0cm 0cm 0cm 0cm, clip=true]{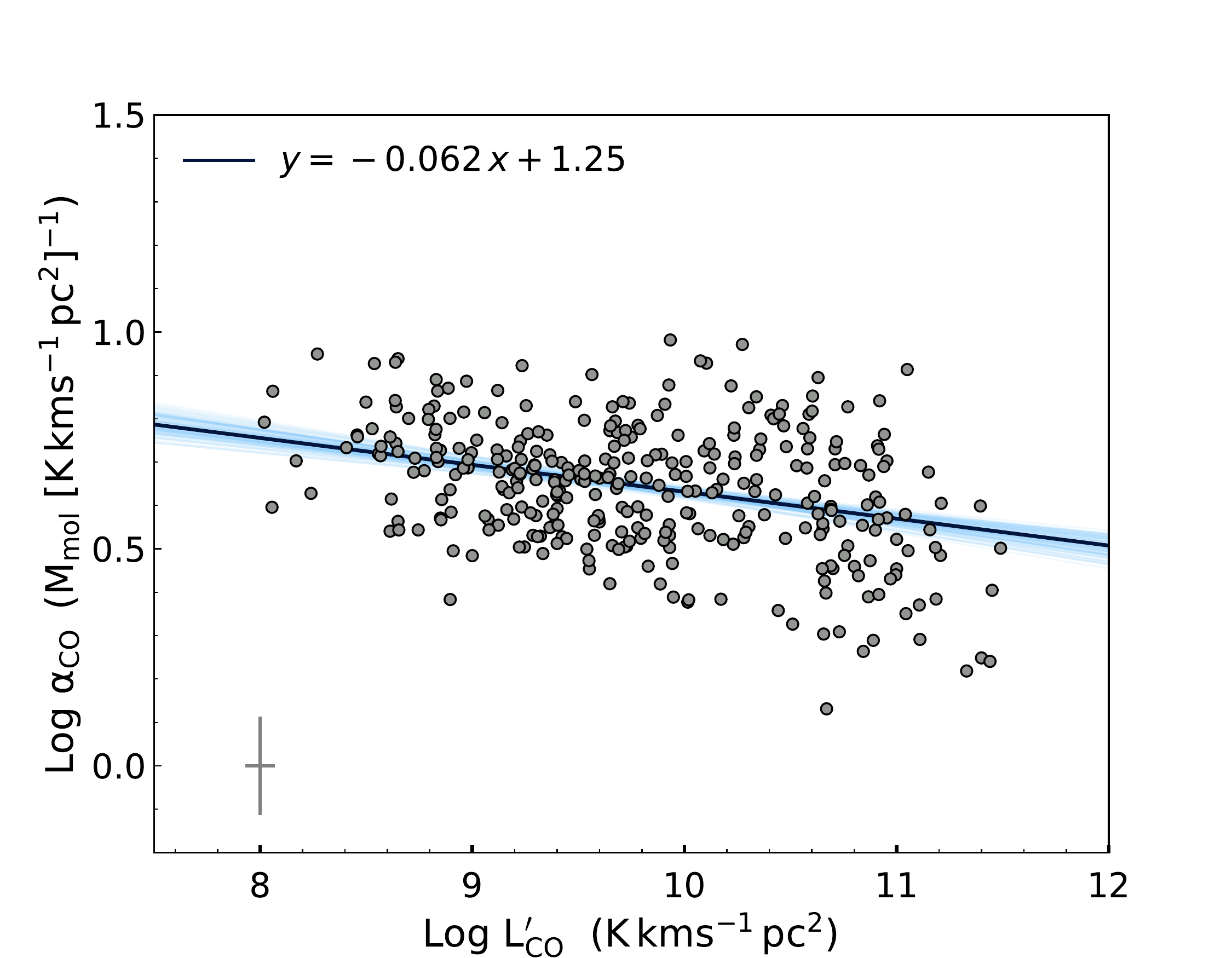}
	\caption{Correlations of the empirical calibration factors
          with their tracer luminosity. The largest sample was used in
          each case: ad ({\bf top and bottom}); Xd ({\bf top centre}).
		For \asub\ (top), we include the relationship from
                \citet{Scoville2016}, exactly as quoted in that
                paper. Parameters for the median posterior fit (dark
                blue line), accounting for co-variance in the errors,
                are quoted in the legend, and we include 100 random
                fits from sampling the posterior as pale blue lines to
                show the scatter. Further details are given in
                Table~\ref{calfitsT}. All tracers have a significant
                correlation with their tracer luminosity.}

\label{empLF}
\end{figure}

\begin{figure*}
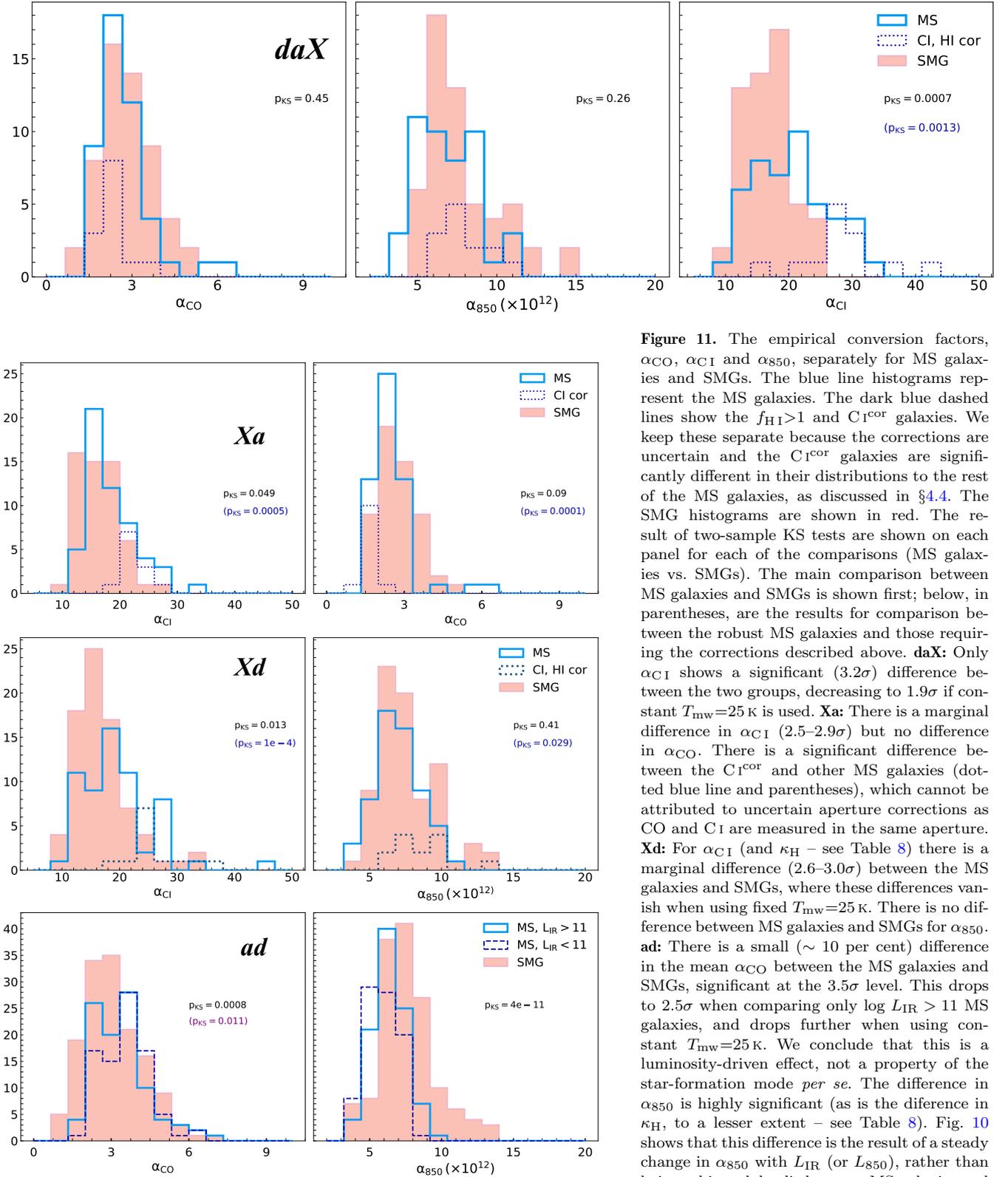

\begin{minipage}{0.98\linewidth}
\begin{lpic}[]{figs/python/cal_empirical_daX(0.39,0.39)}
\end{lpic}
\end{minipage}
\begin{minipage}{0.64\linewidth}
\begin{lpic}[]{figs/python/cal_empirical_Xa(0.33,0.33)}
\end{lpic}
\begin{lpic}[]{figs/python/cal_empirical_Xd_kh_fullMS(0.33,0.33)}
\end{lpic}
\begin{lpic}[]{figs/python/cal_empirical_aod_kh(0.33,0.33)}
\end{lpic}
\end{minipage}
\begin{minipage}{0.35\linewidth}

  \caption{\label{arghF} The empirical conversion factors, \aco, \aci\ and \asub,
    separately for MS galaxies and SMGs. The blue line histograms
    represent the MS galaxies. The dark blue dashed lines show the
    \fhi>1 and \CIcor\ galaxies. We keep these separate because the
    corrections are uncertain and the \CIcor\ galaxies are
    significantly different in their distributions to the rest of the
    MS galaxies, as discussed in \S\ref{J19S}. The SMG histograms are
    shown in red. The result of two-sample KS tests are shown on each
    panel for each of the comparisons (MS galaxies vs.\ SMGs). The
    main comparison between MS galaxies and SMGs is shown first;
    below, in parentheses, are the results for comparison between the
    robust MS galaxies and those requiring the corrections described
    above. {\bf daX:} Only \aci\ shows a significant ($3.2\sigma$)
    difference between the two groups, decreasing to $1.9\sigma$ if
    constant \mwtd=25\,{\sc k} is used. {\bf Xa:} There is a marginal
    difference in \aci\ (2.5--$2.9\sigma$) but no difference in
    \aco. There is a significant difference between the \CIcor\ and
    other MS galaxies (dotted blue line and parentheses), which cannot
    be attributed to uncertain aperture corrections as CO and \CI\ are
    measured in the same aperture. {\bf Xd:} For \aci\ (and \kh\ --
    see Table~\ref{hyp2T}) there is a marginal difference
    (2.6--$3.0\sigma$) between the MS galaxies and SMGs, where these
    differences vanish when using fixed \mwtd=25\,{\sc k}. There is no
    difference between MS galaxies and SMGs for \asub. {\bf ad:} There
    is a small ($\sim 10$ per cent) difference in the mean \aco\
    between the MS galaxies and SMGs, significant at the $3.5\sigma$
    level. This drops to $2.5\sigma$ when comparing only \Lhi\ MS
    galaxies, and drops further when using constant \mwtd=25\,{\sc
      k}. We conclude that this is a luminosity-driven effect, not a
    property of the star-formation mode {\em per se}. The difference
    in \asub\ is highly significant (as is the diference in \kh, to a
    lesser extent -- see Table~\ref{hyp2T}). Fig.~\ref{empLF} shows
    that this difference is the result of a steady change in \asub\
    with \Lir\ (or \lsub), rather than being a bi-modal split between
    MS galaxies and SMGs.}
\end{minipage}
\end{figure*}

\subsubsection{Single band: empirical conversion factors}
\label{empS}

\begin{table}
	\caption{Empirical conversion factors recommended
		for use when only a single gas tracer observation is in hand. These
			 include a factor 1.36 to account for He.}
	\begin{adjustbox}{center}
		\begin{tabular}{lccc}
			\toprule
			Sample & \asub ($\times10^{12}$) & \aco & \aci \\
			& W Hz$^{-1}$ \msun$^{-1}$ & \multicolumn{2}{c}{\aunit}\\
			\midrule
			\Llo &  $5.8\pm0.1$  &  $4.7\pm0.1$      &  ..  \\
			\Lhi &  $6.9\pm0.1$  &  $4.0\pm0.1$      & $17.0\pm0.3$\\ 
			MS       &  $6.2\pm0.1$  &  $4.4\pm0.1$ & $19.1\pm0.6$\\
			SMGs      &  $7.3\pm0.1$  &  $3.8\pm0.1$ & $16.2\pm0.4$\\

			\bottomrule
		\end{tabular}
\end{adjustbox}
\flushleft{Values are the weighted means and errors from each
 of the daX, ad and dX samples which all have the same
 normalisation of $\kh^{\rm N}=1884$\,\khunit. Low luminosity (\Llo) values
 are based only on the ad sample due to the small numbers of
 reliable \CI\ measurements in this luminosity range. The differences in the
 weighted means for SMGs and MS galaxies are significant in
 all cases, but are likely driven by the trends with
 luminosity seen in Figs~\ref{caldaXF}, \ref{calaodF} and
 \ref{empLF}. A more accurate way to determine the
 conversion factor to use would be to use one of the
 relationships from Table~\ref{calfitsT}.}
\label{prescT}
\end{table}

\begin{table*}
\caption{Log means and statistical tests for conversion factors for
  MS galaxies and SMGs.}
\begin{adjustbox}{center}
\begin{tabular}{@l^c^c^c^c^c^c^c^c}
\toprule
\CI & Sample  & $\Xci{\rm (MS)}$ & $\Xci{\rm (SMG)}$ & $\aci{\rm (MS)}$ &
                                                                       $\aci{\rm (SMG)}$& $Z(\sigma)$ & $P_{\rm KS}$ & $P_{\rm KS}$\,(25\,{\sc k}) \\
\midrule
   &       daX & {\boldmath $1.41\pm0.07$} & {\boldmath $1.70\pm 0.06$}  & {\boldmath $19.1\pm0.9$} & {\boldmath $15.8\pm0.5$} & {\bf 3.1} & {\bf 0.016} &  0.074\\
   &        Xd  & $1.44\pm0.05$ & $1.65\pm0.05$   & $18.7\pm0.8$ & $16.4\pm0.5$  & 2.6 & 0.013 & 0.48\\
   &        Xa & {\boldmath $1.44\pm0.04$} & {\boldmath $1.64\pm 0.05$}  & {\boldmath $18.7\pm0.6$} & {\boldmath $16.4\pm0.5$}  & 3.0 & 0.028 & \\ 
\addlinespace[0.5em]
\cmidrule(r){1-1}
\cmidrule(lr){3-4}
CO &   & $\aco{\rm (MS)}$ & $\aco{\rm (SMG)}$ & & & &  & \\
\cmidrule(r){1-1}
\cmidrule(lr){3-4}
   &       daX & $2.56\pm0.14$ & $2.61\pm0.12$  &   &  & 0.3  & 0.35 &  0.017\\
   &        Xa & $2.45\pm0.12$ & $2.70\pm0.10$  &  &  & 1.6 & 0.05 &\\
   &        ad & {\boldmath $3.32\pm0.07$} & {\boldmath $2.87\pm0.09$}  &  &  & {\boldmath $3.8$} & {\boldmath $0.0008$} & 0.04 \\

\addlinespace[0.5em]
\cmidrule(r){1-1}
\cmidrule(lr){3-4}
\cmidrule(lr){5-6} 
Dust &  &  $\kh{\rm (MS)}$ & $\kh{\rm (SMG)}$ & $\asub{\rm (MS)}$ & $\asub{\rm (SMG)}$ & & & \\ \cmidrule(r){1-1}
\cmidrule(lr){3-4}
\cmidrule(lr){5-6} 
&       daX & $1758\pm90$ & $2039\pm113$ & $6.8\pm0.3$ & $7.7\pm0.3$ & 2.2 & 0.04 (0.25) & 0.76 \\
     &        Xd & {\boldmath $1757\pm66$} & {\boldmath $2025\pm58$}  & $6.9\pm0.2$ & $7.4\pm0.2$ & 3.0 & 0.015 (0.41) & 0.58 \\
     &        ad & {\boldmath $1722\pm36$} & {\boldmath $1981\pm65$}  & {\boldmath $6.0\pm0.1$} & {\boldmath $7.3\pm0.2$} & {\boldmath $3.6$} & {\boldmath $0.0017$} & 0.11\\
     &           &   &   &   &   &   \boldmath$(6.5)$ & \boldmath$(4\times10^{-11})$ & \\
\bottomrule
\end{tabular}
\end{adjustbox}
\noindent{\flushleft We compare MS galaxies and SMGs for each subset
  to look for differences in the parameters. The MS group excludes
  \CIcor\ and lo-VALES galaxies, but it does include the \fhi>1
  galaxies, after applying the correction from Appendix~\ref{HIS}. The
  numbers in each subset are: daX: MS=46, SMG=55; Xd: MS=60, SMG=79;
  Xa: MS=54, SMG=55; ad: MS=184, SMG=144. {\bf Bold} indicates
parameters which are significantly different between the MS galaxies
and SMGs in the Z-test and KS tests. The \CI\ parameters, \Xci\ and
\aci, are simply linked due to our adoption of constant $\Q=0.48$,
meaning that the distributions have the same KS results. The dust
parameters, \kh\ and \asub, are related to each other as a function of
\mwtd\ and so they can behave differently, e.g.\ \kh\ can be
indistinguishable between samples but the \asub\ can be significantly
different. Thus in the dust section, there are two $P_{\rm KS}$
values: those for \kh\ and then, in parentheses, those for
\asub. The final column, $P_{\rm KS}$(25\,{\sc k}), is the KS result
when \mwtd\ is fixed to 25\,{\sc k}, which makes
the distributions of \kh\ and \asub\ identical.}
\label{hyp2T}
\end{table*}

\begin{table*}
	\caption{Fits to the conversion parameters and their tracer luminosities. All quantities include He.}
	\begin{adjustbox}{center}
		\begin{tabular}{ccccccc}
			\toprule
			$y$ & $x$ & $m$ & $c$ & $r_{\rm s}$ & $N$ & Sample\\
			\midrule
			log \aco & log \lcoa  & $-0.062$ (0.009) & 1.25 (0.09) & $-0.33$  & 335 & ad\\
			log \asub & log \lsub & $0.052$ (0.008) & 11.60 (0.18) & 0.37 &  335 & ad\\

			log \aci & log \lci & $-0.052$ (0.013) & 1.73 (0.12) & $-0.3$ (0.0003)  & 140 & Xd\\
			\bottomrule
		\end{tabular}
	\end{adjustbox}
	\flushleft{Fits in the form $y=mx+c$ for the empirical
          conversion parameters and their tracer luminosities. The
          \CIcor\ galaxies are excluded from the fits but are shown in
          the plots (Fig.~\ref{empLF}). $r_{\rm s}$ is the Spearman
          rank correlation coefficient with probability of the null
          hypothesis of no correlation in parentheses for
          $p\geq0.0001$. The effects of co-variance in the errors have
          been accounted for.}
              \label{calfitsT}
\end{table*}

The three empirical conversion factors, \asub, \aci\ and \aco,
directly relate the observable tracer luminosity to a gas mass,
according to Eqn.~\ref{MhE}.  If only one tracer (\lcoa, \lci\ or
\lsub) is available, the empirical conversion factor we have estimated in Table~\ref{prescT}
is the best choice. We adopt a convention that the empirical
parameters are referenced to \Mmol, which includes a factor 1.36 for
He.

\begin{table*}
	\caption{Empirical calibration factors derived from
		this study. \Mmol\ columns
		include a factor of 1.36 for He.}
	\begin{adjustbox}{center}
		\begin{tabular}{lccccccc}
			\toprule
			\multicolumn{2}{l}{} & \multicolumn{3}{c}{\Mh} & \multicolumn{3}{c}{\Mmol} \\
			\cmidrule(r){3-5} \cmidrule(l){6-8}
			Sample         &  $N$  & \asub $\,(\times10^{12})$  & \aco   & \aci & \asub $\,(\times10^{12})$   & \aco   & \aci \\
			&     & $\rm{W\,Hz^{-1}\msun^{-1}}$ & \multicolumn{2}{c}{\aunit} & $\rm{W\,Hz^{-1}\msun^{-1}}$ & \multicolumn{2}{c}{\aunit} \\
			\midrule
		daX  & 101 & $9.9\pm0.2$  & $2.6\pm0.1$   & $12.6\pm0.4$ & $7.3\pm0.2$ & $3.5\pm0.1$ & $17.2\pm0.5$\\
			 ad (\Lhi)    & 240 & $9.1\pm0.1$   & $3.1\pm0.1$   &  & $6.7\pm0.1$ & $4.2\pm0.1$ & \\
			ad          & 326 & $8.8\pm 0.1$  & $3.2\pm0.1$   &   & $6.5\pm0.1$  & $4.3\pm0.1 $ & \\
			Xd  & 140  & $9.8\pm0.3$   &               & $12.7\pm0.4$ & $7.2\pm0.2$ &  & $17.3\pm0.5$\\
			Xa    & 109  &                           & $2.5\pm0.1$   & $12.5\pm0.3$ &    & $3.4\pm0.1$ & $17.0\pm0.4$\\
			\bottomrule
		\end{tabular}
	\end{adjustbox}
	\flushleft{The samples which include dust continuum used a normalisation of
		$\kh^{\rm N}=1884$\,\khunit, while for the Xa sample, $\Xci^{\rm N}=1.6\times 10^{-5}$ was used. In this analysis we excluded the \CIcor\ and lo-VALES galaxies (see \S\ref{J19S}).}
\label{empT} 
\end{table*}

Figs~\ref{caldaXF}--\ref{calaodF} show the empirical conversion
factors as a function of \Lir; Fig.~\ref{empLF} shows the empirical
conversion factors as a function of the tracer luminosity, and
Fig.~\ref{arghF} shows their distribution when the sample is split
into MS galaxies and SMGs (see Table~\ref{hyp2T} for details). All empirical factors show significant but shallow
correlations with the tracer luminosity. We have carefully accounted
for the co-variance between the $x$ and $y$ parameters when fitting,
so the correlations we find are not caused by the
involvement\footnote{The inclusion of the co-variance matrix in the
  fit reduces the slope (closer to zero) by 0.05.} of \lsub, \lci\ and
\lcoa\ in the derivation of \asub, \aci\ and \aco.
Table~\ref{calfitsT} lists the fit parameters. The intrinsic scatter
in all of these relationships is very small once the measurement
errors are accounted for. Correlations are also seen between \Lir\ and
\asub, \aci\ and \aco (Fig.~\ref{calXdF}--\ref{calaodF}), albeit
with more scatter.

A correlation between \asub\ and \lsub\ was also noted by Sco16 (shown
as the red dashed line on our plot) but ours is somewhat shallower
($m=0.052$ vs $m=0.07$ from Sco16), although the difference is
unlikely\footnote{Sco16 do not quote an error on their fit, so it is
  difficult to be certain, but the error on Sco16 would likely be
  larger than ours, which means they are consistent to within
  2$\sigma$.} to be significant. These shallow but significant
relationships with their tracer luminosity could be further applied to
give more accurate calibration (see Table~\ref{calfitsT}). The trends
of \aci\ and \aco\ with their respective tracers are explored for the
first time in large numbers here.

\subsection{Discussion of empirical factors relative to the literature.}
\label{litdiscS}
\subsubsection{Submillimetre dust empirical calibration, \asub}

The final calibration factors from this work are provided in Table~\ref{empT}. 
A compilation of \asub\ values from our optimal method\footnote{There
  is no significant difference if we fix $\aco$ to 4.3.} and those
from the literature, referenced to a common $\aco=4.3$: the Galactic
value including He from \citealp{Bolatto2013}) is presented in
Table~\ref{asublitT}. Literature values cover the range, $\asub$ =
3.6--$10.1 \times10^{12}$, comfortably within the range of estimates here:
$\asub^{\rm opt}=$ 6.5--$7.2\times10^{12}$. The lowest value,
$\asub =3.6^{+3.6}_{-1.9}\times10^{12}$, comes from the local sample
of \citet{Orellana2017}, who include H\,{\sc i} as well as \Mmol\
(from \lcoa). Their \asub\ refers to the total gas mass -- sensibly,
since their lower luminosity sample is more H\,{\sc i}-dominated than
the others we compare to -- meaning that a lower value for \asub\ is
required. The highest value, $\asub=(10.1\pm0.3)\times10^{12}$, is
from Sco16.\footnote{The original value of $\asub=6.7\times10^{12}$
  quoted by Sco16 assumed that $\aco=6.5$ to calibrate the gas mass
  from CO. Re-normalising the Sco16 result to the same $\aco=4.3$ as
  our literature comparison increases the Sco16 value to
  $\asub({\aco=4.3})=10.1\times10^{12}$.} There are two reasons
why Sco16 found a significantly higher \asub\ compared to our analysis. The first is simple
mathematics, as Sco16 quote a linear mean for a distribution that has
a significant tail to higher values; in contrast, we quote a log-mean
which is less sensitive to tails. This statistical bias results in a linear mean
for \asub\ which is 20 per cent higher than the log-mean
\citep{Behroozi2013} . Assuming the shape of our \asub\ distribution
is similar to that from Sco16, we adjust their linear mean down by 20
per cent to approximate our log-mean method. Thus our log-mean
estimate of the Sco16 value is
$\asub^{\rm LM}(\rm Sco16) = 8.4\times10^{12}$. Secondly, there have
been changes in the versions of the {\it Herschel} pipeline data used
as the basis for the local portion of the Sco16 sample (see
Appendix~\ref{notesS}). When we re-fitted the local galaxy SEDs to
estimate \mwtd\ and \lsub\ using the most recent {\it Herschel} flux
densities \citep{Chu2017,Clark2018} we found an increase in \lsub\ of
$\sim 0.1$\,dex compared to that reported in Sco16\footnote{This
  difference is not just the photometry change; Sco16 used a different
  method to estimate \lsub\ from the Herschel 500-\mic\ fluxes.}. Once
these factors are accounted for, the Sco16 result is comparable to
ours.

\begin{table*}
	\caption{Summary of our  empirical dust continuum--\Mmol\
		calibration factor, \asub, compared to literature values referenced to $\aco=4.3$ (i.e.\ Galactic
		\aco, including He).}.
	\begin{adjustbox}{center}
	\begin{tabular}{clcll}
		\toprule
		\asub ($\times10^{12}$) & Sample & $N_{\rm gal}$ & Notes & Reference\\
		\asunit  &         &               &       &   \\
		\midrule
		
		$6.4\pm0.1$              &  all  &  328   & log-mean opt & this work\\
		$7.2\pm0.2$              &  SMGs  &  144   & log-mean opt & this work\\
		$5.9\pm0.1$              &  MS  &  184   & log-mean opt & this work\\
		$10.1\pm 0.3$            &  local galaxies and SMGs  &  72 & linear mean & \citet{Scoville2016}\\
		8.3                   &  MS               & 30       &  linear mean  & \citet{Scoville2016}\\
		12.7                  &  SMGs             & 30       &  linear mean & \citet{Scoville2016}\\
		$3.6^{+3.6}_{-1.9}$      & local galaxies &  136  &  median H\,{\sc i} $+$ 4.3\lcoa &  \citet{Orellana2017}\\

		$6.1\pm0.14$           &  $z<0.4$ 160-\mic\ selected   & 41 &  log-mean (ex lovales) &\citet{Hughes2017}\\
		$8.4\pm1.0$           &  $z=1.6$--2.9 unlensed SMGs$^{\dag}$  & 9 & log-mean & \citet{Kaasinen2019}\\
		$11.6\pm1.2$           & $z=1.6$--2.9 unlensed SMGs$^{\dag}$  & 9 & linear mean & \citet{Kaasinen2019}\\
		
		\bottomrule
	\end{tabular}
	\end{adjustbox}
	\flushleft{Errors quoted are the standard error on the mean, from the
		variance of the \lsub/\lcoa\ ratio. Where we have the data
		for \lcoa\ and \lsub, we calculate the mean log \asub\ because
		the distribution of ratios is skewed in linear space \citep{Behroozi2013}, leading
		to a significantly higher value for \asub\ in the linear
		averaging. We also cite the linear average, scaled to $\aco=4.3$
		where that is presented in the original literature
		reference. $^\dag$This small sample may potentially be biased by
		choosing the brightest 850-\mic\ galaxies from the parent sample.} 
\label{asublitT} 
\end{table*}

 \subsubsection*{Atomic carbon empirical factor: \aci}
We compare our optimised \aci\ estimates with others from the literature in Table~\ref{acilitT}, and for reference we also compile the literature values for \Xci\ in Table~\ref{XCIlitT}, the two are related simply by the excitation factor \Q, as given in Eqn.~\ref{aciE}. Our values are a weighted average
of the three samples that contain \CI, where we find
$\langle\aci\rangle=17.3\pm0.3$ (standard error on the mean). This
compares well with the only truly independent measure,
$\aci=21.4^{+13}_{-8}$, from absorber systems across a range of
redshift by \citet{Heintz2020}, but is considerably higher than
reported by many literature studies, e.g.\ $\aci=4.9$--10.3 for a
large study of (U)LIRGs by \citealp{Jiao2017} and
$\aci=7.3^{+6.9}_{-3.6}$ for local disks by \citet{Crocker2019}. Many literature studies assume a
fixed value for either \Xci\ or \aco\ in order to derive \aci\ (typically $\Xci=3\times 10^{-5}$ or $\aco=1$ for high-$z$ SMG or local (U)LIRGs). These assumed values are very different from those we have derived
here under our minimal assumption that metal rich galaxies have similar dust properties. Table~\ref{acilitT} describes the assumptions made for each literature source.

The MS galaxies in this work have a  value of $\aci$ of $19.1\pm0.6$, again significantly higher than that found in \citet{Crocker2019}. However, J19, using a largely overlapping sample, derived a value of $\aci=19.9\pm1.9$ using the same {\em Herschel} FTS \CI\ mapping observations.  
  \citeauthor{Crocker2019} use $L^{\prime}_{\rm [CO](2-1)}$ images and spatially resolved \aco\ estimates from \citet{Sandstrom2013}, which were derived using a robust method which minimises the scatter in the
gas-to-dust ratio (see also \citealp{Eales2012}). While there are no {\em a priori} assumptions about \aco\ in \citet{Crocker2019}, implicit assumptions are required for the average CO $r_{21}$ excitation. The limited sensitivity of the FTS instrument to [\CI](1--0) meant that \CI\ was primarily detected in the brighter nuclear regions, where \aco\ tends to be lower than is typical in spiral disks ($\sim 1$ compared to $\sim 3$--4), \citep[e.g.][]{Sandstrom2013} -- as noted by \citeauthor{Crocker2019} Their measured \lci/\lcoa\ ratios in the resolved regions
are compatible with other MS galaxies in our sample (though still higher than the ratios derived for the same set of sources by J19, see
Fig.~\ref{tdcorrF}), meaning that the \aci/\aco\ ratios are also similar. As \aco\ in these regions is determined to be low in the \citet{Sandstrom2013} analysis, the \aci\ inferred by \citeauthor{Crocker2019} is correspondingly lower as well.

\citet{Jiao2021} (henceforth J21) use CO(1--0), H\,{\sc i}, [\CI](1--0),
dust continuum and metallicity maps to investigate the variation of
\aci\ and \aco\ across the disks of six well-resolved local galaxies
from their J19 study. They use the FIR/submm dust maps from {\it
  Spitzer} and {\it Herschel} and the method of \citet{Draine2007} to
model the dust mass across the galaxy and relate this to a gas mass
via a relationship between dust-to-gas and metallicity
\citep{MM2009,Draine2007,Sandstrom2013}. As they explicitly use the dust mass together with a model of the \gdr\ dependence on metallicity, their results are normalised to the dust properties of
the \citet{Draine2007_kappa} model (hereafter \citetalias{Draine2007_kappa}), which assumes $\gdr=100$ and
$\kd=0.034$\,\kunit\ at solar metallicity. Their self-consistent
DGR(ii) method derives weighted mean values of
$\langle\aci\rangle=19.9\pm 1.9$ (including the information from lower
limits) and $\langle\aco\rangle=2.0\pm0.3$ ($2.6\pm0.4$) over the same
area as the \CI\ observations (the entire CO detection
region)\footnote{We have multiplied the values in J21 by 1.36 to
  include He for consistency with our convention.}. In the central
region, the \aco\ values are significantly lower at
$\aco^{\rm C}=1.5\pm0.3$, while \aci\ is not found to be significantly
different with $\aci^{\rm C}=21.8\pm 0.5$. These values are comparable
to to our average of $\aci=19.1\pm0.6$ for MS galaxies, and to the
average value for $Z_{\odot}$ derived independently by
\citet{Heintz2020} of $\aci(\rm
HW20)=21.4^{+13.3}_{-8.2}$\footnote{While there are differences in
  the normalisation for the dust mass model chosen by J21 and
  ourselves, the introduction of a metallicity dependence
  for the dust-to-gas ratio by J21 means that there is no simple way to scale
  their results to our method. However, we can calculate their average
  `effective \kh', $\kappa_{\rm eff}=1300$, which indicates that J21
  derive a lower \mol\ for a given \lsub\ compared to our
  normalisation (and hence a lower value of \aci\ and \aco). However,
  the six galaxies in J21 are a subset of the \CIcor\ objects, at the
  low luminosity end where there are potential decreases in \kh\ and
  increases in \aci.}.
  
\begin{table*}
	\caption{Summary of our empirical \aci\ calibration compared to work from the literature; \aci\ is quoted including He.}.
	\begin{adjustbox}{center}
	\begin{tabular}{clcll}
		\toprule
		\aci & Sample & $N_{\rm gal}$ & Notes & Reference\\
		\aunit  &         &               &       &   \\
		\midrule
		$17.0\pm0.3$       &  \Lhi           &    & weighted average & this work\\
		$19.1\pm0.6$       &  MS             &    & weighted average & this work\\
		$16.2\pm0.4$       &  SMGs            &    & weighted average & this work\\
		$10.3\pm0.3$       & (U)LIRGs       & 71 &   assuming $\Xci=3\times10^{-5}$       & \citet{Jiao2017} \\
		$4.9\pm0.3$        & (U)LIRGs  & 71 & CO(1--0) with $\aco=1.1$ & \citet{Jiao2017} \\ 
		$7.3^{+6.9}_{-3.6}$  & resolved local disks & 18 & CO(2--1) and resolved $\aco$ from S13  & \citet{Crocker2019} \\
		$19.9\pm1.9$       & resolved local disks & 6 & H\,{\sc i}, CO(1--0), \CI, dust, $Z$ ($\kappa_{\rm eff}\sim 1300$) & \citet{Jiao2021} \\
		$16.2\pm7.9$ & $z=3$ lensed SMGs & 16 & multi-$J$ CO, \CI\, dust modelling & \citet{Harrington2021}\\
		$21.4^{+13.3}_{-8.2}$       & GRB/QSO absorbers & 19 & $H_2$ and \CI\ absorption lines at $\rm{Z_{\odot}}$ & \citet{Heintz2020}\\

		17.6                     & theory  &        &  for $\zcr=5\times10^{-17}\rm{s^{-1}}$      & \citet{Offner2014}\\
		
		\bottomrule
	\end{tabular}
	\end{adjustbox}
	\flushleft{The values from this work are the weighted averages of the
		results from each of the three sub-groups containing \CI\
		information.}
\label{acilitT}
\end{table*}

\begin{table*}
	\caption{Summary of our \Xci\ calibrations compared to other work in the literature.}
	\begin{adjustbox}{center}
		\begin{tabular}{cccccc}
			\toprule
			\Xci ($\times 10^{-5}$) &  Sample & $N_{\rm gal}$ & Notes & Reference\\
			\midrule
			$1.6^{+0.5}_{-0.4}$ & $z=0$--5 \Lhi & 90 & \lsub, CO, \CI\ with $\kh^{\rm N}=1884$\,\khunit & this work\\
			$2.5\pm1.0$ & local SF & 11 & CO(1--0) and $\aco=1$ & \citet{Jiao2019}\\
			$1.3\pm   $ & local SF & 9 & CO(1--0) and \aco\ from S13 & \citet{Jiao2019}\\ 
			$1.6\pm0.7$ & $z\sim1.2$ MS & 11 & CO(2--1) and \aco(Z) ($\langle\aco\rangle=3$) &\citet{Valentino2018}$^\dag$\\
			$2.0\pm0.5$ & $z\sim1.2$ MS & 11 & dust and \gdr(Z) ($\langle\gdr\rangle=134$) &\citet{Valentino2018}$^\dag$\\
			$3.9\pm0.4$ & $z=2-3$ SMGs & 14 & CO(4--3), CO(1--0) and $\aco=1$ & \citet{AZ2013}$^\dag$\\
			$8.4\pm3.5$ & SMGs/QSOs & 10 & CO(3--2) and $\aco=0.8$  & \citet{Walter2011}\\
			$8.3\pm3.0$ & local (U)LIRGS & 23 & CO(1--0) and $\aco=0.8$  & \citet{Jiao2017,Jiao2019}$^\dag$\\
			$0.9\pm0.3$ & $z=1$ ISM selected & 2 & CO(2--1), \CI\ and $\aco=2.6$  & \citet{Boogaard2020}\\
			$2.0\pm0.4$ & $z=1$ ISM selected & 3 & 1.2\,mm, \CI\ and $\asub=6.7\times10^{12}$ from Sco16 & \citet{Boogaard2020}\\
			$^{\ast}1.6^{+1.3}_{-0.7}$ & $z=2$--4 GRB/QSO absorbers & 19 & \mol\ and $\rm{C^0}$ absorption lines for $\rm{Z_{\odot}}$ & \citet{Heintz2020}\\
			$^{\ast}7^{+7}_{-3.5}$ & NGC\,7469 (CND) & 1 & AGN, dynamical mass, \lci\ and \lcoa & \citet{Izumi2020}\\
			$^{\ast}1.4-5$ & NGC\,6240 & 1 & \aco\ from CO--SLED, high-density tracers & \citet{Cicone2018}\\
			\multicolumn{3}{c}{}& and two-phase LVG modelling & \citet{PPP6240}\\
			\bottomrule
		\end{tabular}
	\end{adjustbox}
	\flushleft{$^{\ast}$ indicates estimates of \Xci\ independent
          of assumptions for \aco\ or \kh. $^\dag$ indicates that this
          sample forms part of the literature sample we have used,
          although we have calibrated \Xci\ using the submm luminosity
          and an average normalisation of $\kh^{\rm N}=1884$\,\khunit
          ($\gdr=135$ for $\kd=0.071\kunit$) for the sample, rather
          than \lcoa\ and a fixed \aco. A breakdown of our
          results by intensity of star formation can be found in
          Table~\ref{hyp2T}.}
\label{XCIlitT} 
\end{table*}

\subsubsection*{CO empirical factor: \aco}

We compare our optimised \aco\ estimates with others from the literature in Table~\ref{acolitT}.
Sophisticated LVG modelling with very large datasets which include
high-density gas tracers, optically thin CO isotopologues, full CO
SLEDs, and sometimes the \CI\ lines and dust emission
\citep[e.g.][]{Weiss2007,PPP2012xco,PPP6240,Israel2020,Harrington2021}
can break some of the model degeneracies of the optically thick CO
lines, though the method is still reliant on assumptions for
[CO/\mol], isotopologue ratios, the number of components allowed
(single components give very different results to multiple components)
and the allowed range of velocity gradients in the models.

The best examples are NGC\,6240 \citep{PPP6240} and the {\it Planck}
lensed galaxies \citep{Harrington2021} where detailed LVG modelling
and comprehensive datasets have sufficient constraints to break the
degeneracies which usually be-devil this method. The two-component LVG
result for NGC\,6240 is $\aco=2$--4 \citep{PPP6240} (cf.\ \aco=0.6
when using a single component LVG model \citealp{PPP2012xco}) and we
can further use the ratio of $\lci/\lcoa$ measured by
\citet{Cicone2018} and our relationship,
$\lci/\lcoa=\aco/\aci = 3324\,\aco\Xci$, to infer that
$\Xci=1.4$--$2.9\times10^{-5}$ in the starburst region. In fact, our
optimised values for this galaxy using global fluxes, are
$\aco({\rm daX}) = 2.9\pm0.6$,
$\Xci({\rm daX})=(2.4\pm0.5) \times 10^{-5}$,
$\kh({\rm daX})=2800\pm700$ ($\gdr=200$), in excellent agreement. The
{\it Planck} lensed galaxies analysed by \citet{Harrington2021} do not
have the same degeneracy-breaking lines used by \citep{PPP6240} in
their analysis, but they do have multi-$J$ CO coverage and incorporate
the \CI\ lines and the dust continuum emission in their model fitting,
based on \citet{Weiss2007}. They assume similar dust parameters as we
do for their normalisations ($\gdr=120$--150 with
$\kd=0.08$\,\kunit). With this, they infer an average $\aco=3$--4 and
an average $\aci=16.2\pm7.9$ (incl.\ He), remarkably consistent with
our results, given our very simple approach.  
 
\begin{table*}
	\caption{Summary of our empirical \aco\ calibrations compared to work in the literature, \aco\ is quoted including a factor of 1.36 for He.}
	\begin{adjustbox}{center}
		\begin{tabular}{cccccc}
			\toprule
			\aco  &  Sample & $N_{\rm gal}$ & Notes & Reference\\
			\aunit &     &    &  & \\
			\midrule
			$3.6^{+1.3}_{-1.0}$ & $z=0$--5, \Lhi & 90 & \lsub, CO, \CI\ with $\kh^{\rm N}=1884$\,\khunit & this work \\
			\addlinespace[1pt]
			$4.2^{+1.8}_{-1.1}$ & $z=0$--5, \Lhi & 240 & \lsub, CO with $\kh^{\rm N}=1884\,\khunit$ & this work\\
			\addlinespace[1pt]
			$4.8^{+1.3}_{-1.1}$  & local MS, \Llo & 88 & \lsub, CO with $\kh^{\rm N}=1884$\,\khunit & this work\\
			\addlinespace[1pt]
			$^{\ast}3.1^{+3.1}_{-1.5}$ & local disks & 26 & CO(2--1), $r_{21}=0.7$, H\,{\sc i}, dust & \citet{Sandstrom2013}\\
			$^{\ast}4.2$ (3.5--5.4) & MW large scale &  & $\gamma$-ray various & \citet{Remy2017}\\
			$^{\ast}3.4\pm2.1$ & Planck lensed SMGs & 24 & LVG: multi-$J$ CO, \CI\ and dust & \citet{Harrington2021}\\
			$^{\ast}4.1^{+4}_{-2}$ & NGC\,7469 (CND) & 1 & AGN, dynamical mass, \lcoa & \citet{Izumi2020}\\
			$^{\ast}2-4$ & NGC\,6240 & 1 & LVG: multi-$J$ CO, dense gas tracers & \citet{PPP6240}\\
			$^{\ast}3.8^{+1.0}_{-0.7}$ & $z=0$--5 & 22 & CO, \CI, Z and absorber based \aci      & \citet{Heintz2020}\\ 
			\addlinespace[1pt]
			$^{\ast}4.4^{+2.0}_{-1.4}$ & local galaxies & 24 & C\,{\sc ii}, CO(1--0) and modelling at $\rm{Z_{\odot}}$   & \citet{Accurso2017} \\
			$^{\ast}0.6\pm0.2$ & (U)LIRGs & 28   & LVG:
                                                               Single component, multi-$J$ CO  & \citet{PPP2012xco} \\
			$^{\ast}2-6$ & (U)LIRGs & 28   & LVG: Two-comp, free d$V$/d$R$, dense-gas tracers& \citet{PPP2012xco}\\
			$^{\ast}3.9\pm1.1 $ & local disks & 9 & CO(1--0), H\,{\sc i}, dust & \citet{Eales2012} \\
			$1.8\pm0.5$ & MW local clouds & 6 & H\,{\sc i}, CO(1--0), $\gamma$-ray & \citet{Remy2017} \\
			$2.9\pm0.5$ & MW local clouds & 6 & H\,{\sc i}, CO(1--0), 850-\mic\ dust & \citet{Remy2017} \\
			2.9        & Taurus          & 1 & H\,{\sc i}, CO(1--0) and extinction/reddening & \citet{Chen2015} \\
			$2.4\pm0.4$ & local disks & 7 & CO(1--0), H\,{\sc i}, dust & \citet{Cormier2018} \\
			
			$1.9\pm0.3$ & resolved local disks & 6 & H\,{\sc i}, CO(1--0), [\CI](1--0), dust, $Z$, $\kappa_{\rm eff}\sim 1300$ & \citet{Jiao2021}\\
			$3.2\pm1.0$ & $z=4$ lensed SMGs & 9 & CO(2--1), [\CI](1--0) $\Xci=3\times10^{-5}$ & \citet{Bothwell2017}\\
			\bottomrule
		\end{tabular}
	\end{adjustbox}
	\flushleft{Errors are 1$\sigma$ standard deviations (or 16--84
          percentiles). A breakdown of our results by intensity of
          star formation can be found in Table~\ref{hyp2T}. $^{\ast}$
          indicates estimates which do not rely on assumptions for
          \Xci\ or \kh.}
\label{acolitT} 
\end{table*}

\subsection{Lack of bi-modality in the conversion factors}
\label{calCOS}

Our sample contains normal star forming galaxies -- those obeying the SFR--$M_\star$
correlation that forms as a result of the more intimate relationship
between SFR and H$_2$ -- as well as many extreme star-forming systems, which belong to the (U)LIRG and high-$z$ submillimeter selected samples. Here we remind the reader that we refer to the extreme SF group -- those that supposedly require a
lower \aco\ -- as `SMGs', and the normal star forming sources as `MS galaxies', or
sometimes just `MS'. As mentioned in Section~\ref{obsS}, the assignment of the galaxies to either category is by nature of the data rather `fuzzy' as we do not have a measure of SFR or stellar mass for all sources, nor any homogeneous way to estimate them. We thus rely on the categories used by previous authors where possible, especially for high-$z$ sources. The $z=1$ galaxies from the samples of \cite{Bourne2019,Valentino2018,Valentino2020} are deemed to be `MS', as are the sources from ASPECs \citep{Boogaard2020}. Most low redshift sources with log \Lir<12 are classed as `MS' though there are some exceptional LIRG class sources in the local Universe which have extreme properties as evidenced by their FIR, MIR lines and vibrational HCN \citep{DiazSantos2017,Falstad2021}. 
We note that using a more conservative separation when assigning galaxies into MS and extreme starburst categories does not change any of the results. We therefore conclude that while our assignment of sources into the two SF categories is not perfect, this categorisation is not capable of masking any strong bi-modality in the observable ratios. 

Fig.~\ref{arghF} and Table~\ref{hyp2T} detail the distributions of
conversion factors for each sample, split into MS galaxies and
SMGs. While formally there are significant differences in the
parameters for some samples, these are very small -- around 10--20 per
cent in the mean, rather than the factor $\sim 3$--$4\times$ often assumed for
\aco\ \citep[e.g.][\aco=0.8, derived for four ULIRGs]{Downes1998}. In fact, only the ad sample
shows any difference in \aco\ between the MS galaxies and SMGs, while
the estimates based on \CI\ and CO, or on all three tracers, show no
significant difference. This is partially explained by the larger
luminosity range in the ad sample, combined with the previously noted
negative correlation between \aco\ and luminosity (Figs~\ref{calaodF}
and \ref{empLF}), with a factor $\sim 2\times$ reduction in \aco\ for
a factor $\sim 100\times$ increase in \lcoa. We cannot rule out that
the correlation of \aco\ with luminosity is the true reason that the
ad sample shows a significant difference between MS galaxies and
SMGs\footnote{$\Lir{\rm (ad)}{\rm (MS)}=10.95$
  ($\lcoa{\rm (ad)}{\rm (MS)}=9.31$) while
  $\Lir{\rm (Xa)}{\rm (MS)}=11.22$ ($\lcoa{\rm (Xa)}{\rm
    (MS)}=9.52$). Using the relation in Table~\ref{calfitsT}, the
  expected $\Delta \aco=\aco{\rm (MS)} - \aco{\rm (SMG)}=0.36$ for the
  Xa sample and -- due to the lower numbers in the Xa and daX samples
  -- such a difference would not be detected at a significant level,
  if it existed.}.

This is not the first time\footnote{However our current dataset is
  more homogeneous, using only CO(1--0) or CO(2--1) and a consistent
  approach to modelling the dust with our empirical relations for
  \mwtd.}  that lack of bi-modality in \aco\ has been reported when
compared to dust-based determinations
\citep[e.g.][]{Magdis2012,Rowlands2014,Genzel2015}.  The range of
\aco\ we find for SMGs (see Fig.~\ref{calaodF}) is well within the
framework set out by \citet{Papadopoulos2012}, who noted that galaxies
with a highly turbulent ISM (e.g. ULIRGs and SMGs) can have
\aco\ similar to galaxies with a much more quiescent ISM, the only
difference being that in a turbulent ISM, the distribution of gas mass
as a function of density is weighted to higher densities
than in a less-turbulent ISM. 

Recent joint SLED/SED modelling of an
exquisite dataset that includes CO, \CI\ and dust continuum for lensed
SMGs \citep{Harrington2021} finds a mean $\aco =3.4$--4.2 for these highly
turbulent galaxies (albeit with a large dispersion). The \citeauthor{Harrington2021} radiative transfer models 
employ a continuum distribution of molecular gas mass as a function
of average Mach number (and average density of the molecular cloud ensemble), making them better equipped to `capture' any re-distribution of
the underlying molecular gas mass towards higher densities. 
While important for other
issues (e.g.  the initial conditions of star formation in
SMGs/ULIRGs), such a re-distribution in a highly turbulent ISM may actually leave \aco\ statistically unaffected. The initial reports of a bimodal \aco\ factor in
the local Universe, with $\sim $4-5$\times $ lower values for ULIRGs
than LIRGs and ordinary spirals, can possibly be explained by a CO-luminous, strongly unbound,
low-density molecular gas component found preferentially in ULIRGs. Such a component can dominate the global CO(1--0) line luminosities of
ULIRGs/SMGs (even if containing only small fractions of their total
molecular gas), while its large $\rm K_{vir}$ values will yield systematically low \aco\ factors, under
one-component LVG modelling  (Equation 9).\footnote{Also we must consider the size of the original sample -- four ULIRGs in the first study by
\citet{Downes1998}.}

For individual galaxies, only multi-component models of SLED/SED (that
also include molecules/transitions tracing the dense gas) can properly
account for this effect \citep[e.g.][]{PPP6240,Harrington2021}, while for large galaxy samples, our
cross-calibration of \aco\ against the other two molecular gas mass
tracers, is the most economical method.  In that regard it is worth
noting that {\it dust continuum is immune to the gas-dynamics effects
  described above,} i.e.  a diffuse low-density, unbound, $\rm H_2$
gas component will contribute very little to the total dust continuum
if its gas/dust mass is indeed low.  The optically thin \CI\ line
emission will also be much less sensitive than CO(1--0) to such
gas-dynamics effects exactly because of its low optical depths.  These
are perhaps the reasons why our cross-calibration of \aco\ against
dust and \CI\ emission has not uncovered any obvious bimodality of its values in
MS galaxies compared to SMGs.

The range of values we find for \aco\ is consistent with expected values 
for $Z>0.5\,Z_{\odot}$ galaxies (\citealt{Accurso2017}, based on calibrating \aco\ using C{\sc ii}). Using their predictions, we would expect
$2.7<\aco<15.2$ for the likely range of metallicity and offset from the MS in our sample.

\begin{figure*}
	\includegraphics[width=0.67\textwidth,trim=0cm 0cm 0cm 0cm, clip=true]{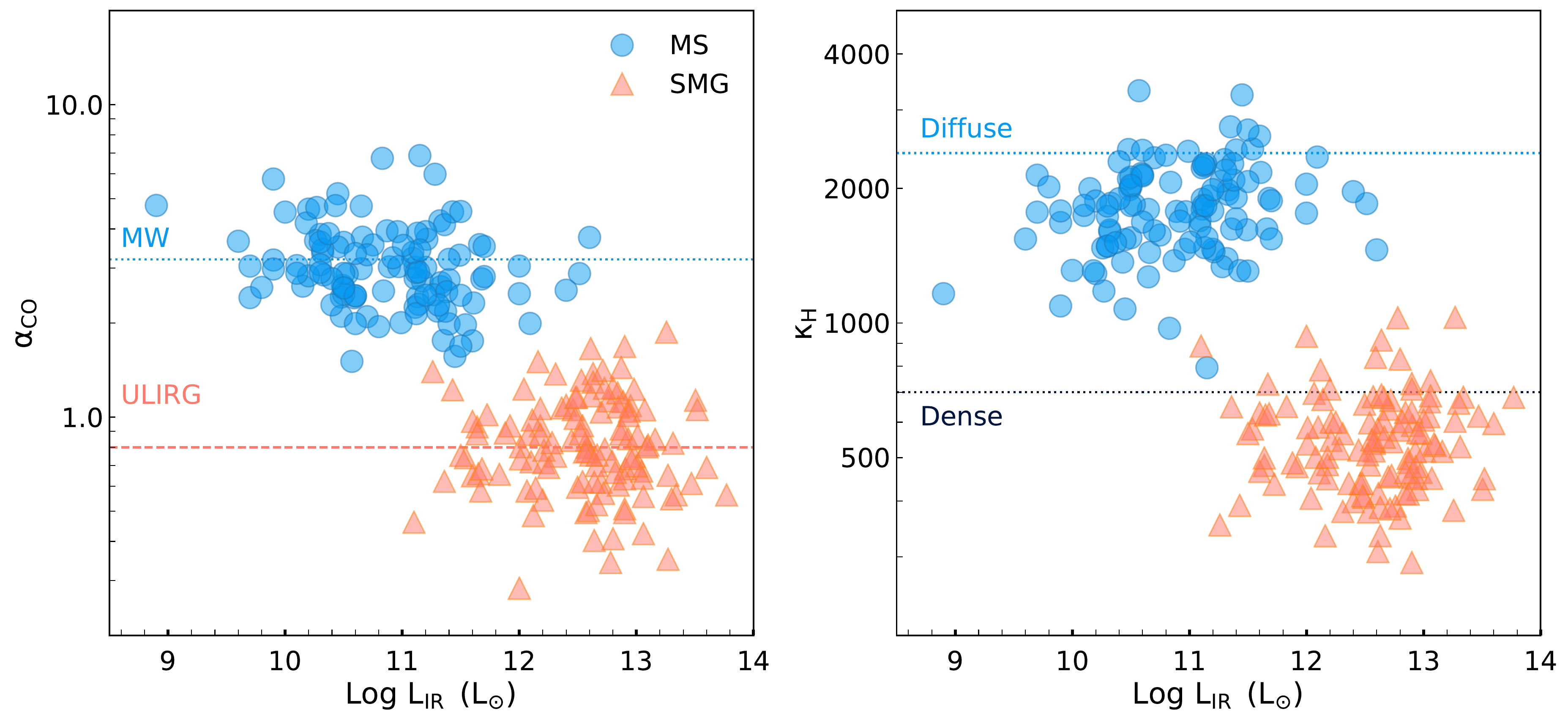}
	\includegraphics[width=0.31\textwidth,trim=0cm 0cm 0cm 0cm, clip=true]{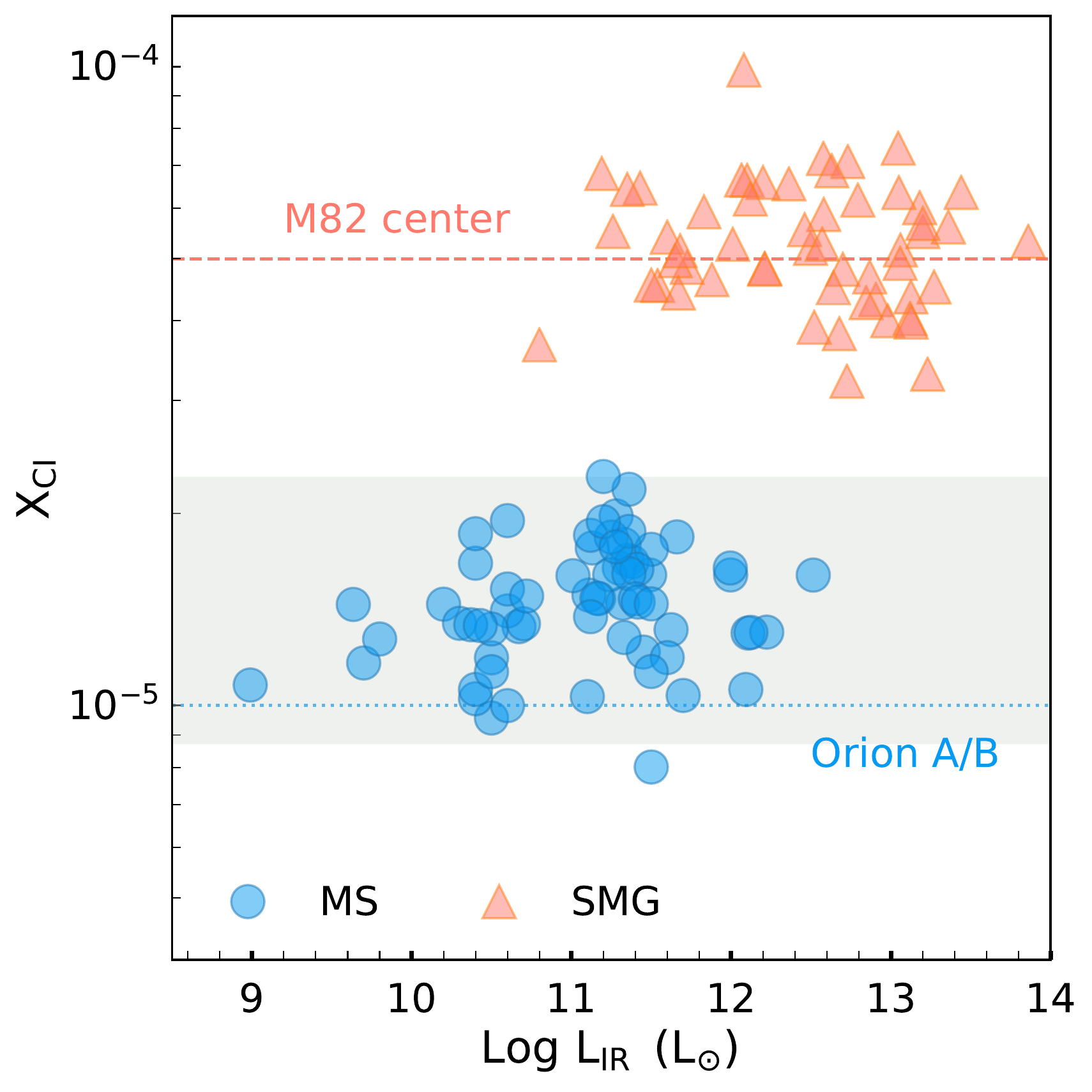}
	\caption{Results assuming a bi-modal behaviour for \aco. We run
          the optimisation process with the normalisation for SMGs set
          to $\aco^{\rm N}(\rm SMG)=0.8$, keeping MS galaxies at
          $\kh^{\rm N}(\rm MS)=1884$\,\khunit (equivalent to
          $\aco$=2.8). This creates the strong bi-modality in \aco\
           but also induces a similarly strong or even
          stronger effect in \kh\ and \Xci, which would be mirrored in
          \asub\ and \aci.}
 \label{smgF}
 \end{figure*}

Our underlying assumption: that  the dust--gas  properties of MS galaxies and SMGs can be described as a uni-modal distribution with well defined mean and scatter, is based on our finding that the luminosity ratios (Fig.~\ref{LhistsF}) -- the most basic observables used in deriving  the empirical  conversion  factors -- have such a distribution. They show no evidence for the strong bi-modality advocated for \aco\ in some of the literature. That the distribution of the observed luminosity ratio is, to first order, similar to the distribution of the conversion parameters, is the simplest `Occam's Razor' assumption we can make.  
 
To see what a different initial assumption would mean for the conversion factors, we repeated our analysis, this time
 inserting the popular bi-modal behaviour in \aco\ \citep{Greve2005,Weiss2005,
   Tacconi2006,Tacconi2008,Genzel2010,Walter2011,AZ2013,Jiao2017,Valentino2018}
 as our prior, such that the sample mean normalisations for SMGs and
 MS galaxies are set to be different: $\aco^{\rm N}(\rm SMG)= 0.8$ and
 $\kh^{\rm N}(\rm MS)=1884$\,\khunit. Our optimal method then allows
 the data to return the most likely values for the other
 parameters\footnote{We did not re-calculate the intrinsic scatter
   split into MS galaxies and SMGs, only the values of the \Xci/\aco\
   and \aci/\kh\ pairs.}  under these assumptions.

 Fig.~\ref{smgF} shows  the results with this  bi-modal normalisation,
 (blue  points: MS  galaxies, red  points: SMGs).  By design,  we have
 reproduced  the  extreme  bi-modality  $\aco(\rm  MS)\sim  3$--4  and
 $\aco(\rm  SMG)\sim 1$,  but Fig.~\ref{smgF}  clearly shows  that the
 same extreme bi-modality  has to be present in \kh\  (\gdr) and \Xci,
 giving a clear prediction that $\Xci(\rm SMG)\geq 4\times 10^{-5}$ if
 the bimodality in \aco\ really exists, with essentially no overlap in
 \Xci\ between the MS galaxies and  SMGs. To test this will require an
 independent determination of \Xci\ in SMGs, without reference to dust
 or CO calibration. To date, there  is no such determination of \Xci, although \citet{Izumi2020} observed the nearby LIRG NGC~7469 with ALMA, using kinematic data to derive $\rm{M_{dyn}}$, which is the sum of \Mh, stellar mass and dark matter. This method has promise, but the systematic uncertainties in \Mh\ from this analysis are too large (0.3\,dex) to answer our question. While the \citeauthor{Izumi2020} study clearly indicates\footnote{via the extremely high observable ratio $\lci/\lcoa=0.92$ in the CND} that the [$\rm{C^0}/CO$]
abundance can be enhanced in extreme environments, the CND is only a tiny region and
the {\em global ratio} for this source is very similar to other LIRGs, with $\lci/\lcoa = 0.20\pm0.04$. Any study which wishes to test the bi-modality hypothesis must also be representative of the galaxy global properties.

Here  we  must  stress  again that  for  individual  galaxies,  joint
 SLED/SED  radiative transfer  models of  well-sampled SLEDs  and dust
 emission  SEDs  do  recover  Galactic-valued \aco\  factors  even  in
 (U)LIRGs or SMGs  \citep{PPP6240,Harrington2021}. However
 such results cannot be used in  a statistical sense, i.e.  as typical
 of the  respective galaxy  populations for  obvious reasons,  and our
 statistical approach remains the sole  avenue.

Indeed, the only way that a bi-modal \aco\ for MS galaxies and SMGs can be
 reconciled with \Mh\ estimates using dust or \CI\ is to impose
 the same bi-modality on their conversion parameters (\kh, \Xci, \asub, \aci). A reduction of
 \aco\ by a factor $3\times$ necessitates a decrease [increase] in
 \kh\ [\Xci] by the same factor. Thus, if $\aco=0.8$ is preferred for
 extreme star-forming galaxies \citep[e.g.][]{Walter2011}, then
 $\kh=600$ ($\gdr=43$ for $\kd=0.071$\,\kunit) and
 $\Xci=5.3\times10^{-5}$ must also be adopted (statistically, for this
 galaxy population). This discrepancy was previously noted by
 \citet{Bothwell2017} and \citet{Valentino2018} who found that using
 $\Xci(\rm SMG)=3\times10^{-5}$ with \lci\ as a tracer resulted in
 larger gas masses than using \lcoa\ with the `ULIRG' value of
 $\aco=0.8$. {\em Therefore, the popular `choices' of $\aco=0.8$ and $\Xci=3\times10^{-5}$ are incompatible with each other.} 

Based on our current understanding, there are two plausible physical mechanisms which may cause an increase in \Xci\ and a decrease in \kh\ in extreme ISM conditions. The effect of enhanced cosmic ray densities on carbon chemistry \citep{Bisbas2015,Bisbas2021,Glover2016,Gong2020} favour a higher $\rm [C^0/CO]$ abundance, however, this mechanism is density dependent and is less effective in dense regions, which typify the ISM of SMGs. Thus while extreme environments with elevated cosmic rays or X-rays would certainly act to increase \Xci\ at a fixed density, it does not simply follow that extreme SF activity will produce high \Xci\ since those same regions (CRDR/XDR) are typically found in regions with increased density. 

The higher dense gas fractions common in SMGs may favour higher rates of grain growth, or mantling, both of which would reduce the value of \kh\ -- i) by decreasing the \gdr\ and ii) by increasing
the dust emissivity, \kd. Our results imply, however, that any such changes must act in harmony with each other so as to maintain the same observable ratios, so an increase in
\Xci\ must correlate directly with a decrease in \kh\ and \aco. This
prediction is a clear challenge to models, and full astro-chemical simulations for the extreme physical conditions expected in the ISM of SMGs and ULIRGs will be needed to explore how the three tracers can vary in the exact same way through very different physical mechanisms.

\subsection{On the robustness of our choices}

\subsubsection{Impact of uncorrected \fhi>1 galaxies}

Statistically, the effect of having uncorrected \fhi>1 galaxies is
small, since there are only 15 such galaxies, where
$\langle\aco\rangle$ increases\footnote{Note that this offset is
  linear, not logarithmic.}  by +0.27 in their luminosity bin compared
to when they are removed altogether (by +0.18 compared to when they
have been corrected).  We are thus confident that dust associated with
\HI\ is not biasing our overall determination of conversion factors
and their trends, at least in this sample. For individual galaxies,
however, the difference in \aco\ can be very large. When using this
method for low-redshift galaxies with significant \HI\ within the
dust-emitting region, corrections are needed.

\subsubsection{Impact of using a constant \mwtd}
\label{fixTS}

The strong correlation found between \mwtd\ and luminosity
(Fig.~\ref{mwtdzF}) has not been considered in previous works. In
Appendix~\ref{QTS}, we show a comparison of results using our
empirical \mwtd\ relations to those for constant, \mwtd=25\,{\sc
  k}. Summarising these findings:

\begin{enumerate}
\item{The median offsets between parameters when using constant vs.\
    variable \mwtd\ are $<0.015$\,dex. The scatter in a parameter is
    generally within 0.1\,dex (Fig.~\ref{delcalqu_histF}). Thus the
    global averages we present in this paper are not affected by a
    change to constant \mwtd=25\,{\sc k}.}
\item{Allowing \mwtd\ to vary with \Lir\ is more realistic and leads
    to a shallow but significant trend with luminosity, such that
    \aco\ decreases with increasing \Lir, while \kh, \asub\ and \Xci\
    increase slightly. Using a constant \mwtd\ of 25\,{\sc k} produces
    no trends of any conversion factor with \Lir\
    (Fig.~\ref{fixQT_lirF}).}
\item{Using a constant \mwtd\ of 25\,{\sc k} results in gas masses up
    to 0.1\,dex lower at \Llo\ and 0.1\,dex higher at $l_{\rm IR}<12.0$
    compared to the variable \mwtd\ used in the main analysis
    (Fig.~\ref{delMhQT_lirF}).}
\end{enumerate}

\section{Discussion}
\label{DiscS}

The diverse galaxies in this study show a remarkable consistency in
their gas mass tracers, with linear relationships between all three
pairs of observables, \lsub, \lci\ and \lcoa.

We find weak trends in the conversion factors with \Lir; decreasing \aco,
\aci\ and increasing \asub, \kh, \Xci. These trends are very shallow, amounting to a factor $<2\times$ change
over 2--3 orders of magnitude in luminosity. The intrinsic variation
in \kh\ and \Xci\ (the physical quantities encompassing most of the
uncertainties in the corresponding conversion factors) is likely very small,
and approximating them with a single constant value should be robust.
For the sub-samples with a \CI\ tracer (daX, Xa, Xd), we see decreases in all three tracer conversion factors at \Llo: \Xci\ (15--25 per cent), \aco\ (10--30 per cent) and
\kh\ (\gdr: 20--25 per cent). However, the data indicating a drop in conversion factors originates from the J19 sample (see earlier discussions). More
\CI\ studies of normal star-forming galaxies in the local Universe are
urgently required to further explore any such trend, in particular using the global
\CI\ line emission, rather than that of the central few kpc of a
galaxy.

The average values of \Xci, \aco\ and \kh\ (\gdr) for galaxies with
all three tracers (the daX sample) and \Lhi\ are our `reference'
values, $\acoR =3.7$ (including He), $\XciR = 1.6\times10^{-5}$, $\kh^{\rm R} =
  1990\, (\gdrR=141)$. These agree within the errors with the mean
values determined using only two tracers.  These reference values are
{\em not unique} because only the ratios and products of the conversion
factors are constrained by the observables, \lci/\lcoa,
\lci/\lsub\ and \lsub/\lcoa.

Once a conversion factor is known  or assumed, however, the others can
be   determined    by   the    self-consistent   ratios    listed   in
Table~\ref{methodT}. For   example, using the ad sub-sample and normalising   to   $\kh^{\rm
  N}=2800\,\rm m^2\,kg^{-1}$ would produce $\Xci=1.1\times10^{-5}$ and
$\aco=4.7$,  in  reasonable  agreement  with  \citet{Accurso2017}  for
$Z=0.6\,Z_{\odot}$,  while  normalising  to $\aco^{\rm  N}=0.8$  gives
$\Xci=5.8\times 10^{-5}$ and $\gdr=34$.  While the data are consistent
with  both of  these possibilities,  or any  other combination  of the
above  ratios, we  must caution  that the  low values  of \aco\  often
recovered from CO-only methods (and after modeling only a few low-J CO
lines) may be an artifact of well-known gas-dynamics effects, which are
expected to have very little impact on the global \CI\ line emission and
none whatsoever on the corresponding dust continuum.

For galaxies at \Llo\ we can only use the ad sample (CO and dust
continuum) because of the uncertainties surrounding the \CIcor\
galaxies. For the 88 galaxies at \Llo\ with CO and dust
measurements, $\aco = 4.8^{+1.4}_{-1.1}$ (including He), with
$\langle\kh\rangle=1718$\,\khunit (\gdr=122) but note that this is
still a sample of massive and metal-rich galaxies, just at lower log $L_{\rm IR}\sim 9-11$.
This study is not applicable to low mass metal-poor
galaxies.

\section{Conclusions}

We have cross-calibrated the three mainstays of molecular gas
measurements in extra-galactic astronomy: $^{12}$CO(1--0), \CIfull\
and submm continuum emission from dust. This analysis uses galaxy
samples spanning $0<z<5$ and more than four orders of magnitude in
\Lir. All the galaxies are metal rich and/or massive, to remove the
need for large corrections for metallicity effects.

\begin{itemize}
\item{We present a new method of optimising gas mass estimation when
    multiple tracers are observed, making use of the intrinsic scatter
    in all three pairs of gas tracers. We demonstrate its
    effectiveness compared to the simpler method used previously in
    the literature, and give examples and prescriptions for its use.}
\item{In a purely empirical analysis, we show that \lci\ is the
    molecular gas tracer with the least intrinsic scatter,
    particularly at \Lhi. In such galaxies, \lci\ should be
    the preferred tracer, all other considerations being equal.}
\item{Using our optimised method, we determine the mean empirical
    conversion factors for \Mmol\ (including He). For \Lhi\
    these are: ${\aci^{\rm R}=\acir\pm\eacir}$,
    $\asub^{\rm R}=(\asubr\pm\easubr)\times10^{12}$,
    $\aco^{\rm R}=\acorm\pm\eacorm$, with a scatter of 0.11-0.15 dex. These values are for an overall normalisation set to the average dust properties of local galaxies and diffuse dust in the Milky Way (\kh = \gdr / \kd = 1884 kg $\rm m^{-2}$). A change in this choice of normalisation will affect \aco\ and \aci\ in a proportional manner and \asub\ in an inversely proportional way}. Our reference conversion values
    can be applied to any metal-rich galaxy with $Z>0.5\,Z_{\odot}$ in
    the range $0<z<6$.
\item{Using the same method we determine the principal mean
    physical parameters on which these conversion values depend. For galaxies at \Lhi:
    ${\Xci^{\rm R}=\xcir\times10^{-5}}$, $\kh^{\rm R}=\khr\, (\gdr^{\rm R}=141)$.}
\item{The relationships between the observables, \lsub, \lcoa\ and \lci\ are consistent with being linear and the ratios of these
    observables do not show a strong dependence on IR luminosity, dust
    temperature, redshift or the intensity of star formation.}
\item{The ratio of \lci/\lcoa\ is marginally (3$\sigma$) different for
    MS galaxies and SMGs, with the latter having higher \lci/\lcoa,
    broadly consistent with expectations from astro-chemical cloud
    models that include enhanced cosmic rays.}
\item{We find $\Q=0.48$ to be a reasonable choice for the excitation
    function (required to convert \lci\ to $M_{\rm C\,I}$), based on
    recent analysis showing that \Q\ has a super-thermal behaviour in
    non-LTE conditions \citep{PPPDunne2022}. For a range of plausible
    galaxy ISM density and gas temperatures, the 99th percentile
    confidence interval on this value is $\pm 16$ per cent.}
\item{We present empirical relations for the mass-weighted dust
    temperature, \mwtd, to allow observers to better estimate their
    dust calibration factors. We find a significant trend, where
    \mwtd\ increases with \Lir. The median \mwtd\ for SMGs at
    $z\sim 2.5$ is $\mwtd^{\rm SMG}=\tmwSMG\pm\etmwSMG$\,{\sc k},
    while for MS galaxies $\mwtd^{\rm \ms}=\tmwMS\pm \etmwMS$\,{\sc
      k}.}
\item{We find a weak trend for \kh\ and \Xci\ to increase with \Lir,
    and a similar trend for \aco\ to decrease. The empirical
    conversion factors (\aco, \aci\ and \asub) also show a shallow
    but significant correlation with their tracer luminosities. These
    trends are not apparent if a constant \mwtd\ is adopted. They are
    therefore driven by the change in \mwtd\ with luminosity.}
\item{Using an Occam's Razor assumption that metal-rich galaxies have similar dust emissivity per unit gas mass, we find no evidence for the factor 3--4 \aco\ bi-modality between SMGs and MS galaxies often adopted in the literature. The shallow trends we do find reflect the common assumption that extreme SF systems have lower \aco\ and higher \Xci, albeit at a far more subtle level, with only a $\sim 15$\,per\,cent difference in the sample mean \asub\ (higher), \aci\ (lower) and \aco\ (lower) for extreme star-forming galaxies versus `normal' MS star-formers.}
o overall

\item{With the Occam's Razor assumption, we also find no evidence to support the extremely high global estimates of
    $\Xci$ ($\sim 6\times 10^{-5}$) reported in some literature for
    ULIRGs/SMGs -- the high reported values are a consequence of assuming a low $\aco\sim 1$. High \Xci\ values may be expected, and indeed have been measured in small ($<500$pc) regions such as M82 (nuclear starbursts) and XDR regions around AGN, but the extent to which a global estimate would be enhanced depends on the dominance of that extreme environment in the galaxy's \mol\ reservoir.}
\item{One can, however, still postulate a different prior for the normalisation assumption and impose the popular bimodality in \aco. The constancy of the measured tracer luminosity ratios then forces the conversion factors for the other two tracers (dust and \CI) to become bi-modal in the same way.}
\end{itemize}

We conclude  by noting that  lacking a direct \Mh\  measurement method
(i.e. via  the $\rm H_2$ lines  themselves), one must assume a
normalisation for one of the sample mean conversion factors in statistical studies like ours. In the present study we choose to benchmark to the dust   emission,   with   $\kh^{\rm   N}=1884$\,\khunit. Other
normalisation choices can of course be  made, but currently dust emission is the simplest and best understood tracer, and has the advantage of being totally insensitive to the
gas-dynamic effects that affect  the \aco\ conversion factor  (e.g.  unbound molecular gas components in the winds that exist in actively star-forming galaxies; winds which can be CO-bright
while  carrying little  mass).
The  [\CI](1--0)  line emission  will also be largely unaffected by these
gas-dynamics effects, and as  such the corresponding conversion factor, \aci,
shows promise as a good benchmark, borne out by the empirical finding that it has the least intrinsic scatter of the three tracers. With
more extensive observational and theoretical studies of \CI\ line emission (particularly in  galaxies of lower IR luminosity), the limits of its usefulness as a gas tracer can be determined. 

\section*{Data Availability}
Data tables based on the samples used in this paper are available via anonymous ftp to cdsarc.u\-strasbg.fr (130.79.128.5), alternatively via \url {http://cdsarc.u-strasbg.fr/viz-bin/qcat?J/MNRAS/}. The datasets were derived from sources in the public domain, which are listed in Table~\ref{SampleT}.

\section*{Acknowledgments}
The authors thank the referee for their careful reading and insightful comments on the original version of the paper.  
LD thanks P.~Clark, S.~Glover, Q.~Jiao and T.~Bisbas for helpful discussions. LD, SJM
and HLG acknowledge support from the European Research Council
Consolidator grant, Cosmicdust. 

This paper makes use of the following software available publicly from
github: corner.py, emcee.py \citep{DFM2013,corner}.

\bibliographystyle{mnras/mnras}
\bibliography{masterbib}

\bsp
\appendix

\section{Notes on the literature fluxes}
\label{notesS}

In order to produce a homogeneous and up-to-date set of fluxes, we have
applied the following corrections.

\paragraph*{Corrections to previously published work:}

\begin{enumerate}
\item{Since Sco16 was published, the 500-\mic\ flux densities used for
    their local sample \citep{Dale2012} were updated following the
    latest {\it Herschel} calibration. To estimate \mwtd, \td\ and
    \lsub\ we fitted the photometry presented by \citet{Chu2017} and
    \citet{Clark2018} using the method described in
    \citet{Dunne2001}.}
\item{The 850-\mic\ photometry for local galaxies in the SLUGS sample
    \citep{Dunne2000} is contaminated by the CO(3--2) line. We have
    corrected for this using the results of \citet{Seaquist2004},
    where for galaxies with $D<148$\,Mpc we reduce the 850\,\mic\ flux
    density by 25 per cent.}
\item{It appears that the CO(2--1) data from \citet{Aravena2016}, as
    reproduced in \citet{Bothwell2017}, has been incorrectly converted
    to $L^{\prime}_{10}$ (\lcob\ appears to have been multiplied by
    0.9 instead of being divided by it). We have corrected this error
    and applied our chosen value of $r_{21}=0.8$ for the conversion.}
\end{enumerate}

\paragraph*{Homogenisation of distances:}

The most local galaxies ($D<30$\,Mpc) often have a variety of
distances used in the literature. As we have often taken \Lir, \lci,
\lcoa\ and \lsub\ from different papers, we have had to homogenise the
literature luminosities to correspond to a common distance. The
distance chosen is that listed in \citet{Dale2017} and presented in
Table~\ref{SampleT}.

\paragraph*{Updating local CO data:}

The Sco16 local galaxy sample used CO(1--0) fluxes from the FCRAO
single-dish survey of \citet{Young1995}, which has significant and
uncertain extrapolations to total fluxes for extended galaxies. We
have updated the CO data for these very local galaxies to use CO(1--0)
maps from the COMING survey \citep{Sorai2019} where possible as well
as from other mapping datasets from the literature
\citep{Gao2004,Kuno2007,Young2008,Galametz2011,Koda2011,Schruba2012,Ueda2014}.

\paragraph*{New CO measurement for ID141:}
We use an unpublished CO(1--0) flux for ID141, which was observed with
the Jansky Very Large Array and has ${S_{10}=0.61\pm0.09\,\rm Jy\,\kms}$.

\section{Required corrections}

\subsection{H\,{\sc i}-dominated galaxies at lower \Lir}
\label{HIS}

There is a potential source of bias when deriving calibration factors
involving \lsub\ for galaxies with large ratios of $\fhi= \HI/\mol$,
as the dust may be tracing \HI\ as well as \mol. If we apply our
method from \S\ref{optS} to such H\,{\sc i}-dominated galaxies, we
will infer the presence of more \mol\ due to the dust which resides
only in the \HI\ phase. Because we calibrate in pairs of tracers, this
leads to an over-estimate of \aco\ or \aci\ as well as a bias in the
dust-based calibration factor.

To investigate this, we estimated \fhi\ in the same regions as the
submm flux densities for the local galaxies we could find in the
literature \citep{Dunne2000,Spekkens2004,Wong2013,Groves2015,
  Thuan2016,Dale2017,Koribalski2018,Jiao2021}. As \fhi\ correlates
inversely with \Ms, metallicity and \Lir\
\citep[e.g.][]{Bothwell2014,Saintonge2016}, this issue affects more of
the low \Lir\ galaxies (mostly in the ad sample). For any galaxies
with $\fhi>1$ within the optical disk, we make a correction to \lsub,
removing that portion of the dust emission which is likely associated
with the excess \HI. This correction is designed to produce the same
\lsub/\mol\ ratio as a galaxy with $\fhi=1$.
\begin{equation}
	\lsub^{\rm cor} = \lsub \left(\frac{2}{\fhi+1}\right)   \label{HIE}
\end{equation}

\noindent Galaxies with $\fhi>1$ are shown with this correction
applied as cyan diamonds in the figures. The higher luminosity
(U)LIRGs and SMGs are dominated by molecular gas
\citep[e.g.][]{Yao2003} so we do not need to correct these.

\subsection{Discussion of local \CI\ data}
\label{J19A}

For the {\it Herschel} FTS measurements of local (U)LIRGs
\citep{Lu2017}, we only include local galaxies with $D>27$\,Mpc to avoid
issues with mis-matched beams. We also rejected galaxies where there was
a large discrepancy between the measurement of \citet{Lu2017} and that
of \citet{Kamenetzky2016} (using the same data). 

The set of local galaxies which were mapped by the {\it Herschel} FTS
and presented by J19 are shown in the figures, but not included in the
averages for the following reasons:
\begin{enumerate}
\item{The \CI\ and CO measurements are made in matched apertures,
    however the area mapped in \CI\ is sometimes much smaller than
    that used for the 500--850\,\mic\ flux densities reported in the
    literature. Any analysis which involves both \lci\ and \lsub\
    requires a correction to \lci\ to address the mis-match in
    apertures. We attempted to do this by taking the global CO
    luminosities (which are equivalent global fluxes to the submm
    continuum measurements) and assume that the deficit between the
    global \lcoa\ and that measured in the same aperture as the \CI\
    by J19 is the same as the deficit in \lci:
\begin{equation}
\lci^{\rm cor}  = \lci^{\rm J19} \frac{\lcoa^{\rm global}}{\lcoa^{\rm J19}} \label{CIcorE}
\end{equation}  
These corrections (JC) range from $\rm JC = 0.00$--0.74\,dex, and the
pink diamonds in the figures indicate those galaxies that have
$\rm JC>0.07$\,dex. Even after applying the corrections, the J19
galaxies have different average properties in the \lsub/\lci\ ratio
(see Fig.~\ref{LhistsF}). We therefore, do not have confidence in our
comparison of \lci\ to \lsub\ for these galaxies and so exclude them
from the statistics.}
\item{Although the CO and \CI\ luminosities from J19 are measured in
    the same apertures, there is a trend for these resolved galaxies
    to have lower \lci\ for a given \lcoa\ compared to galaxies which
    have more global flux measurements. There could be a sampling bias
    because \CI\ is only detected over the inner kpc or so of the
    larger galaxies. The CO luminosity per mass of gas (\aco) has been
    found to be lower in the central regions of many galaxies
    \citep{Sandstrom2013}, which would produce a decrease in
    \lci/\lcoa. Since we wish to compare the same averaged global
    fluxes across all galaxies, we remove these `centrally-biased'
    galaxies from our statistical analysis, but we show them in the
    figures for completeness.}
\item{Finally, a more recent paper by \citet{Jiao2021} did produce
    matched dust and \CI\ measurements for a subset of the J19
    galaxies. The results are shown in Fig.~\ref{LcorF}(d) where it
    can be seen that the J21 galaxies are still deficient in \CI\
    compared to the higher luminosity galaxies. This cannot be due to
    a mis-matched aperture but the same sampling bias is present
    toward the inner regions of the resolved galaxies. An offset to
    lower \lci\ per \lsub\ implies either depressed \lci\ (lower \Xci)
    or increased \lsub\ per unit gas mass (lower \gdr, or
    higher dust emissivity).}
\item{Unfortunately, this is the only published set of \CI\ fluxes for
    galaxies with $\log \lci<8$ and the only set of fluxes published
    for the mapping mode of the {\it Herschel} FTS. There is no
    description in the literature of how the processing for this mode
    should be made, and there are differences in the results of J19
    and \citet{Crocker2019}, who analyse some of the same mapping
    data. Despite our best attempts to contact the relevant team, we
    have not been given the details of their flux measurements.  We
    can only note that the \CI\ fluxes from {\it Herschel} FTS mapping
    are not necessarily repeatable when analysed by different teams
    and so elect to exclude the resolved J19 galaxies from the
    statistical analysis. }
\end{enumerate}

\noindent Excluded galaxies are denoted as `\CIcor' and they are
shown as pink diamonds on the relevant figures.

\section{Dust mass opacity and the relationship of dust to gas}
\label{kappaA}

The dust mass opacity coefficient, $\kappa_{\rm d}(\lambda)$, is
proportional to the emissivity per unit mass of dust. It is related to
the calibration parameter we use in our analysis,
$\kh=\gdr/\kappa_{\rm d}$, where \kh\ refers to the dust emission per
H mass, thus encompassing the two unknowns of dust optical properties
and gas-to-dust ratio (\gdr).

The dust optical properties are not easily measured, and can vary
enormously from laboratory-based studies to theoretical dust models
and from those inferred by observations (for a review see
\citealp[e.g][]{Dunne2003,Clark2019}).

\begin{table*}
  \caption{Summary of our physical dust calibrations (\kd, \gdr)
    compared to other work in the literature, where \gdr\ and \kh\
    refer to the mass of hydrogen in all forms, excluding He.}
\begin{adjustbox}{center}
\begin{tabular}{cccc}
\toprule
$\kh = \gdr/\kd$  &  Sample  & Notes & Reference\\
\khunit & & & \\
\midrule
1884 (1500--2200)  & ex-gal  & average of extragalactic estimates & this work\\
\multicolumn{4}{c}{}\\
\multicolumn{4}{c}{\bf Milky Way diffuse and atomic regions}\\
\midrule
$2352\pm198$ & diffuse  & 850\,\mic, H\,{\sc i} very diffuse sight lines & \citet{Planck2014xvii}\\
$1988\pm710$  & all sky   & 850\,\mic, H\,{\sc i}, CO(1--0) with $\aco=3.2$ & \citet{Planck2014xi}\\
$1380\pm251$      & Taurus H{\sc i}  & H\,{\sc i} with 25\% opacity correction, Planck, scaled $\beta=1.8$ & \citet{Planck2011xix}\\
$1518$ &   & 250\,\mic\ scaled to 850\,mic\ with $\beta=1.8$, H\,{\sc i} & \citet{Boulanger1996}\\  
\multicolumn{4}{c}{}\\
\multicolumn{4}{c}{\bf Milky Way molecular/higher density regions}\\
\midrule
$1392$ & $\rm{log(N_H)>20}$ &  850\,\mic, H\,{\sc i}, CO(1--0) with $\aco=3.2$ & \citet{Planck2014xi}\\
$1392$ & $\rm{log(N_H)\sim21}$ & 850\,\mic, H\,{\sc i}, CO(1--0) with $\aco^\gamma$ & \citet{Remy2017}\\
$1012-1044$ & DNM &  Dark neutral medium, 850\,\mic, $\gamma$-rays & \citet{Remy2017,Remy2018}\\ 
$700\pm200^{\dag}$ & local clouds (\mol) & 850\,\mic, CO(1--0), \aco\ from $\gamma$ & \citet{Remy2017}\\
$1210\pm184^{\dag}$ & local clouds (H\,{\sc i}) & 850\,\mic, H\,{\sc i} & \citet{Remy2017} \\
$654\pm85$        & Taurus \mol &  NIR extinction, Planck, scaled $\beta=1.8$ & \citet{Planck2011xix}\\
\multicolumn{4}{c}{}\\
\multicolumn{4}{c}{\bf Local galaxies}\\
\midrule
$1663\pm333$ &  9 & CO(1--0), H\,{\sc i}, 500\,\mic\ dust scaled to 850\,\mic\ with $\beta=1.8$ & \citet{Eales2012}\\
2296 (163/0.071) & 101 Sab--Sbc & CO(1--0) with $\aco=3.2$, H\,{\sc i}, dust SED fits & \citet{Casasola2020}\\
$1692-2169$ & 130 Sa--Sc & CO(1--0), H\,{\sc i}, dust MBB, \aco(Z) & \citet{Bianchi2019}\\
$2096$ &  26 & CO(2--1), H\,{\sc i}, dust \citetalias{Draine2007_kappa} fits & \citet{Sandstrom2013}\\
2402 (92/0.0383) & 189 & CO(1--0), H\,{\sc i}, dust \citetalias{Draine2007_kappa} fits $\aco=3.2$   & \citet{Orellana2017}\\
$1500-2200$ & M74, M83  & $Z$, H\,{\sc i}, CO(2--1), 500\,\mic\ with \citet{James2002} method & \citet{Clark2019}\\           
\multicolumn{4}{c}{}\\
\multicolumn{4}{c}{\bf Physical dust models commonly used in the literature.}\\
\midrule
3232 (109/0.034) & theoretical  & physical dust model producing too 
                                  much $A_v/N_{\rm H}$ & \citet{draine2003}; \citetalias{Draine2007_kappa}  \\
                   &              &                                                  & \citet{Planck2016xxix}\\
$1972$             & theoretical  & up-dated \citetalias{Draine2007_kappa} dust model   & \citet{DH2020}\\
1901 (135/0.071) & theoretical  & physical dust model THEMIS & \citet{Jones2017,Jones2018}\\

\bottomrule
\end{tabular}
\end{adjustbox}
\flushleft{The first column is \kh, the ratio of the
    gas-to-dust ratio (\gdr) and the dust mass opacity coefficient. Where
    there is an explicit assumption for \gdr\ or \kd\ in a reference, we
    include it in parentheses. $^\dag$The clouds in these rows are
    the same; \citeauthor{Remy2017} have calculated the dust opacity
    for each gas phase separately. \aco(Z) from \citet{Amorin2016}.}
\label{kappalitT}
\end{table*}

Commonly adopted extragalactic estimates range from
$\kappa_{850}=0.03$--$0.08\,\rm m^2\,kg^{-1}$
\citep{Li2001,Dunne2000,James2002,draine2003,
  Planck2011xix,Eales2012,Clark2016,Bianchi2019}, though higher values
(by factors of several) are inferred for the very densest and coldest
environments where grains can grow icy mantles and coagulate
\citep{Kohler2015,Remy2017,Ysard2018}. These changes in opacity have
also been correlated with a loss of PAH and stochastically heated
small grains \citep{Flagey2009,Ysard2013}.  \citet{Remy2017} suggest
that regions of the ISM with dust opacities a factor $\sim 2$ higher
than the diffuse ISM (and with cold dust, $\td\sim16$--18\,{\sc k}),
would be those where grains are accreting carbonaceous mantles, as in
the THEMIS dust model \citep{Jones2017,Jones2018}. This carbon
mantle-accreting regime is largely assumed to be the dark neutral
medium (close to the atomic-molecular transition, where there is low
CO emission and high H\,{\sc i} opacity). Deeper within clouds, where
the temperature drops to $\td<16$\,{\sc k}, the dust begins to
aggregate and accrete ice mantles, which increases the opacity
further. These very dense, cold environments do not, however, contain
the bulk of the ISM mass and certainly do not emit a dominant fraction
of \lsub\ in a galaxy \citep{Draine2007,Bianchi2019}. The increase in
dust emissivity (\kd) from atomic to moderately dense molecular
material is in the range 1.2--2.0 \citep{Remy2017}.

In fact, it is \kh\ -- the parameter relating the dust emissivity to
the gas mass -- that can be measured in astrophysical situations,
since we have no absolute knowledge of \gdr. Table~\ref{kappalitT}
lists a comprehensive set of observational and theoretical values for
\kh\ from the literature. Estimates of \kh\ in the Milky Way are made
across a number of sight-lines, from H\,{\sc i}-only (diffuse) to
H$_2$-dominated clouds (dense) where CO emission is used with
assumptions about \aco\ in order to determine $N_{\rm H}$. Independent
confirmation is provided by studies \citep[e.g.][]{Remy2017} using
$\gamma$-ray observations to determine the gas column; the resulting
values of \kh\ are in good agreement (see Table~\ref{kappalitT}), with
\kh\ being higher along diffuse sight-lines (1800--2400), dropping to
700--1500 in denser molecular or dark neutral media.

In extragalactic studies, a similar method is used, although with
larger uncertainties as it is less straightforward to decompose the
atomic and molecular components along the line of sight. These studies
find a range of $\kh=1500$--2200\,\khunit, closer to the diffuse ISM
measurements in the Milky Way.

For a given dust model, we can also calculate the theoretical \kh\
given the assumed dust optical properties, chemical abundances and
depletions. The theoretical values are also listed in
Table~\ref{kappalitT} where the current consensus is for
$\kh \sim 1900$--2000\,\khunit. The popular
\citet{draine2003} model has a significantly
higher $\kh=3200$\,\khunit\ (lower $\kd=0.034$\,\kunit\ for
$\gdr=109$) than all of the empirical measurements. This was noted by
\citet{Draine2014} and \citet{Planck2016xxix} and has been updated in
the more recent version of this model by \citet{Hensley2021}. We
encourage readers to use the updated version in order to produce
dust-based measurements which are consistent with what we know about
dust from observations.

\section{Deriving gas mass from observations of \CIfull}

\label{QA}
\begin{figure*}
	\includegraphics[width=0.9\textwidth,trim=0cm 0cm 0cm 0cm, clip=true]{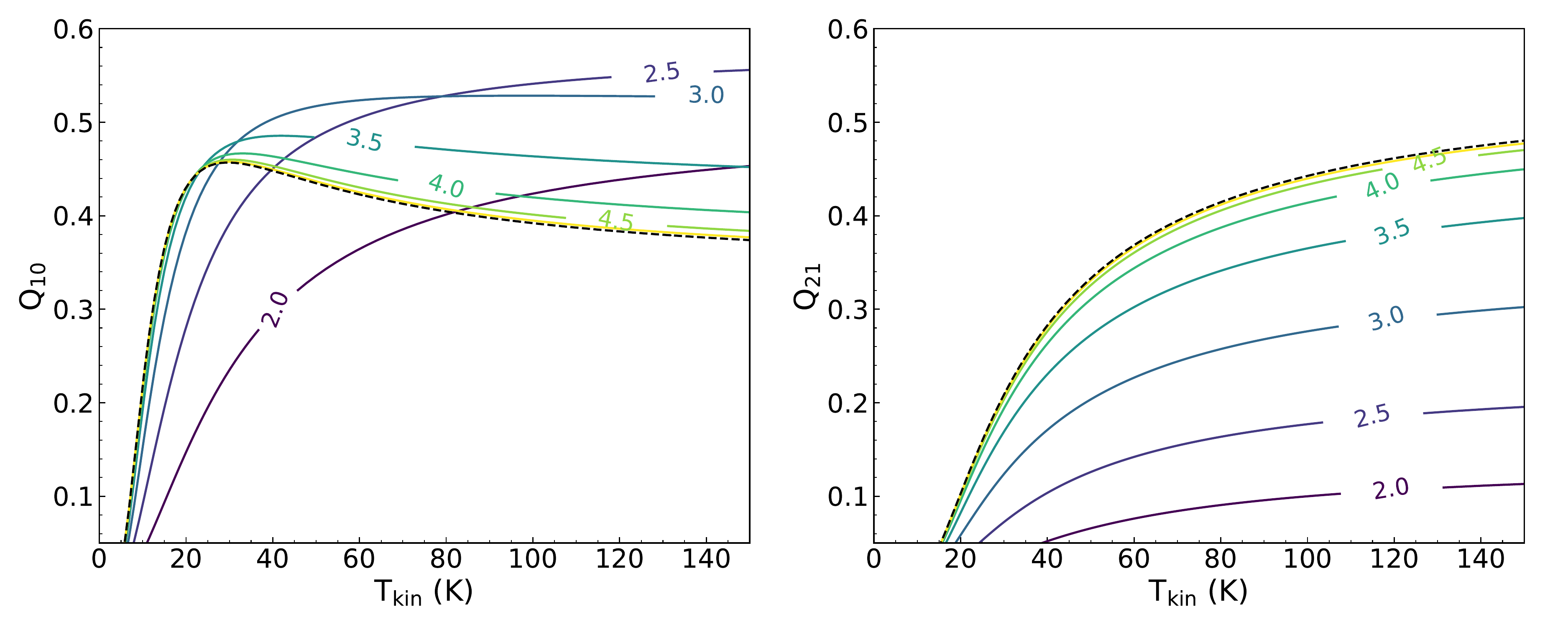}
	\caption{$Q_{10}$ (left) and $Q_{21}$ (right) as function of
          gas temperature for a range of densities, where lines are
          labelled with the values of log $n$. The LTE value is shown
          as the black dash-dot line.}
              \label{QulF}
\end{figure*}

The excitation term, $Q_{\rm ul}$, which describes the fraction of C
atoms in each excited state, is a function of ($n$,\tk) in non-LTE
conditions, and is derived analytically in the Appendix to
\citet{PPP2004}. A recent study of the [\CI](2--1)/(1--0) line ratio
found that [\CI](2--1) is strongly sub-thermally excited, and [\CI](1--0)
presents interesting super-thermal behaviour in the range of density
and temperature expected for galaxies. We illustrate the dependence of $Q_{\rm ul}$ on ($n$,\tk) in
Fig.~\ref{QulF}. As discussed by \citet{PPPDunne2022}, the value of
\Q\ for lower densities ($n=300$--3000\,\cc) can exceed the LTE value
at \tk>20\,{\sc k}, but Fig.~\ref{QulF} shows that for a reasonable
range of $n$ and \tk\ ($300<n<10,000$\,\cc, $25<\tk<80$\,{\sc k}) \Q\
does not go outside the range 0.35--0.53. In fact, for a uniform
probability of ($2.4< \log n <4.0$) and ($25<\tk<80$\,{\sc k}) the 99
per cent range for \Q\ is 0.40--0.54, median=0.48. The relative
uncertainty on the calibration of \CI\ mass from the lack of knowledge
of ($n$,\tk) is thus $<\pm 16$ per cent. We will therefore use the
median value of $\Q=0.48$ throughout, because even though we may be
able to use the measured \td\ to infer the galaxies with higher or
lower \tk\ (assuming $\tk=\alpha^{\rm TD} \, \td$ -- see
\citealp{PPPDunne2022}), the lack of knowledge of the density and the
super-thermal behaviour in the $J=1$ state means that there is no
direct correlation between \Q\ and \tk. Using sensible average
parameters for MS galaxies [and SMGs], so $n$=500 [5,000]\,\cc and 
\tk\ = 40 [80]\,{\sc K} we find only a small ($\sim$ 10 \,per\,cent) difference in the \Q\ values expected.

The LTE expressions for \Q\ and \tx\ should not be used \citep{PPPDunne2022} as $\rm Q_{10}^{LTE}$ is actually {\em lower} than the non-LTE \Q\ for densities higher than a few hundred \cc, and thus its use would lead to a systematic bias - e.g. for \tk=60\,{\sc K} and n=1000\cc, the LTE value for \Q\ is 18 \,per\,cent lower than the appropriate non-LTE value. This would lead to an 18 \,per\,cent over-estimate of the \mol\ mass using \CI.

For the [\CI](2--1) line, things are not so promising
(Fig.~\ref{QulF}(right)). The range of possible values of $Q_{21}$ are
large, ranging from 0.07--0.37 for the 99 per cent range. The median
is $Q_{21}=0.22$, giving an uncertainty range of $\pm 68$ per cent for
reasonable values of ($n$,\tk). Because of the sub-thermal behaviour,
the [\CI](2--1) line is a sensitive indicator of density
\citep{PPPDunne2022} and galaxies with strong [\CI](2--1) emission will
have a larger fraction of their \mol\ in a dense state.

\section{From pairwise variances  to individual variances}
\label{pairwiseS}

We have measurements of different tracers of gas mass for several
galaxies, but no direct measurements of \Mh\ itself.  Hence, it is not
possible to measure directly how well each tracer follows the gas
mass. However, we do have measurements of the different tracers for
each galaxy, so we can estimate the scatter in the difference between
the tracers. Under some assumptions this allows us to infer the
scatter between each tracer and the gas mass.

To simplify the notation, we write the log of observed quantities
and corresponding standard deviation of errors as 
\begin{equation}
  \label{eqn:x_errs}
	\begin{aligned}
		x_1& = \rm \log(L_{850}),  &  \quad& \sigma_1 = \rm \log(1+\sigma_{850}/L_{850}), \\       
		x_2 &= \rm \log(L'_{\rm CO}),  &   &\sigma_2 = \rm
                \log(1+\sigma_{\rm CO}/L'_{CO}),\\ 
		x_3 &= \rm \log(L'_{\rm C\,I}),  &   & \sigma_3 = \rm
                \log(1+\sigma_{\rm C\,I}/L'_{\rm C\,I}).   
	\end{aligned}
\end{equation}
If true value of the log of the gas mass is
$ \hat m = \log(\Mh) $, and the true values of the observed
quantities are $\hat x_i$, where $i=1 ... 3$, then we can write
\begin{equation}
	\hat m  =  \hat x_i + \hat a_i,
	\label{eqn:cal-const}
\end{equation}
where $\hat a_i $ are the  true calibration factors for each galaxy,
\begin{equation}
	\begin{aligned}
		\hat a_1 = &- \rm \log(\hat \alpha_{850}) \\
		\hat a_2 = &\quad \rm \log(\hat \alpha_{CO}) \\
		\hat a_3 = &\quad \rm \log(\hat \alpha_{C\,I})\\
	\end{aligned}
\end{equation}
Note that the true values $\hat a_i$ may be different for each galaxy,
depending on the individual physical conditions within the galaxies.

If we choose a particular set calibration factors for all galaxies,
say $\tilde{a}_i$, this provides three estimates of the gas mass for
each galaxy,
\begin{equation}
	m_i  =  x_i + \tilde{a}_i 
\end{equation}
The error in each mass estimate is 
\begin{equation}
	\begin{aligned}
		m_i -\hat{m} & =  x_i - \hat{x_i}  + \tilde{a}_i - \hat{a}_i  \\
		& = \delta x_i + \delta a_i
	\end{aligned} 
\end{equation} where $\delta a_i$ is the difference between the true
factor for this galaxy and the value we have chosen, and $\delta x_i$
are the measurement errors of the observations. 
If we assume that the errors on  $x_i$ are not correlated 
with the errors on $a_i$, then the variance of the mass errors is given by:
\begin{equation}
	\var(m_i -\hat{m}) =  \sigma_i^2 + s_i^2  
\end{equation} where $s_i^2$ is the variance of the
true calibration factors.

The value of $s_i^2$ gives a direct measure of how accurate
the particular tracer is when using a universal calibration factor
for all galaxies. Without knowing the true gas mass, we do not have a direct
measure of this value, but we can obtain an estimate by considering the
differences between the mass measurements:
\begin{equation}
	\begin{aligned}
		m_i - m_j &=  x_i - x_j  + \tilde{a}_i - \tilde{a}_j \\
		&= \delta x_i -\delta x_j + \delta a_i -\delta a_j 
	\end{aligned} 
\end{equation}
If we ignore all co-variance terms, the variance of the differences is
given by:
\begin{equation}
	\label{pair_var}
	v_{ij} = \var(m_i - m_j) =  \sigma_i^2 + \sigma_j^2 + s_i^2 + s_j^2
\end{equation}
It is straightforward to re-arrange these equations to find the
intrinsic variance of the calibration factors as: 
\begin{equation}
	s_0^2  =  \left(v_{01} - v_{12} + v_{20} \right)/2 - \sigma_0^2
\end{equation}
with similar equations for $s_1^2$, and $s_2^2$. So long as we have
good estimates of the measurement errors, $\sigma_i$, for the observed
quantities, we can estimate the scatter in calibration constants for
each tracer. Using our dataset we have measured the variance for each
pair of factors in Eqn.~\ref{pair_var}. Assuming that the co-variance
between the calibration factors is zero, we use the three pair
variances to estimate the intrinsic variance of the three individual
calibration factors. The resulting standard deviations are
$s_{\kappa} = 0.1294$, $s_{\alpha} = 0.1436$ and
$s_{\mathrm{X}} = 0.1125$, using all galaxies except the
\CIcor\footnote{When restricting the analysis to \Lhi\ galaxies,
  \CI\ produces notably less scatter than both CO and dust continuum, with
  $s_{\kappa} = 0.1339$, $s_{\alpha} = 0.1646$ and
  $s_{\mathrm{X}} = 0.082$.}. Values are listed in
Table~\ref{methodT}.

This analysis shows that \Xci\ has the smallest scatter between
galaxies, especially when considering \Lhi\ galaxies, which is a new
result, independent of any assumptions.

\section{A Bayesian approach to combining gas mass estimates}
\label{bayesS} 

Our method of combining the three gas mass tracers is based on the
idea that the conversion factors for any particular galaxy come
from parent distributions with variances as derived in
Appendix~\ref{pairwiseS}. This means that we should allow for the
expected scatter in conversion factors as well as the observational
error when combining estimates from the different tracers. Using a
Bayesian approach to the problem, we show the most likely mass
estimate is simply the inverse variance weighted mean of the tracers,
where the weights include both measurement error and the variance in
conversion factors.

We continue to use the the notation as in Appendix~\ref{pairwiseS}, where the
observed quantities are $x_i$ and errors $\sigma_i$.
Assuming the measurement errors are Gaussian the probability of
measuring the observed value of $x_i$ is
\begin{equation}
	P(x_i| \hat x_i, \sigma_i) = {\cal N}(x_i | \hat x_i, \sigma_i^2) 
\end{equation}
where $\cal N$ represents the normal distribution centred on $\hat x_i$
and with variance $\sigma_i^2$. 
Now, for each observation we can use Bayes theorem to estimate the
posterior probability that the gas mass is $m$, 
\begin{equation}
	P(m,\hat a_i | x_i) = P(x_i| m, \hat a_i) P(\hat a_i)P(m) / P(x_i) 
\end{equation}
where we have assumed $m$ and $\hat a_i$ are independent. For the
prior on $\hat a_i$, we assume a normal distribution with mean
$\bar a_i$ and variance $s_i^2$, as discussed in
Appendix~\ref{pairwiseS} . We assume a flat prior on $m$, implying
that $P(m)$ is constant. Since $P(x_i)$ is also constant, the position
of the maximum posterior probability does not depend on the actual
value of $P(m)/P(x_i)$, and for convenience we set this to
1. Therefore:
\begin{equation}
	\begin{split}
		P(m,\hat a_i | x_i)  &\propto P(x_i| m , \hat a_i) P(\hat a_i)\\
		&= {\cal N}(x_i|  m - \hat a_i, \sigma_i^2)
		{\cal N}(\hat a_i | \bar a_i,
		s_i^2)\\
		&= {\cal N}(\hat a_i|  m - x_i, \sigma_i^2)
		{\cal N}(\hat a_i | \bar a_i,
		s_i^2) .
	\end{split}
\end{equation}
Here we have used Equation~\ref{eqn:cal-const} to go from $m-\hat a_i$
to $m-x_i$.  Since we are interested primarily in the value of the gas
mass, and not explicitly in the values of the calibration factors, we
can marginalise over the values of $\hat a_i$. Ignoring the
uncertainties on the variances, $\sigma_i^2$ and $s_i^2$, leads to:
\begin{equation}
	P(m | x_i)   = {\cal N}(m | x_i + \bar a_i, \sigma_i^2+s_i^2) .
\end{equation}

Including all three observations for the galaxy this becomes
\begin{equation}\begin{split}
		P(m | \{x_i\})  &= \prod_{i=1}^{3} {\cal N}(m | x_i + \bar a_i, \sigma_i^2+s_i^2) \\
		&\propto 
		\exp\left( -\sum_{i=1}^3 \frac{(m-x_i-\bar
                    a_i)^2}{2(\sigma_i^2+s_i^2)}\right) .
	\end{split}
\end{equation}
So maximising the posterior probability with respect to $m$ is
equivalent to minimising $\chi^2$, where:
\begin{equation}
  \chi^2 = \sum_{i=1}^{3}  \frac{(m-x_i-\bar a_i)^2}{ 2( \sigma_i^2+s_i^2)} .
\end{equation}
The minimum with respect to $m$ is given by
\begin{equation}
  \begin{split}
    m^{\rm opt} &= \left(\sum_{i=1}^3
          \frac{x_i+\bar{a_i}}{\sigma_i^2+s_i^2} \right) \Bigg/
        \left(\sum_{i=1}^3 \frac{1}{\sigma_i^2+s_i^2} \right) \\
        &= \left(\sum_{i=1}^3 (x_i+\bar{a_i})w_i \right) \Bigg/
          \left(\sum_{i=1}^3 w_i \right),
  \end{split}
\end{equation}
\noindent where $w_i = 1/(\sigma_i^2+s_i^2)$. So the optimal mass estimate is simply the inverse variance-weighted
mean of the three estimates, where each uses the mean conversion
factor, and where the variance for each measure is the sum of the
measurement error and the expected variance of the conversion
factor.

The uncertainty on $m^{\rm opt}$ is  the uncertainty on the weighted mean,
\begin{equation}
  \label{eqn:var_m}
  \sigma_{m^{\rm opt}} = 1 \Bigg/
  \left(\sum_{i=1}^3 w_i \right).
\end{equation}

The corresponding estimates of the conversion factors for a
particular galaxy are then simply given by: 

\begin{equation}
	a_i = m-x_i, \quad i=1...3  
\end{equation}

The uncertainty on the factor $a_i$ depends on the uncertainty on
$m^{\rm opt}$, from equation \ref{eqn:var_m}, and the uncertainty on the
measurement $x_i$, from equation \ref{eqn:x_errs}. Since the
estimate of $m$ depends on the measurements $x_i$, there is a non-zero
covariance between $m$ and $x_i$. Allowing for this covariance, the expected
uncertainty on $a_i$ is given by:
\begin{equation}
  \label{eqn:a_i_errs}
  \sigma_{a_i}^2  =  \sigma_{\rm m^{opt}}^2 + \sigma_i^2 \left( 1-
    \frac{2 w_i}{\sum_{i=1}^3 w_i } \right).
 \end{equation}

\section{Sensitivity of tracer to SFR and radiation field intensity}
\label{tdcorrA}

\begin{figure*}
	\includegraphics[width=0.45\textwidth,trim=0cm 0cm 0cm 0cm, clip=true]{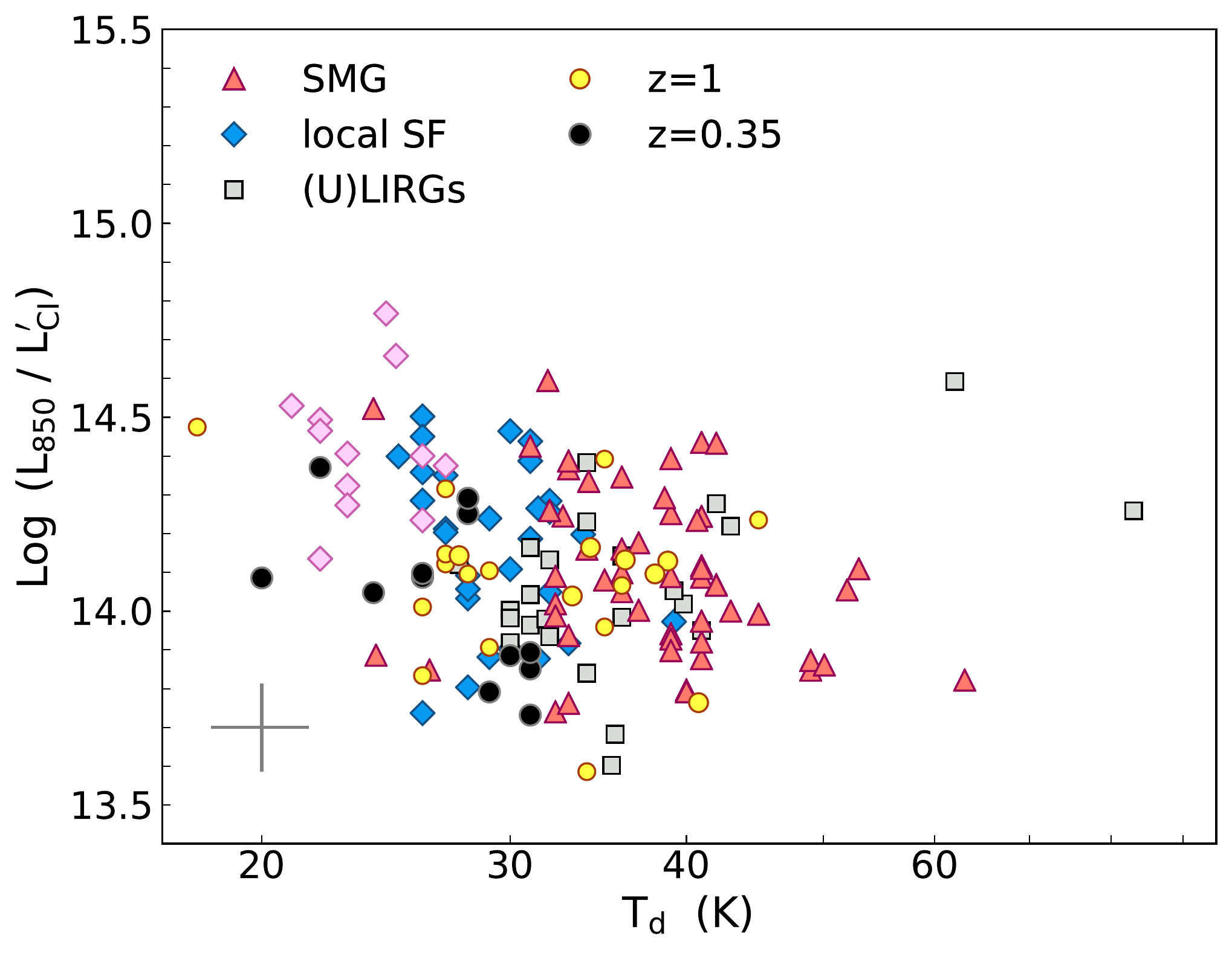}
	\includegraphics[width=0.45\textwidth,trim=0cm 0cm 0cm 0cm, clip=true]{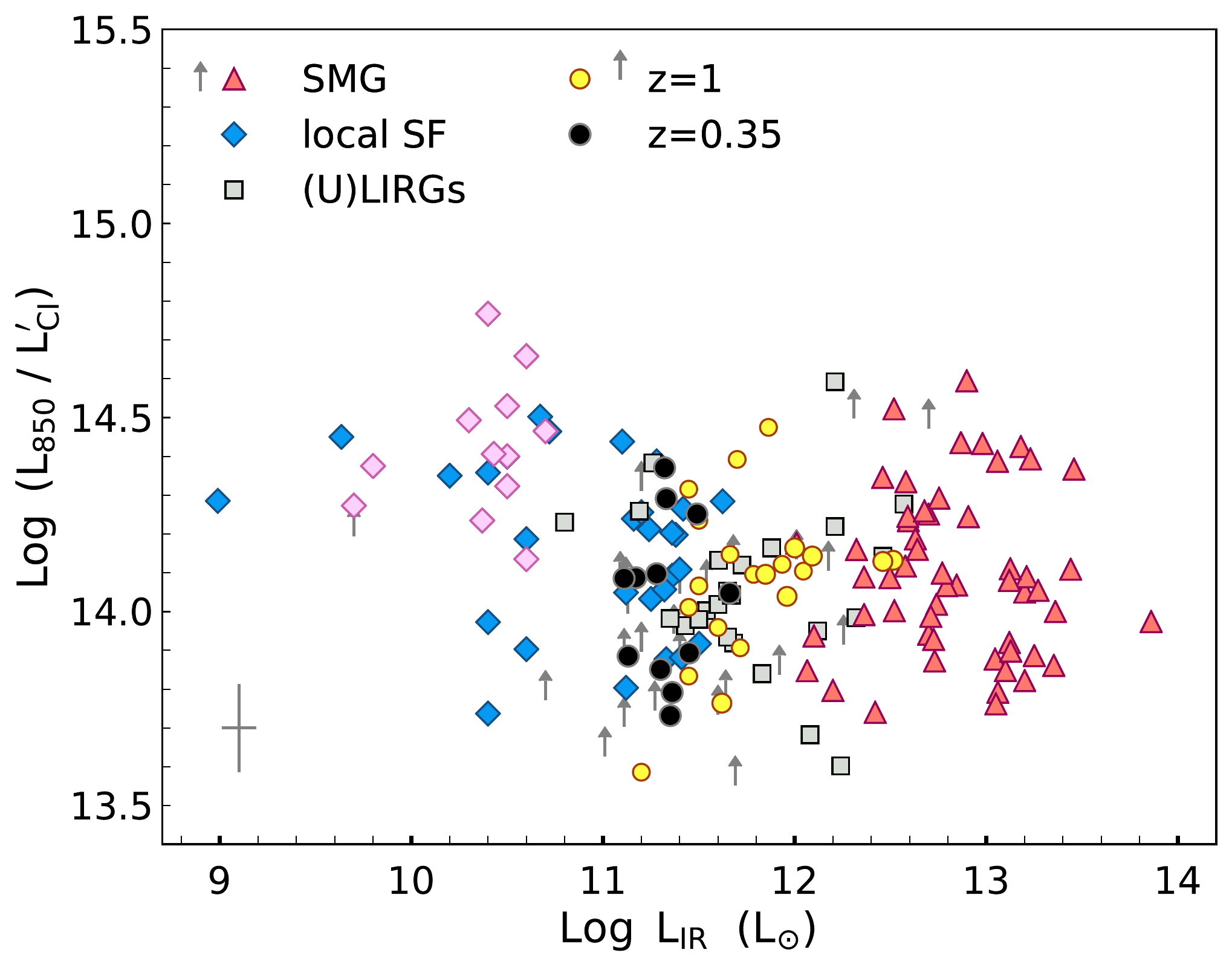}\\
	\includegraphics[width=0.45\textwidth,trim=0cm 0cm 0cm 0cm, clip=true]{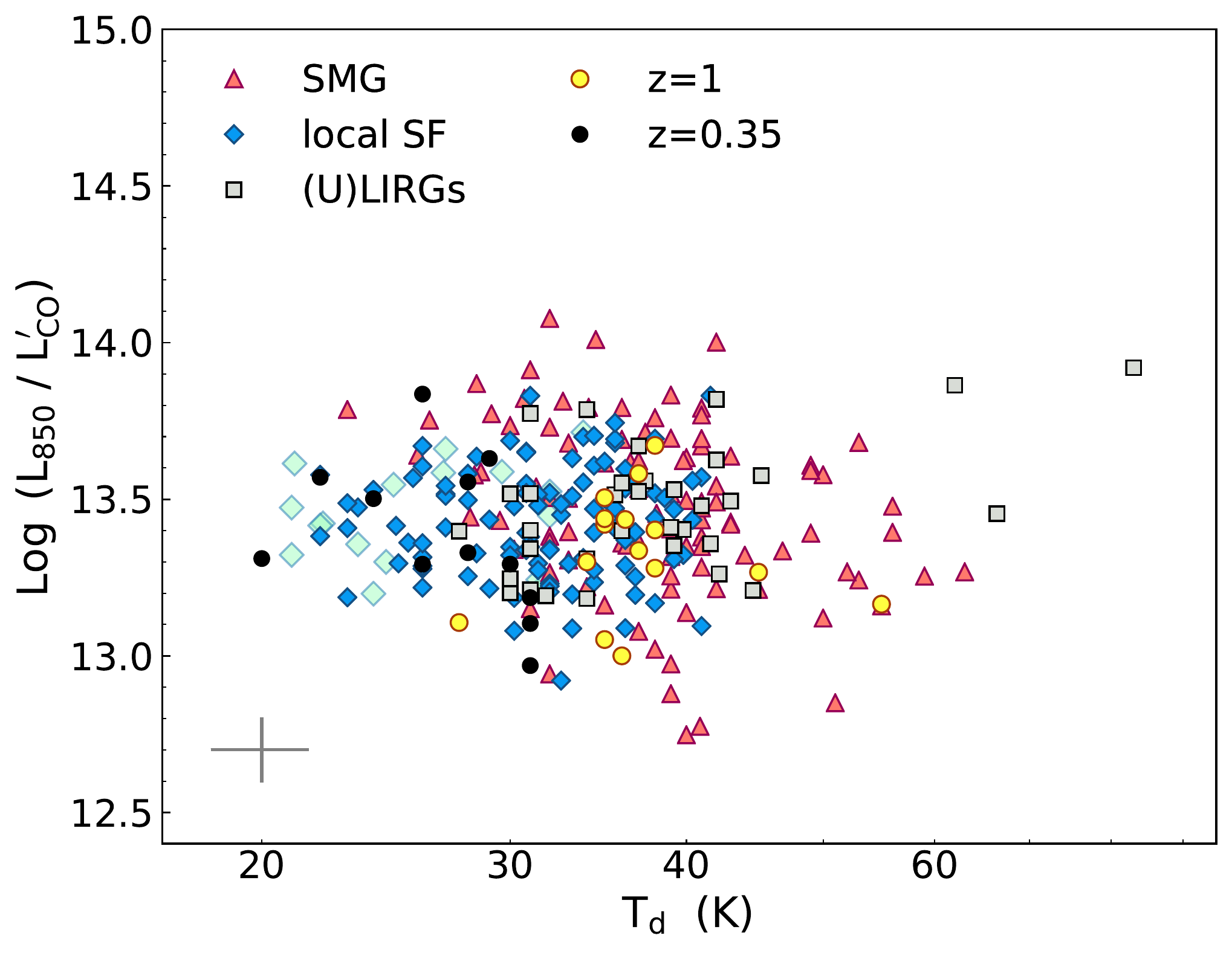}
	\includegraphics[width=0.45\textwidth,trim=0cm 0cm 0cm 0cm, clip=true]{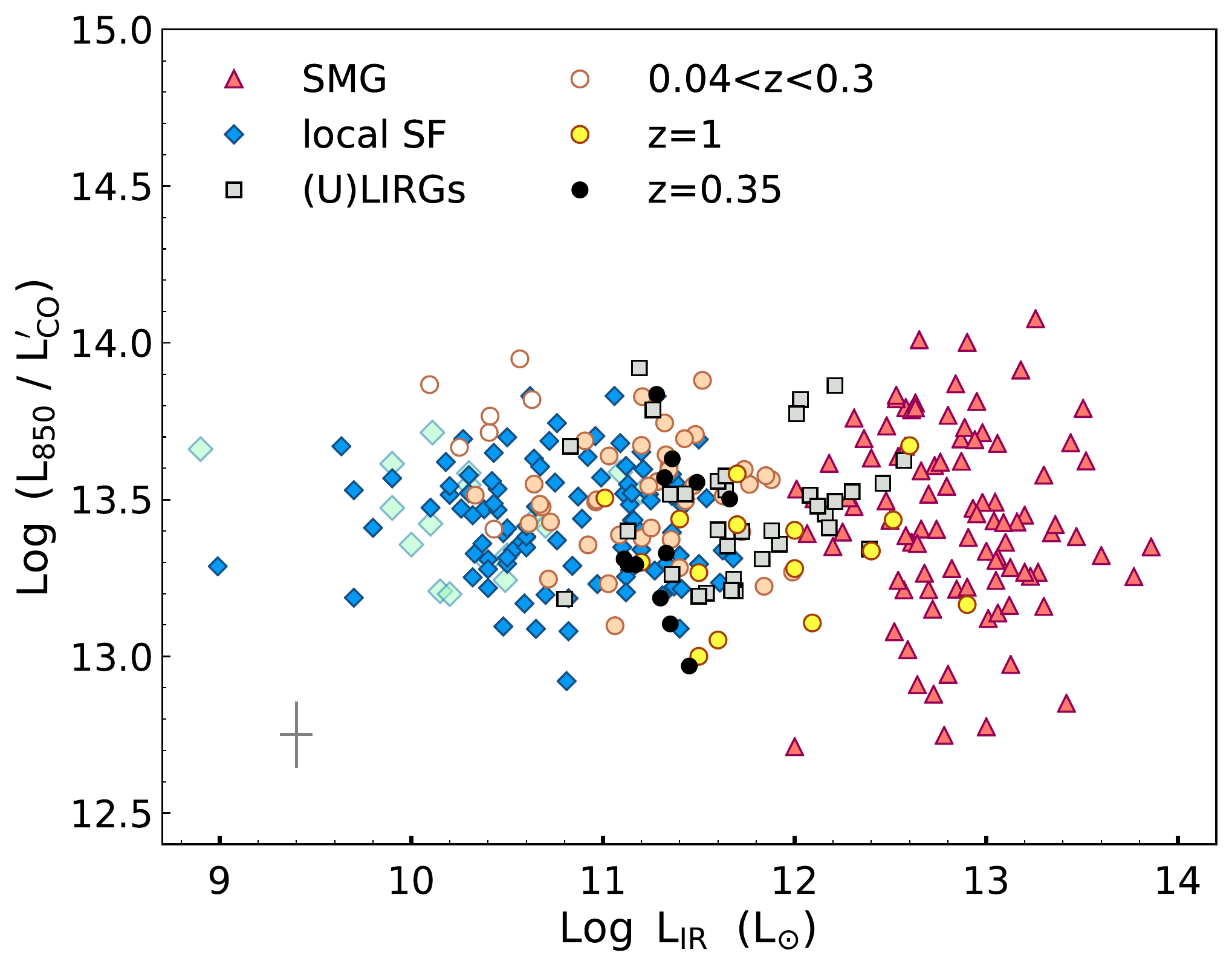}\\
	\caption{Luminosity ratios as a function of
          luminosity-weighted (peak SED) dust temperature ({\bf left})
          and \Lir ({\bf right}). In the top row, the resolved local
          galaxies from J19 which require aperture correction are
          shown as pink diamonds. In the bottom row, the galaxies with
          $\fhi>1$ are shown as cyan diamonds, after correction
          following Eqn.~\ref{HIE}. The \lci/\lcoa\ ratio has a
          significant correlation with \td\ and with \Lir, and is
          shown in Fig.~\ref{tdcorrF} in the main text.}
\label{tdcorrAF}
\end{figure*}

Fig.~\ref{tdcorrAF} shows the observable ratios, \lsub/\lci\ and
\lsub/\lcoa, as a function of \td\ (left) and \Lir\ (right). There is
no significant trend for \lsub/\lci\ or \lsub/\lcoa\ with either \td\
or \Lir. There is a noticeable offset to higher \lsub/\lci for the
\CIcor\ galaxies (pink diamonds), which also have lower \td\ and \Lir\
than the other samples. As these galaxies require large corrections to
\lci\ in order to compare to \lsub, we cannot be sure if this is a
real effect, or just an under-estimate of the required correction. A
larger sample of low-temperature, low-luminosity galaxies with matched
apertures will be required to investigate this.

\section{Tests of robustness}
\label{testsA}

\begin{figure*}
\includegraphics[width=0.49\textwidth,trim=0.0cm 0cm 0cm 0cm, clip=true]{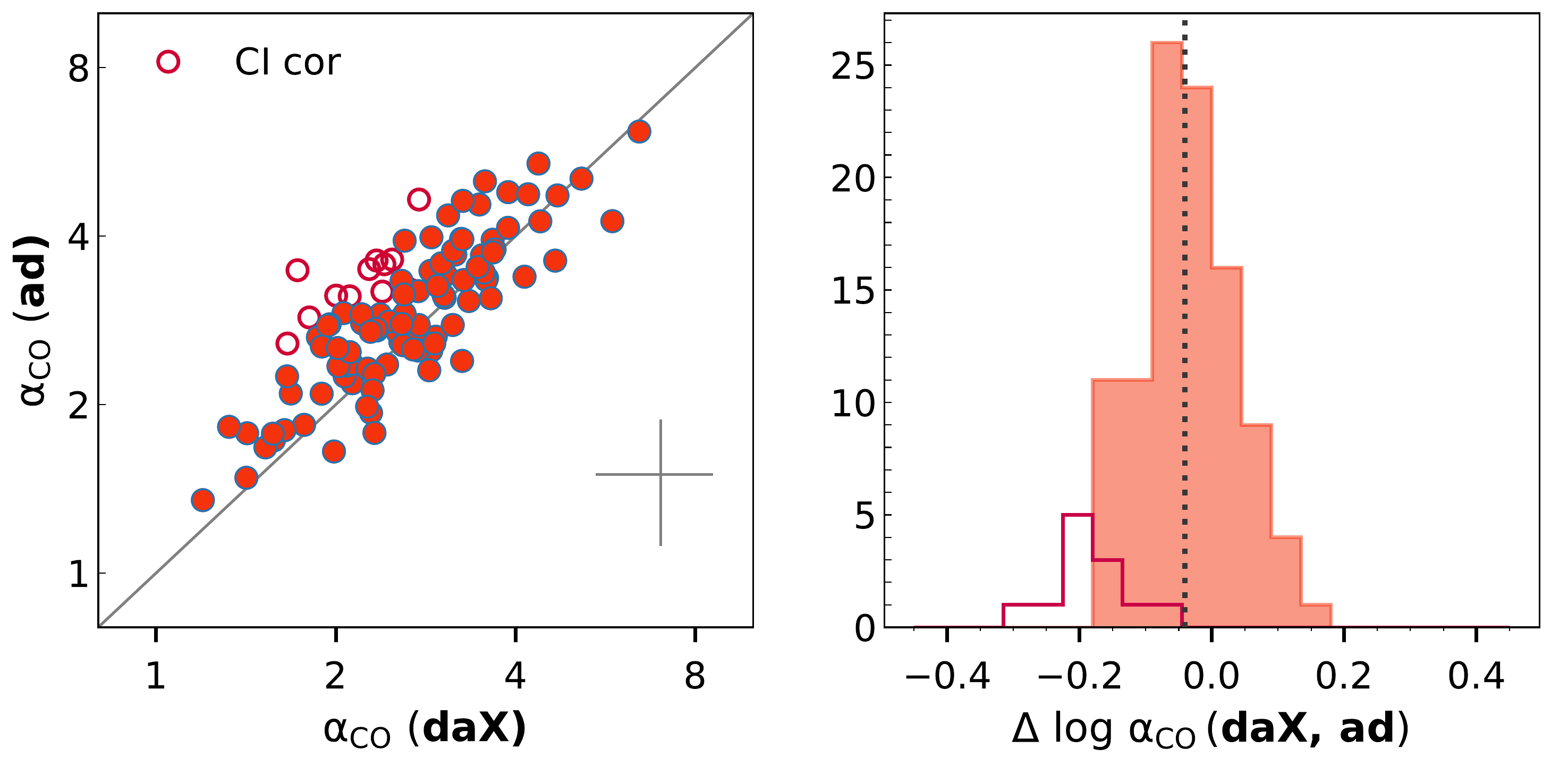}
\includegraphics[width=0.49\textwidth,trim=0.0cm 0cm 0cm 0cm, clip=true]{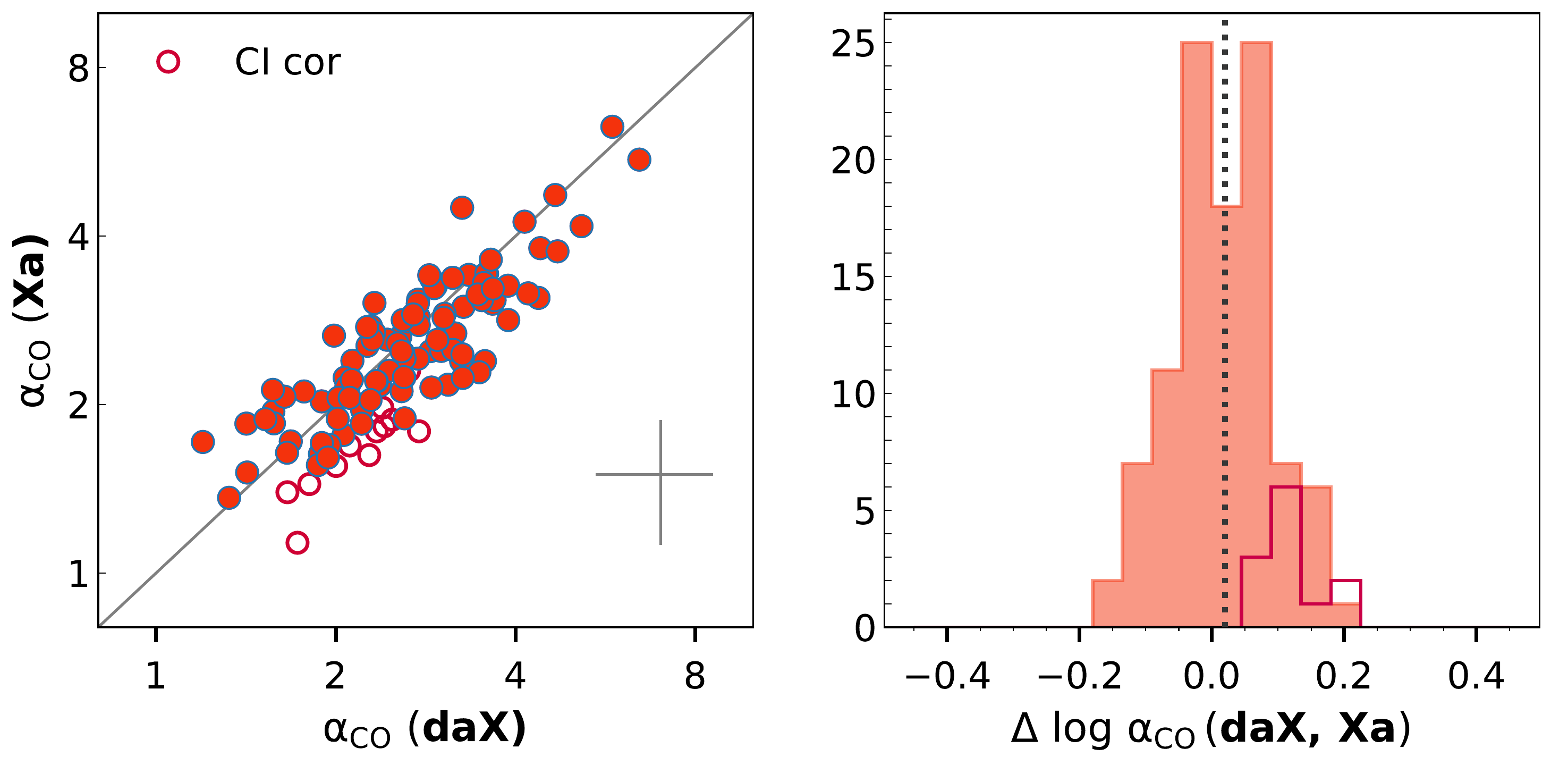}\\
\includegraphics[width=0.49\textwidth,trim=0.0cm 0cm 0cm 0cm, clip=true]{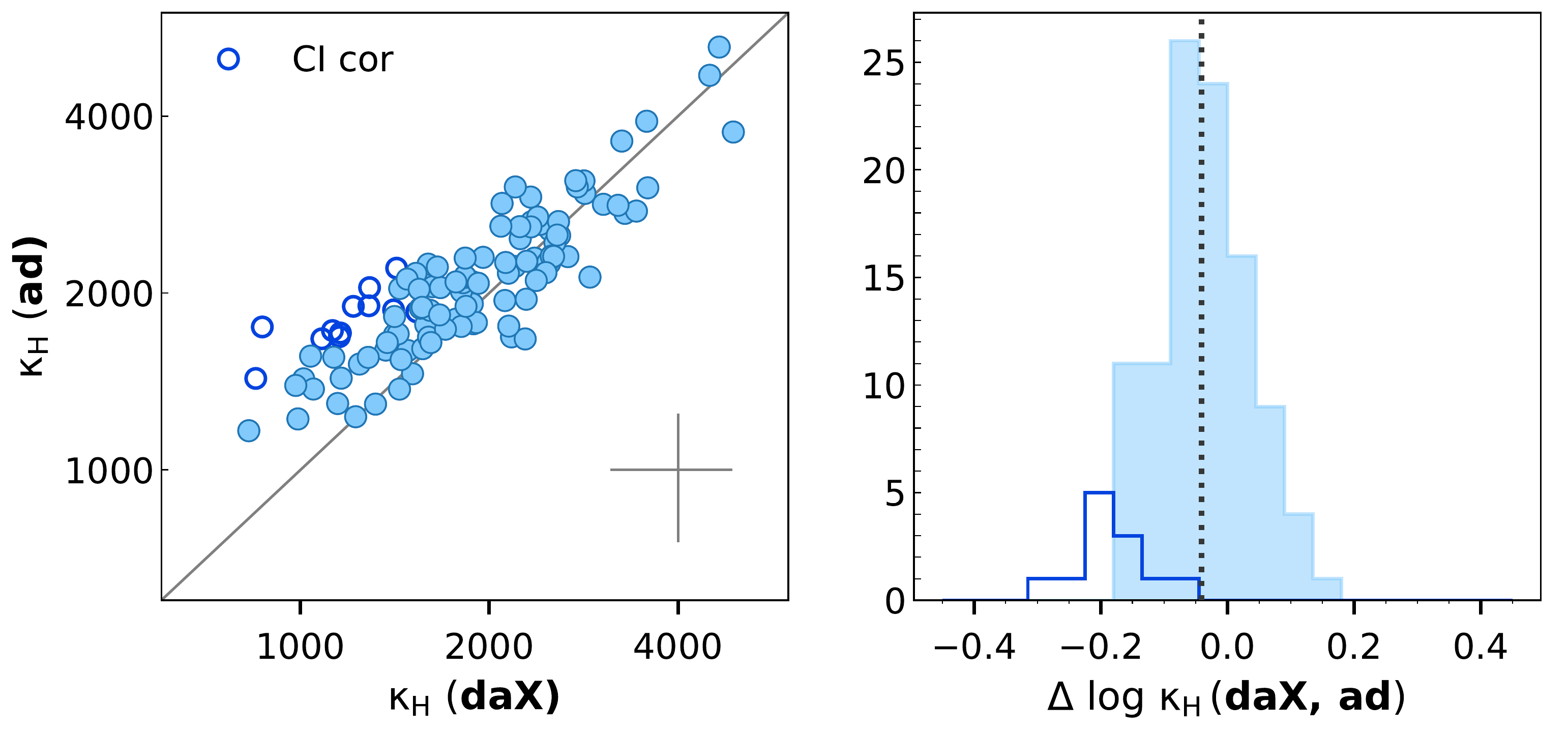}
\includegraphics[width=0.49\textwidth,trim=0.0cm 0cm 0cm 0cm, clip=true]{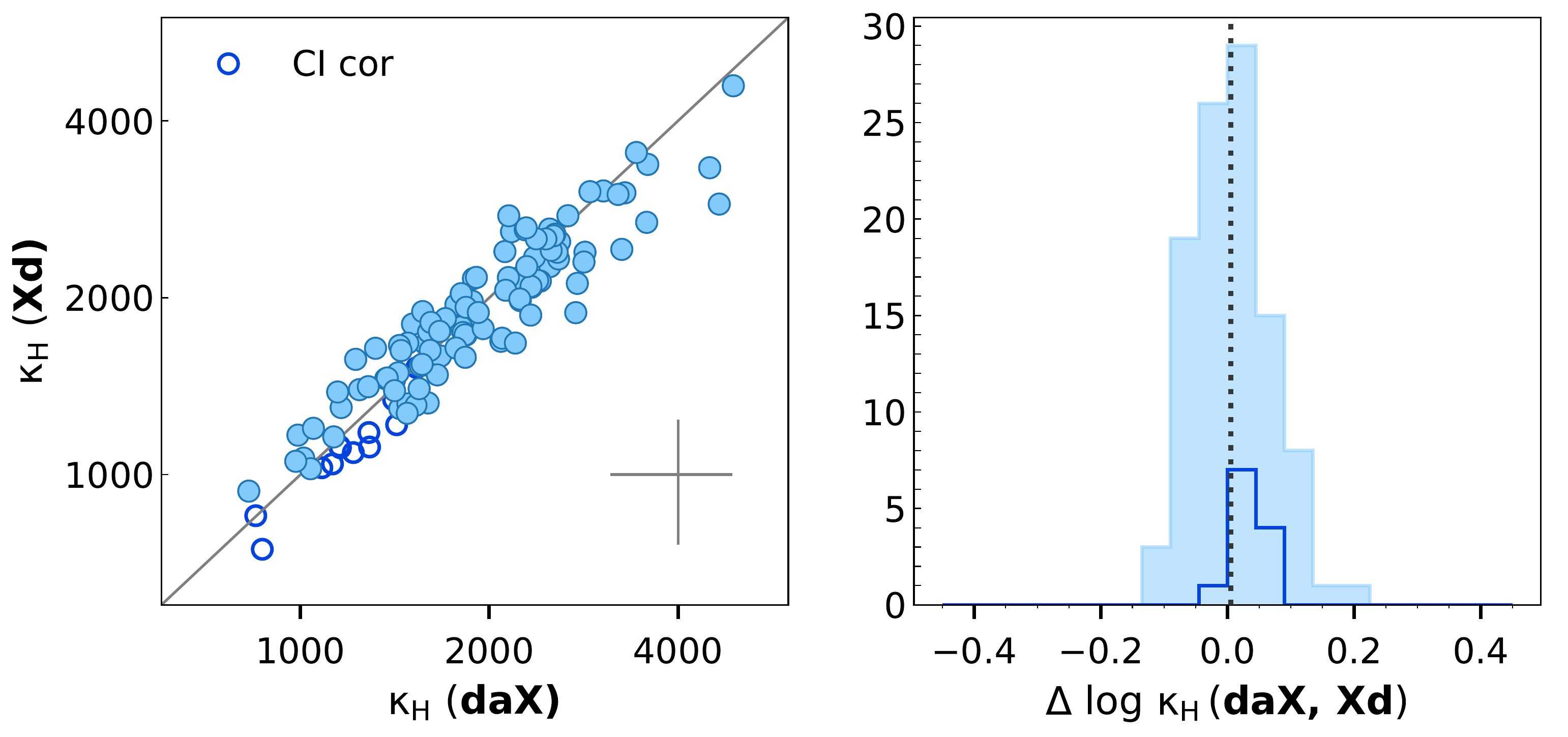}\\
\includegraphics[width=0.49\textwidth,trim=0.0cm 0cm 0cm 0cm, clip=true]{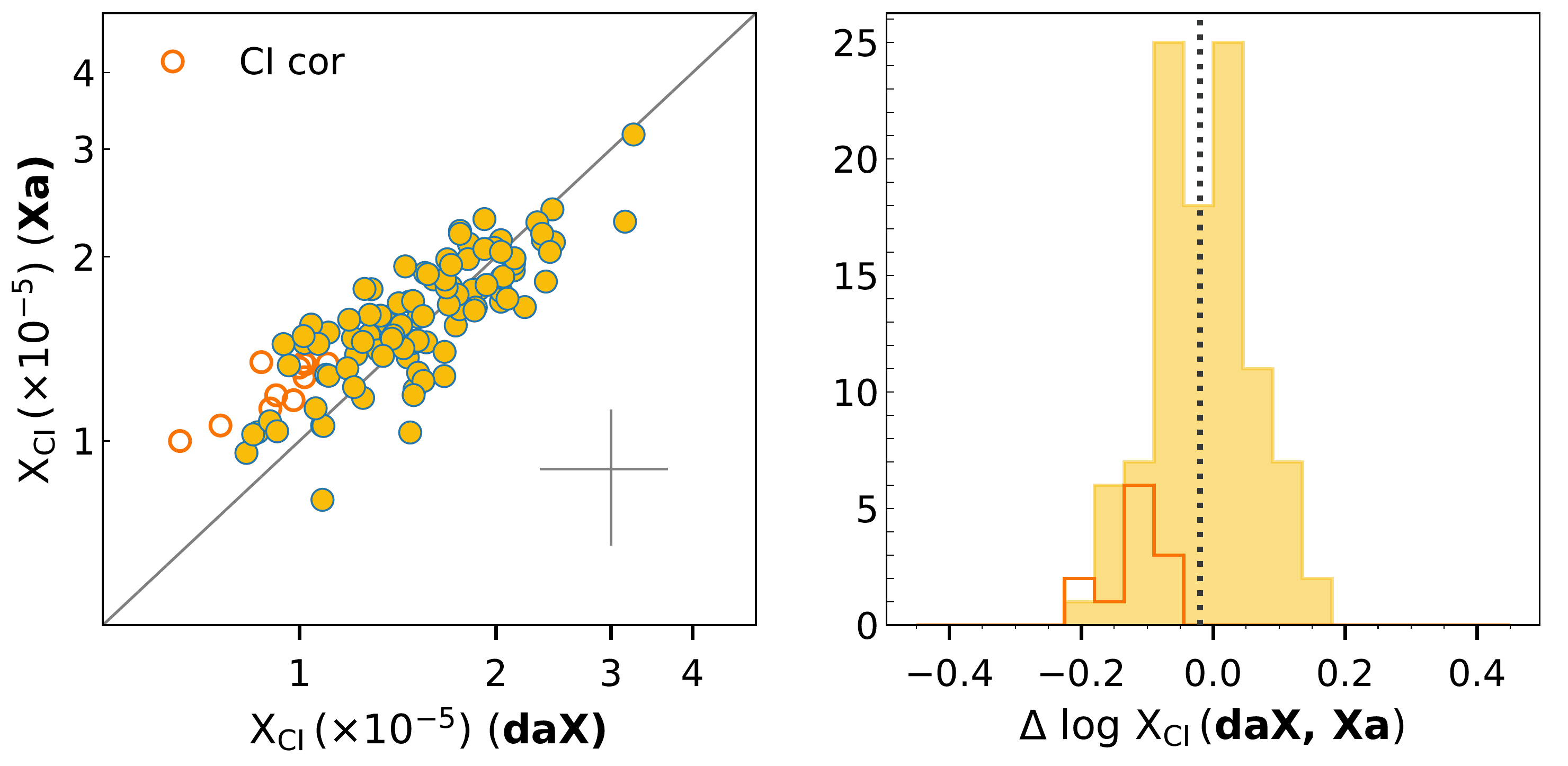}
\includegraphics[width=0.49\textwidth,trim=0.0cm 0cm 0cm 0cm, clip=true]{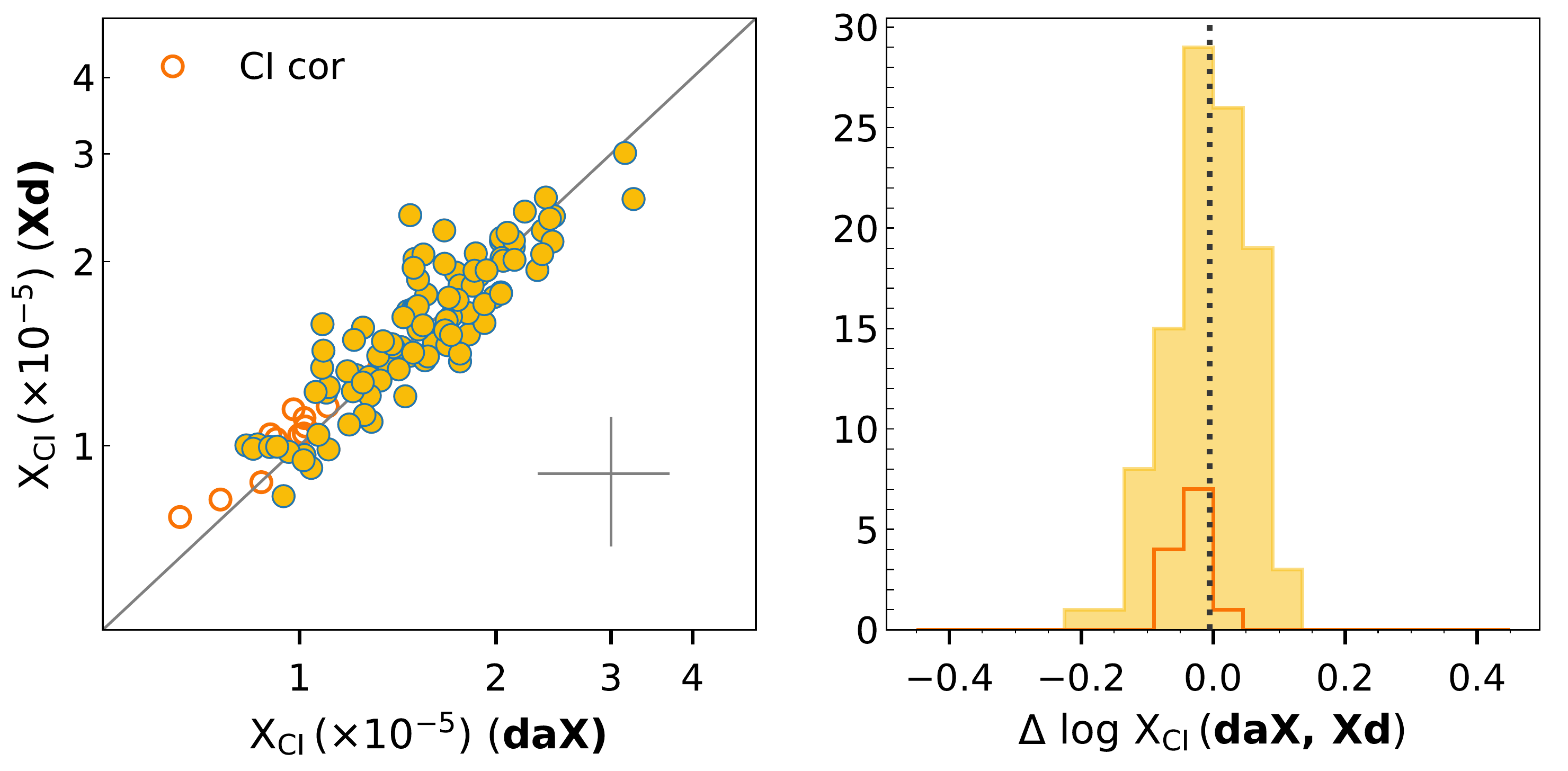}\\
\caption{Comparison of optimised conversion factors derived using
  three pairs of tracers, daX ($x$ axes of scatter plots) compared to
  using just one pair. The histograms show the offsets the parameters
  ($P$) $\Delta P = \log P_{\rm daX} - \log P_{S}$. The filled
  histogram represents galaxies free from systematic uncertainties in
  \CI\ fluxes for the J19 sample (the affected \CIcor\ galaxies are
  shown as the line histogram). The median offsets are shown as the
  grey dotted lines on the histograms (calculated excluding the
  \CIcor\ galaxies) and are very small (0.02, 0.006, 0.04\,dex for Xa,
  Xd, ad respectively). The offset histograms have a maximum of
  0.2\,dex with a r.m.s.\ of 0.06--0.08\,dex. The scatter plots show
  the robust galaxies as filled circles, with \CIcor\ as open
  circles. The larger median offset for the daX--ad comparison is
  apparent as a non-linear trend in the \kh\ values: for
  $\kh({\rm daX})<2000$ there is a persistent trend for the ad pair to
  return a higher \kh\ than the daX three-pair method. The offset is
  also present in the \aco\ parameter (top-left panels) but looks more
  like a constant offset in the calibration rather than a
  non-linearity. The \CIcor\ galaxies (line histogram, open circles)
  are biased in the sense that the daX and one-pair methods produce
  very different values for the same parameters. The difference is
  smallest for the Xd sample because both daX and Xd are affected by
  the same corrections to the \CI\ fluxes. The difference is more
  pronounced when comparing daX to ad or Xa, the reason being that the
  uncertain correction only affects the daX results (because the
  correction is between \CI\ and dust continuum) while the ad and Xa
  samples do not have the C\,{\sc i}--dust continuum pair.}
\label{acocompF}
\end{figure*}

\subsection{Consistency of parameter estimates}

We investigated the consistency of our parameter estimates for the
same galaxies when three tracers are used compared to only
two. Fig.~\ref{acocompF} shows that there is a reasonable correlation
between the three-tracer and two-tracer estimates, with only small
differences in the sample medians when different numbers of tracers
are used. The Xd pair produces the closest match to the method with
three pairs (Fig.~\ref{acocompF} centre and lower-right panels), with
no bias and a small scatter. If restricted to choosing only one pair
to observe, the best choice seems to be \lsub\ and \lci.

\subsection{Impact of using fixed vs. variable \mwtd}
\label{QTS}

In this section we test a different approach to \mwtd, one of the main
physical dependencies that impacts on the calibration of gas
masses\footnote{This is not to suggest that \aco\ is not dependent on
  the physical properties of the gas, but being optically thick, this
  line does not have any simple relationship with anything we can
  empirically determine. Similarly, we have shown that \Q\ is not easy
  to determine per galaxy, but its range is small enough to have no
  significant impact on our calibration study.}. To estimate gas mass
from \lsub, the mass-weighted dust temperature, \mwtd, is required.
\mwtd\ has been set to 25\,{\sc k} in previous studies
\citep[e.g.][]{Scoville2014,Scoville2016,Hughes2017}, adding to the
uncertainty in gas-mass estimates for individual galaxies. However, as
we wish to study trends in the conversion factors, we are concerned
about the possible effects of systematic trends in \mwtd, since these
may affect the resulting behaviour of the conversion factors if
ignored.

Having determined empirical relationships between $z$, \Lir, SED
colour (\Lir/\lsub) and \mwtd\ in \S\ref{dustS}, we compare the
calibration results using these empirically determined \mwtd\ to the
standard assumption of constant \mwtd=25\,{\sc k} made in the
literature. Fig.~\ref{fixQT_lirF} shows the impact of using our
empirical relations (coloured points), versus keeping \mwtd\ fixed
(grey points). Each panel shows one of the affected conversion
factors derived from either the ad or Xd samples. The trends with
luminosity -- visible for our default prescription -- disappear when
a constant $\mwtd=25$\,{\sc k} is used.

\begin{figure}
	\includegraphics[width=0.48\textwidth,trim=0cm 0cm 0cm 0cm, clip=true]{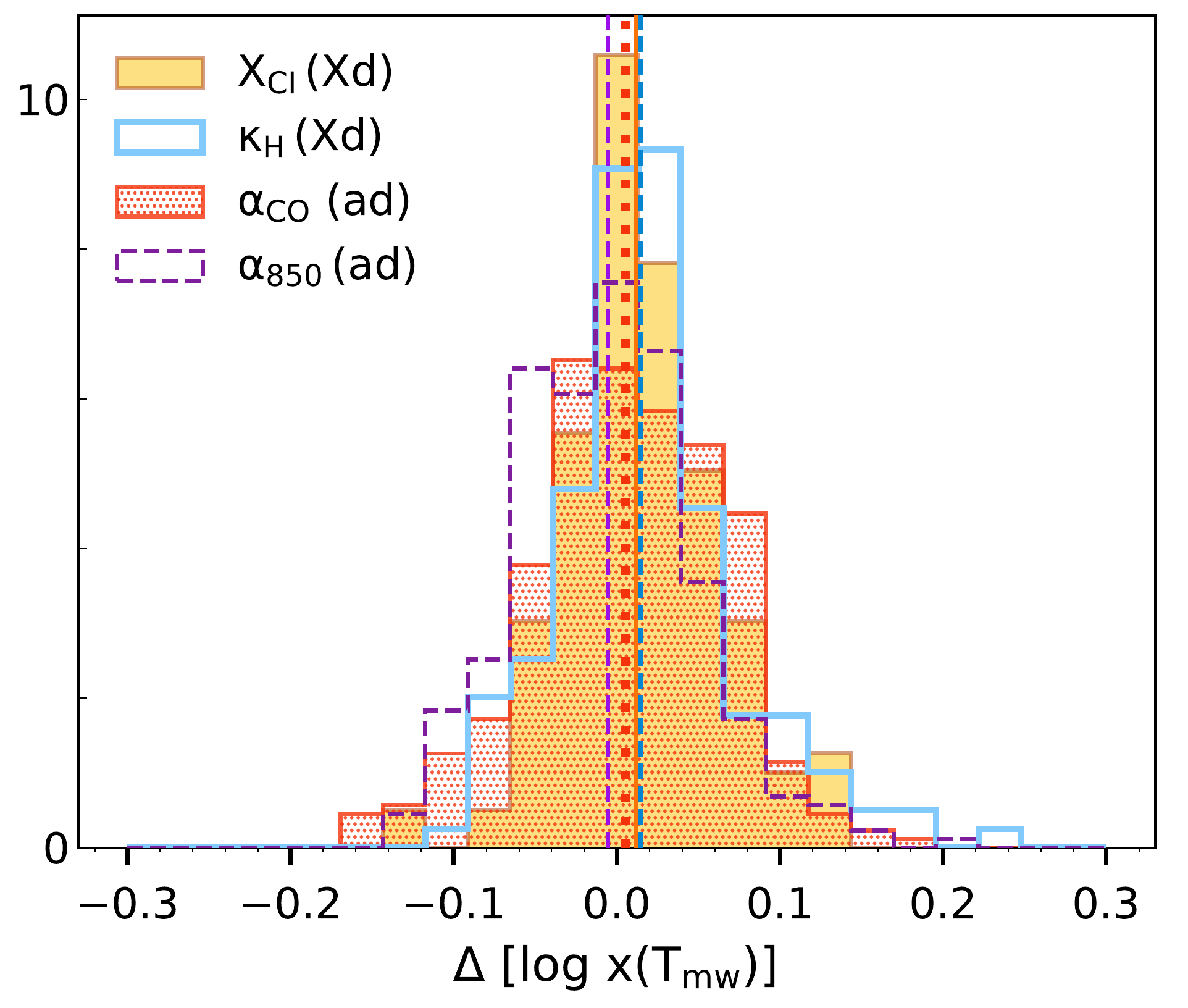}
	\caption{Log difference between the conversion factors
          (\Xci,\gdr,\aco, \asub) using the empirical relation for
          \mwtd, compared to a constant \mwtd\ of 25\,{\sc k}. The median
          log difference for each conversion factor is shown as a
          vertical line, all are $<0.015$\,dex, meaning the choice of
          \mwtd\ for the dust does not have a significant impact on
          the overall average values derived from this study. }
\label{delcalqu_histF}
\end{figure}

\begin{figure}
\includegraphics[width=0.48\textwidth,trim=0cm 0cm 0cm 0cm,
  clip=true]{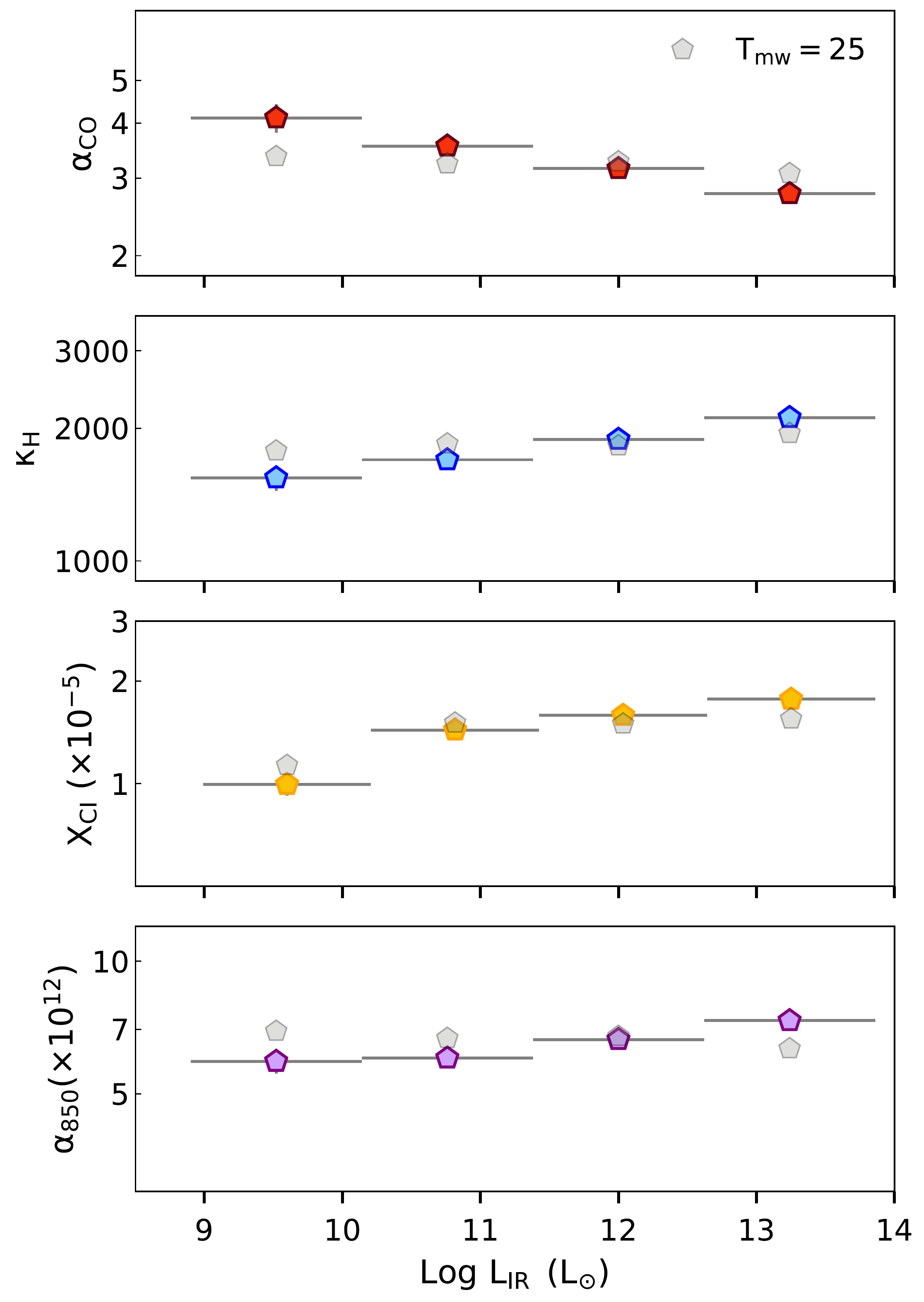}
  \caption{Running means of the conversion factors as a function
    of \Lir. The solid coloured points use the variable \mwtd, as
    described in \S\ref{dustS}. The grey points represent the same
    galaxies and the same method but this time with $\mwtd=25$\,{\sc
      k}. Errors are standard errors on the mean. To better sample the
    full luminosity range, we have used the ad sample for this
    analysis, except for the \Xci\ panel which uses the Xd sample,
    excluding the \CIcor\ galaxies. The important finding is that the
    trends in conversion factors with \Lir\ disappear when a constant
    \mwtd\ is assumed.}
\label{fixQT_lirF} 
\end{figure}

The histogram of the offsets in each conversion factor when using
the empirical \mwtd\ compared to constant $\mwtd=25$\,{\sc k}
(Fig.~\ref{delcalqu_histF}) shows that the choice of \mwtd\ makes no
significant difference to the median values of the parameters
($<0.015$\,dex). For individual galaxies, the average uncertainty
introduced by using a constant \mwtd\ is 0.046--0.06\,dex (1$\sigma$),
with a maximum of $\sim 0.2$\,dex.

\begin{figure}
\includegraphics[width=0.48\textwidth,trim=0cm 0cm 0cm 0cm, clip=true]{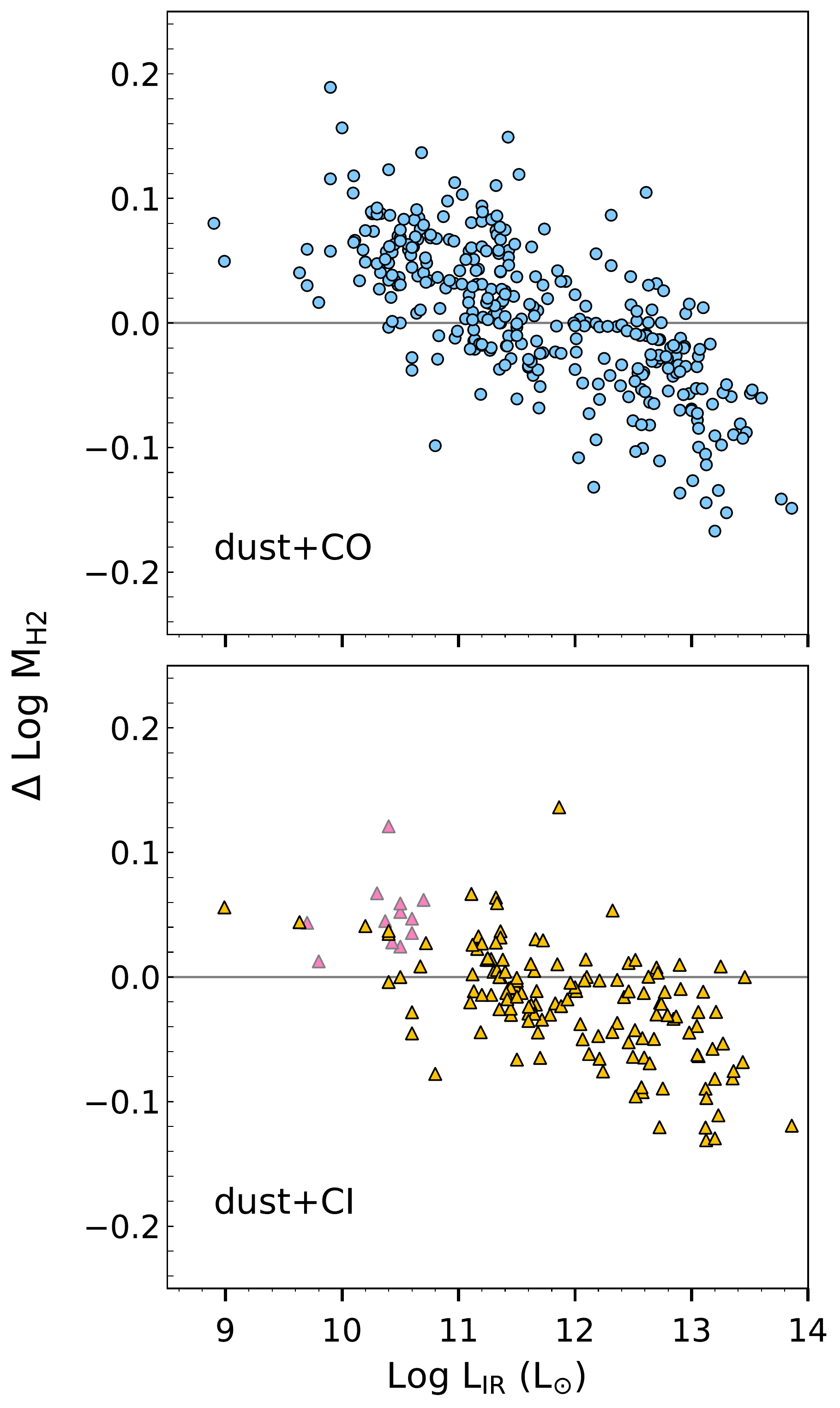}
\caption{Difference in inferred \Mh\ when using the empirical
  prescription for \mwtd\ (\S\ref{dustS}) compared to constant
  $\mwtd=25$\,{\sc k}. The top panel represents gas mass derived for
  the ad sample; the lower panel uses Xd. The pink triangles represent
  the \CIcor\ galaxies. Using a variable \mwtd\ compared to a constant
  25\,{\sc k} produces higher [lower] gas masses at \Llo\ [log $L_{\rm IR}$ > 12]
  by up to 0.1\,dex. This is not a significant issue given all other
  uncertainties affecting gas mass estimates.}
\label{delMhQT_lirF} 
\end{figure}

Finally, the difference in the gas-mass estimates, \Mh, when using
constant \mwtd\ versus our empirical prescription is shown in
Fig.~\ref{delMhQT_lirF}. At lower \Lir, a constant \mwtd\ produces
lower \Mh\ compared to our empirical method, because these galaxies
are local disks which tend to have colder diffuse dust
temperatures. At higher log \Lir$>12$, the trend reverses as the
diffuse dust temperatures increase to $\sim30$\,{\sc k}.

\section{A robust Orthogonal Distance Regression algorithm}
\label{ODRS}

In order to fit the most robust linear model to the data, we have
employed an Orthogonal Distance Regression and included intrinsic
scatter. \footnote{These ideas are outlined in \citet{Hogg2010} and
\citet{dfmplane}, however both of their Bayesian implementations
result in biases in the estimate slope. The biases are quite
pronounced when the range sampled by the data is not much larger than
the errors on the data, but are significant even when the range
sampled is $\sim$10$\sigma$. The biases also depend on which axis is
chosen as the ``true'' independent variable and whether the errors are asymmetrical ($\sigma_x \ll \sigma_y$, or $\sigma_x \gg \sigma_y$). We found that an
ODR which does not use the Bayesian likelihood formalism is the only
one which does not have such biases; hence our choice to use it here.}

We use the {\sc emcee} MCMC sampler \citep{DFM2013} to explore the
$\chi^2$ space and compute robust confidence intervals. Our algorithm
results in parameters which are symmetric under transformation of $x$
and $y$, allowing us to utilise the full co-variance matrix,
including the intrinsic scatter as a third variable.

The MCMC is set up to explore the following likelihood function:

\begin{equation}
Ln L = -0.5\sum^{N}_{i=1}(\Delta^2/\sigma^2 + \ln(\sigma^2/S_2))
\end{equation}

\[
\Delta = {\bf v.Z} - b\, \cos(\theta)
\] 
with ${\bf Z}$ as the data array of $x$ and $y$ values, $b$ as the intercept
and $\theta$ related to the slope as $m = tan(\theta)$. $v$ is a
matrix to rotate to find the perpendicular distances, given by ${\bf v}=
[-sin(\theta), cos(\theta)]$.

\[
\sigma^2 = ({\bf S+\Lambda_{\rm m}.v}).{\bf v} 
\] 
where ${\bf S}$ is the co-variance matrix. To include intrinsic
scatter in the orthogonal direction, as well as measurement errors
into the fitting, we add a term to the co-variance matrix, as
suggested in \citet{dfmplane}:

\begin{equation}
{\bf \Lambda_{\rm m}} = 
\begin{pmatrix}
\tan(\theta)^2 & -\tan(\theta)\\
-\tan(\theta) & 1.0\\
\end{pmatrix}
\times \cos(\theta)^2 \times e^{2\,\ln(\lambda)}
\end{equation}

\[
S_2 = ({\bf S.v}).{\bf v}
\]

The initial conditions were given by the ordinary least-squares fit
parameters for variance in the $y$ direction. The run was checked to
ensure adequate burn-in and independence between samples. We used 32
random walkers with 6,000 steps each.
\label{lastpage}
\end{document}